\documentclass[12pt]{JHEP3}
\newcommand{\intd}{\mathrm{d}}
\newcommand{\ex}{\mathrm{e}}

\usepackage{amsmath,epsfig}
\usepackage{amssymb,amsfonts}
\usepackage{latexsym}
\usepackage{slashed}
\usepackage{empheq}
\numberwithin{equation}{section}
\usepackage{dsfont}
\usepackage{cancel}
\usepackage{overpic}
\usepackage{subcaption}
\usepackage{subeqnarray}
\usepackage{xcolor}
\let\normalcolor\relax
\usepackage{graphicx}
\usepackage{amsmath}
\usepackage{longtable}
\usepackage{color}
\usepackage{bbm}

\usepackage[force]{feynmp-auto}

\allowdisplaybreaks

\usepackage[multiple]{footmisc}

\newcommand{\exclude}[1]{}
\def\nn{\nonumber}

\def\d{\mathrm{d}}

\def\a#1{\alpha_{#1}}

\def\beq{\begin{equation}}
\def\eeq{\end{equation}}
\def\be{\begin{equation}}
\def\ee{\end{equation}}
\def\bea{\begin{eqnarray}}
\def\eea{\end{eqnarray}}
\def\bal{\begin{align}}
\def\eal{\end{align}}

\def\td{\tilde}

\newcommand{\bigzero}{\mbox{\normalfont\Large\bfseries 0}}

\def\2b2[#1,#2][#3,#4]{\left( \begin{array}{cc} #1 & #2 \\ #3 & #4 \end{array}
\right)}
\def\3b3[#1,#2,#3][#4,#5,#6][#7,#8,#9]{\left( \begin{array}{ccc} #1 & #2 #3 \\
#4 & #5 & #6\\#7&#8&#9\end{array} \right)}

\newcommand{\ka}{\kappa}

\newcommand\fverb{\setbox\pippobox=\hbox\bgroup\verb}
\newcommand\fverbdo{\egroup\medskip\noindent%
                        \fbox{\unhbox\pippobox}\ }
\newcommand\fverbit{\egroup\item[\fbox{\unhbox\pippobox}]}

\newcommand{\bear}{\begin{eqnarray}}

\newcommand{\eear}{\end{eqnarray}}

\newcommand{\bsea}{\begin{subeqnarray}}
\newcommand{\esea}{\end{subeqnarray}}
\newbox\pippobox

\def\d{\delta}
\def\g{\gamma}

\def\6{\partial}
\def\a{\alpha}
\def\nn{\nonumber}

\def\e{\epsilon}
\def\m{\mu}
\def\n{\nu}

\def\s{\sigma}

\def\sp{\;\;\;,\;\;\;}

\def\sq
\def\a{\alpha}
\def\b{\beta}
\def\l{\lambda}

\def\hri#1#2{\href{http://arxiv.org/abs/#1}{[ArXiv:#1]#2}}
\def\hre#1#2{\href{http://arxiv.org/abs/#1/#2}{[ArXiv:#1/#2]}}
\def\hrj#1#2{\href{https://doi.org/#1}{#2}}

\def\e{\epsilon}

\def\d{\delta}

\def\D{\Delta}

\def\OO{{\cal O}}

\title{Holographic neutrino transport in dense strongly-coupled matter}
\author{
 M. J\"arvinen$^\sharp$, E. Kiritsis$^\natural$$^\flat$,  F. Nitti$^\natural$, E. Pr\'eau$^\natural$
~\\
~\\
$^\sharp$ \href{https://www.apctp.org}{Asia Pacific Center for Theoretical Physics}, Pohang 37673, Republic of Korea, and
 \href{https://pheng.postech.ac.kr/}{Department of Physics}, Pohang University of Science and Technology, Pohang 37673, Republic of Korea\\
~\\
$^\natural$ \href{http://www.apc.univ-paris7.fr}{Universit\'e  Paris Cit\'e, CNRS, Astroparticule et Cosmologie,  F-75006 Paris, France}\\
~\\
$^\flat$ \href{http://hep.physics.uoc.gr}{Crete Center for Theoretical Physics}, Institute for Theoretical and Computational Physics,
Department of Physics,  P.O. Box 2208,\\
University of Crete, 70013, Heraklion, Greece
}

\preprint{APCTP Pre2023 - 004\\
CCTP-2023-3\\
ITCP-2023/3}

\abstract{A (toy) model for cold and luke-warm strongly-coupled nuclear matter at finite baryon density, is used to study neutrino transport.
The complete charged current two-point correlators are computed in the strongly-coupled medium and their impact on neutrino transport is analyzed.
The full result is compared with various approximations for the current correlators and the distributions, including the degenerate approximation, the hydrodynamic approximation as well as the diffusive approximation and we comment on their successes. Further improvements are discussed.
  }

\begin{document}
\maketitle

\section{Introduction}

\label{Intro}

Neutrino transport plays a pivotal role in various astrophysical processes involving dense QCD matter.
The most studied class of such processes are the core collapse supernovae which occur in the last stages of the lifetime of massive stars.  Neutrino-driven heating and turbulence are crucial ingredients in the complex dynamics that leads to the explosion of the star (see the reviews~\cite{Janka:2012sb,Burrows:2020qrp}). Strong explosions, similar to what is observed in nature, are only obtained in simulations that properly take into account these ingredients.

Neutrino interactions are also important in the physics of neutron stars.
Right after a neutron star is formed in a supernova explosion, its temperature is comparable to the QCD scale, and subsequently cools down due to neutrinos emitted by various processes~\cite{Yakovlev:2004iq,Potekhin:2015qsa}.
Neutrino cooling is the main mechanism for the first $\sim 10^5$ years (after which photon cooling dominates).

Apart from a supernova remnant, a hot neutron star can also be formed as a product of a binary neutron star collision. The observation of gravitational waves from the merger event GW170817 by the LIGO and Virgo collaborations together with the analysis of the electromagnetic signal from the kilonova has boosted the interest in neutron stars 
recently~\cite{LIGOScientific:2017vwq,LIGOScientific:2017ync}. State-of-the-art neutron star binary merger simulations are now developing towards a stage where the effects of neutrino transport are included~\cite{Sekiguchi:2011zd,Foucart:2016rxm}. While it can be estimated that this effect is relatively small in the actual merger phase, neutrino emission affects significantly the evolution of the hypermassive neutron star after the merger within timescales accessible in simulations~\cite{Foucart:2022bth}.
This is expected to hold even for a class of events where a black hole is formed: analysis of the electromagnetic signal from GW170817 suggests that a collapse to a black hole took place about one second after the merger in this event~\cite{Gill:2019bvq}, which is easily long enough for neutrino effects to matter. Moreover, neutrinos affect drastically the composition of the ejecta, the evolution of the torus, and the jet formation after the merger~\cite{Cusinato:2021zin}.

The importance of neutrino transport has sparked a wide literature studying the neutrino emission rates and opacities in dense matter.  Most theoretical studies of neutrino transport focus on the nuclear matter phase, which is natural as most of the matter in neutron stars and in the collapsing core in the supernova process is known to be in this phase, and various effective theory tools are available for nuclear matter. There is a vast literature on this topic, see~\cite{Yakovlev:2000jp,Chamel:2008ca, Schmitt:2017efp, Sedrakian06}.

Computing the emissivities and opacities at high densities boils down to computing the correlators of the currents of the weak interactions in the strongly interacting QCD matter. Standard methods for estimating these correlators include the use of mean-field theory~\cite{Reddy:1997yr} and the addition of correlation effects through ring resummation, i.e., the random phase approximation~\cite{Burrows:1998cg,Reddy:1998hb} and its improvements (see, e.g.,~\cite{Margueron:2003fq,Margueron:2006wh}).
Results in various limits and approximation schemes, such as the degenerate limit and the ``elastic'' approximation where the recoil of the nucleon is neglected~\cite{Bruenn:1985en}, have been worked out.

However, for the highest densities reached in core collapse supernovae or in the cores of massive neutron stars, other phases than regular nuclear matter may appear. Phase transitions may play an important role in neutrino transport: while the equation of state typically changes modestly at phase boundaries (e.g. densities may jump by an $\mathcal{O}(1)$ factor at a first order transition), observables related to transport can easily change by orders of magnitude. Perhaps the most natural transition to consider is the transition from nuclear matter to quark matter, where neutrino emissivities are expected to be larger than in regular nuclear matter by orders of magnitude. This is indicated by analyses both in the ungapped regime~\cite{Iwamoto:1980eb,Burrows:1980ec,Iwamoto:1982zz,Dai:1995uj,Schafer:2004jp}, and in color-superconducting phases \cite{Reddy:2002xc,Jaikumar:2002vg,Jaikumar:2005hy,Alford:2007xm}.
But estimates for neutrino transport are also available in phases with pion~\cite{Bahcall:1965zz,Bahcall:1965zzb,Maxwell:1977zz} or kaon condensates~\cite{Brown:1988ik,Thorsson:1993bu} and nuclear matter with superfluidity~\cite{Flowers:1976ux,Steiner:2008qz}.

At high densities in QCD, i.e., densities well above the nuclear saturation density $n_s \approx 0.16$ fm$^{-3}$, all these results include however sizable or uncontrolled uncertainties. This happens because first-principles methods are not reliable in this region of the phase diagram. For the equation of state, loop expansions in chiral perturbation theory for pure neutron matter converge below $n \approx 2n_s$~\cite{Tews:2018iwm}, while perturbation theory requires densities above $n \approx 40 n_s$ to be reliable~\cite{Kurkela:2009gj}, which is clearly higher than the densities reached in neutron star cores. The uncertainty of the equation of state~\cite{Annala:2017llu,Annala:2021gom,Altiparmak:2022bke} readily affects the estimates of neutrino opacities and emissivities, and approximations used in the computation of the current-current correlators bring in additional uncertainty. The importance of the uncertainties in the densest regions is enhanced because the neutrino interactions with QCD matter become significantly stronger with increasing density.  In the absence of reliable first-principle methods, it is therefore useful to analyze neutrino transport in this region by alternative and complementary approaches, such as the gauge/gravity duality.

The gauge/gravity duality (or ``holography'') is a general tool for analyzing strongly coupled gauge theories such as QCD. In this method, following the original AdS/CFT conjecture, the strongly coupled regime of QCD is mapped to a classical higher dimensional gravitational theory. While the precise correspondence between QCD and such a higher dimensional theory is not known, the method has proved to be useful to study the properties of QCD, in particular the properties of hot QCD plasma produced in heavy ion collisions. Examples include the description of the far-from-equilibrium dynamics right after the collision~\cite{Heller:2011ju,Heller:2013fn}, and the famous estimate for the shear viscosity in strongly coupled plasma~\cite{Policastro:2001yc,Kovtun:2004de}. 
Naturally, transport in hot quark gluon plasma at low densities has been studied also in holographic models that mimic properties of QCD more closely, see for example~\cite{I1}-\cite{Iatrakis:2014txa}.

There has also been a considerable interest in studying dense matter by using the gauge/gravity correspondence (see recent reviews~\cite{Jarvinen:2021jbd,Hoyos:2021uff}).
The equation of state of QCD matter at high density has been analyzed both in the nuclear matter~\cite{Ishii:2019gta,Ghoroku:2021fos,Kovensky:2021kzl,Demircik:2021zll,Bartolini:2022rkl} and quark matter~\cite{Hoyos:2016zke,Jokela:2018ers,BitaghsirFadafan:2019ofb,Mamani:2020pks} phases as well as in more exotic phases~\cite{Ghoroku:2019trx,Kovensky:2020xif,Pinkanjanarod:2020mgi,BitaghsirFadafan:2020otb,Kovensky:2023mye}, aiming at applications in neutron star physics. Also transport coefficients in quark matter, i.e., viscosities and conductivities, have been estimated~\cite{Hoyos:2020hmq,Hoyos:2021njg}.

In this article, we initiate the holographic study of neutrino transport. We consider a simple holographic model, based on an Einstein-Yang-Mills action. In this model, charged black hole geometries (Reissner-Nordstr\"om black holes) are interpreted as the dual of dense unpaired quark matter in QCD. We analyze charged current interactions in holographic matter, leading to estimates for the emission and absorption of neutrinos. Neutral current interactions (neutrino scattering) will be discussed in future work.

\subsection{Summary of results}

The transport of neutrinos is described by the Boltzmann equation\footnote{Written in flat space here, for simplicity.} for the neutrino distribution $f_\n$
\be
\nn (K_\n\cdot \partial)f_\n = j(E_\n) (1-f_\n) - \frac{1}{\l(E_\n)}f_\n \, ,
\ee
with $K_\n$ the on-shell neutrino 4-momentum and $E_\n$ the neutrino energy. The radiative coefficients $j$ and $\l$ are properties of the medium: $j$ is the neutrino emissivity and $\l$ the mean free path. As reviewed in the first section of this work, the calculation of the neutrino radiative coefficients in a neutron star requires the knowledge of the chiral current two-point function in dense QCD matter. Computing this correlator is a strongly-coupled issue, which remains unsolved.  As mentioned in the introduction, several approximations have been considered in the literature but these remain highly model-dependent.

In this work, the approach that we follow is to compute the chiral current two-point function holographically, using the simplest holographic model where this calculation can be done. We focus here on the charged current contribution, leaving the analysis of the neutral current for future work.
The model contains many of the properties that are expected from a quark-gluon plasma at finite density but also has simplifications that are unphysical. It has an underlying scaling symmetry in the absence of baryon density, and the mechanism of chiral symmetry breaking is not (yet) implemented. Therefore our calculation should be considered as a first step towards performing this calculation in a successful theory, like holographic V-QCD, \cite{V1}.

An important property of holographic strongly coupled theories at finite density is the following: although scaling symmetries are broken by the finite density, there is an emergent one-dimensional scaling symmetry at zero temperature which is also accompanied by a large density of states at very low energies, \cite{malda} and can even be responsible for glassy behaviour, \cite{Biroli}. This symmetry is associated to an AdS$_2$ factor\footnote{AdS$_d$ stands for the anti-de Sitter geometry in d space-time dimensions. AdS$_d$ is a constant negative curvature manifold with infinite volume and maximal O(2,d-1) symmetry.}  in the geometry of the relevant black hole. Such a regime exists in our theory  and it is the one that  controls most of the  calculation. It has been also seen in the phenomenologically successful and more complete model of V-QCD, in \cite{V4}.
It is an interesting question,  that we do not address in this paper, to investigate what are the signals of this behavior,
 in both neutrino transport as well as the dynamics of neutron star mergers.

The holographic toy model that we consider is a bottom-up model where, in addition to the metric, the 5-dimensional bulk contains  gauge fields belonging to the $U(N_f)_L\times U(N_f)_R$ flavor group
  dual to the field theory chiral current operators. The bulk holographic action which controls the dynamics of these fields, is the Einstein-Yang-Mills action
\be
\nn S = S_c + S_f \, .
\ee
\be
\nn S_{\text{c}} = M^3N_c^2 \int\mathrm{d}^5x\sqrt{-g}\, \left(R + \frac{12}{\ell^2}\right) \, ,
\ee
\be
\nn S_f = - \frac{1}{8\ell}(M\ell)^3w_0^2 N_c \int\mathrm{d}^5x\sqrt{-g} \left(\text{Tr}\,\mathbf{F}_{MN}^{(L)}\mathbf{F}^{MN,(L)} + \text{Tr}\,\mathbf{F}_{MN}^{(R)}\mathbf{F}^{MN,(R)} \right) \, ,
\ee
where $\mathbf{F}^{(L/R)}$ is the field strength for the chiral gauge fields, $N_c$ the number of colors and $\ell$ the AdS length. The model has two dimensionless parameters, $M\ell$ and $w_0$ and these enter in the physics of the dual, strongly-coupled quantum field theory. As detailed in Appendix \ref{Sec:param}, the parameters are fixed to match the lattice result for the QCD thermodynamics in the deconfined phase at low baryon density.

The background solution is the gravitational  dual of a medium composed of quark matter at equilibrium, at finite temperature $T$ and quark number chemical potential $\m$. We consider isospin symmetric matter, with isospin chemical potential
\be
\nn \m_3 = 0 \, .
\ee
Neutron star matter is known to be far from isospin symmetry, as it contains many more neutrons than protons. This isospin asymmetry may have a significant influence on the transport of neutrinos, as can be seen for example in the condition for $\b$-equilibrium below. In this work, we restrict to the isospin symmetric case for the sake of simplicity\footnote{As we discuss later in this introduction, a non-zero $\mu_3$ may change non-trivially the background solution and phase, and makes the computation of current correlators more involved.}, leaving the study of isospin imbalance  for future work. The corresponding gravitational dual corresponds to a charged AdS black-hole with a charge proportional to $\m$.

To this strongly-coupled medium, we add neutrinos and electrons so that we have full charge neutrality. The neutrinos that scatter in this medium are assumed to be sufficiently close to equilibrium for the chemical potential $\m_\n$ to be well defined, and at $\b-$equilibrium with the quarks and electrons
\be
\nn \m_\n = \m_e + \m_3 = \m_e \, .
\ee
However, the distribution of neutrinos is generically different from the equilibrium distribution. In particular, it is expected that there is generically no Fermi surface with  chemical potential  $\m_\n$ associated to it.

Following the usual holographic procedure, the charged current two-point function is then evaluated from the solution to the equations of motion for gauge field perturbations on the black-hole background. The correlators are computed numerically for energies $\omega$ and momenta $k$ between $0$ and a few times $r_H^{-1}$, $r_H$ being the horizon radius.

From the numerical solution for the chiral current correlator, the neutrino charged current radiative coefficients can be computed as a function of the neutrino energy. This calculation is the main result of this work. When completed with the neutral current coefficients, it can be used to simulate numerically the transport of neutrinos, in the kind of quark matter described by our holographic model. The numerical results are discussed in detail in Section \ref{Sec:res}. Here, we give a summary of this analysis.

\subsubsection*{Approximations}

We considered in this work several approximations, in which the radiative coefficients have simpler expressions. Apart from academic interest, these approximations are useful to obtain a better qualitative understanding of the exact numerical results. We list them below
\begin{itemize}
\item \emph{The degenerate approximation}, where the expressions in the limit of $\mu \gg T$ are used for the Bose-Einstein and Fermi-Dirac equilibrium distributions $n_b$ and $n_f$. In this limit, the weak processes which contribute to neutrino transport are clearly identified. This is summarized in table \ref{tab:cEn}.

\item \emph{The hydrodynamic approximation}, where the radiative coefficients are computed by expanding the charged current 2-point function $G^R_c(\omega,k)$ at leading order in the hydrodynamic expansion, that is at leading order in $r_H\omega$ and $r_Hk$. Note that, at $\m\gg T$, the parameters of the expansion are $\omega/\m$ and $k/\m$, rather than $\omega/T$ and $k/T$.  As reviewed in Section \ref{Sec:cfT}, this emergence of a hydrodynamic behavior at low temperature is a consequence of the AdS$_2$ geometry of an extremal horizon.

 In relation to this, a cautionary note is relevant here.  Typically, as $T\to 0$, like in the case studied here, hydrodynamics is known to break down as the non-hydrodynamic poles of the energy-momentum tensor correlators are moving towards zero energy and momentum and eventually collide with the hydrodynamic poles. However, in an AdS$_2$ regime, like the one encountered here, there are indications that there is   a kind of hydrodynamics that survives,~\cite{DP13}. The relevant poles of the correlators were studied in a toy model in  \cite{DP14}. An infinite lattice of equidistant poles were found that seemed to collide once in a while with the hydrodynamic pole.
    However, we have found that the presence of such an infinite lattice of poles do not seem to affect our two-point correlator of currents and its leading hydrodynamic behaviour. We suspect that this is because the residues of these poles are very small in this regime. A calculation of these residues,  in a large d limit, using the framework of  \cite{Gursoy:2021vpu} seems to corroborate this expectation. It is however a topic that may be important more generally, and more is needed to understand it fully.

The leading order hydrodynamic expression for the retarded correlator of the charged currents, is given by
\be
\nn G^{R}_{c,\l\s}(\omega,k) = -i\s \left(P_{\l\s}^\perp(\omega,k) \omega + P^\parallel_{\l\s}(\omega,k)\frac{\omega^2-k^2}{\omega+iDk^2}  \right) \, ,
\ee
where $P^\perp$ and $P^\parallel$ are respectively the projectors transverse and longitudinal to the 3-momentum $k$.  In this approximation, the only strongly-coupled calculation required to determine the charged current correlator, is that of the two transport coefficients: the conductivity $\s$  and the diffusivity $D$. In the simple holographic model considered in this work, analytic expressions can be derived for the transport coefficients
\be
\nn \s = \frac{|M_{ud}|^2}{8r_H}N_c(M\ell)^3w_0^3 \sp D = \frac{1}{2}r_H \, ,
\ee
where $M_{ud}$ is the $ud$ component of the Cabibbo-Kobayashi-Maskawa (CKM) matrix, $M\ell$, $w_0$ and $N_c$ are the parameters of the theory and $r_H$ is the horizon radius of the dual black hole, which is an explicit function of the baryon density and temperature.

The hydrodynamic approximation approaches the exact result  in the limit where all the leptonic energies $\mu_e, \mu_\n, E_\n$ are much smaller than $r_H^{-1}$. At $\m \gg T$, $r_H\m_e$ (and $r_H\m_\n$) is found to asymptote  to a constant close to 1, \eqref{A1}. For $r_H E_\n$ also smaller than one, the leptonic energies are therefore at the limit of the regime of validity of the hydrodynamic approximation. As we summarize in more details below, this approximation typically gives a good qualitative description of the radiative coefficients, but its accuracy is around a few tens percent.

\item \emph{The diffusive approximation}, where the hydrodynamic approximation is used, and it is further assumed that the dominant contribution to the radiative coefficients comes from the time-time component of the retarded 2-point function. We show in Appendix \ref{Sec:DiffApp} that this approximation is valid in the degenerate and hydrodynamic regime
\be
\nn T \ll \m_e,\m_\n,E_\n \ll \m \, .
\ee
We use the diffusive and degenerate approximation to derive approximate expressions for the opacities
\be
\nn \kappa(E_\n) \equiv j(E_\n) + \frac{1}{\l(E_\n)} \, .
\ee
The results are shown in \eqref{A8} for neutrinos, and in \eqref{A11}-\eqref{A12} for anti-neutrinos. When it comes to describing the actual numerical results, these approximate expressions were found to be inaccurate. However, the expression at $E_\n = 0$ \eqref{A13}
\be
\nn \ka_{e,0} = \frac{G_F^2\s}{\pi^2}
\m_\n^4 \, ,
\ee
which originates fully in the transverse part of the correlator, was found to be in good agreement with the exact result for baryon densities $n_B \gtrsim 10^{-2}\,\mathrm{fm^{-3}}$. \eqref{A13} is therefore a good estimate of the typical scale of the opacities as a function of the baryon density.

\end{itemize}
\subsubsection*{Properties of the radiative coefficients}

We now summarize the main properties of the numerical solution for the neutrino radiative coefficients, that are presented in Section \ref{Sec:res}. The radiative coefficients depend on several parameters: the neutrino energy $E_\n$, but also the parameters of the theory $M\ell$ and $w_0$, as well as the environmental parameters $\m$ and $T$. As detailed in Appendix \ref{Sec:param}, $M \ell$ and $w_0$ are fixed by matching the zero-density thermodynamics of the model to the free quark-gluon plasma result. As for the environmental parameters, the temperature is fixed to values typical of young neutron stars\footnote{Of course, the temperature is not constant as a function of the distance to the center of the star. However, the typical relative variation is of order 1.} \cite{Oertel:2020pcg}
\be
\nn T = 10\,\text{MeV} \, ,
\ee
and we investigate the remaining 2-dimensional parameter space, of neutrino energy $E_\n$ and baryon density $n_B$. We consider regimes of energy and density that are typical of transport in a neutron star, i.e. energies below ten times the temperature \cite{Oertel:2020pcg}, and densities much larger than the thermal scale
\be
\nn E_\n \lesssim 10\,T \sp n_B \gg T^3 \, .
\ee

We first discuss the charged current polarization functions, which are the direct outcome of the holographic calculations. The polarization functions at $n_B = 1\,\mathrm{fm^{-3}}$ are shown in figure \ref{Fig:ImP}. The comparison with the leading order hydrodynamic prediction reveals that the polarization functions look qualitatively similar to the hydrodynamic expressions. We also evaluate quantitatively the difference between the two, focusing on the region of the energy-momentum space which is relevant to the calculation of the neutrino opacities. As a result, we find that the error from the hydrodynamic approximation to the transverse polarization function is comprised between $0$ and about $50\%$, whereas for the longitudinal part it ranges from $0$ to about $100\%$.

The next step of the analysis in section \ref{Sec:res} is the discussion of the neutrino opacities themselves. The latter were computed numerically for a whole range of baryon densities $n_B$ and neutrino energy $E_\n$, for values typical of transport in a neutron star
\be
\nn 10^{-3}\,\mathrm{fm^{-3}} \leq n_B \leq 1\,\mathrm{fm^{-3}} \sp 0 \leq E_\n \leq 100\,\text{MeV} \, .
\ee
The solutions at the two extreme values of the density, $n_B = 10^{-3}\,\mathrm{fm^{-3}}$ and $n_B = 1\,\mathrm{fm^{-3}}$, are shown as a function of the neutrino energy in figure \ref{Fig:op12}. The general qualitative behavior depends mainly on the statistical factors, so that it agrees with other calculations discussed in the literature: the opacities increase with both the density and the neutrino energy. When the baryon chemical potential is much larger than the temperature, a dip is observed in the neutrino opacity for energies close to the neutrino chemical potential $\m_\n$.

\begin{figure}[h]
\begin{center}
\includegraphics[width=0.49\textwidth]{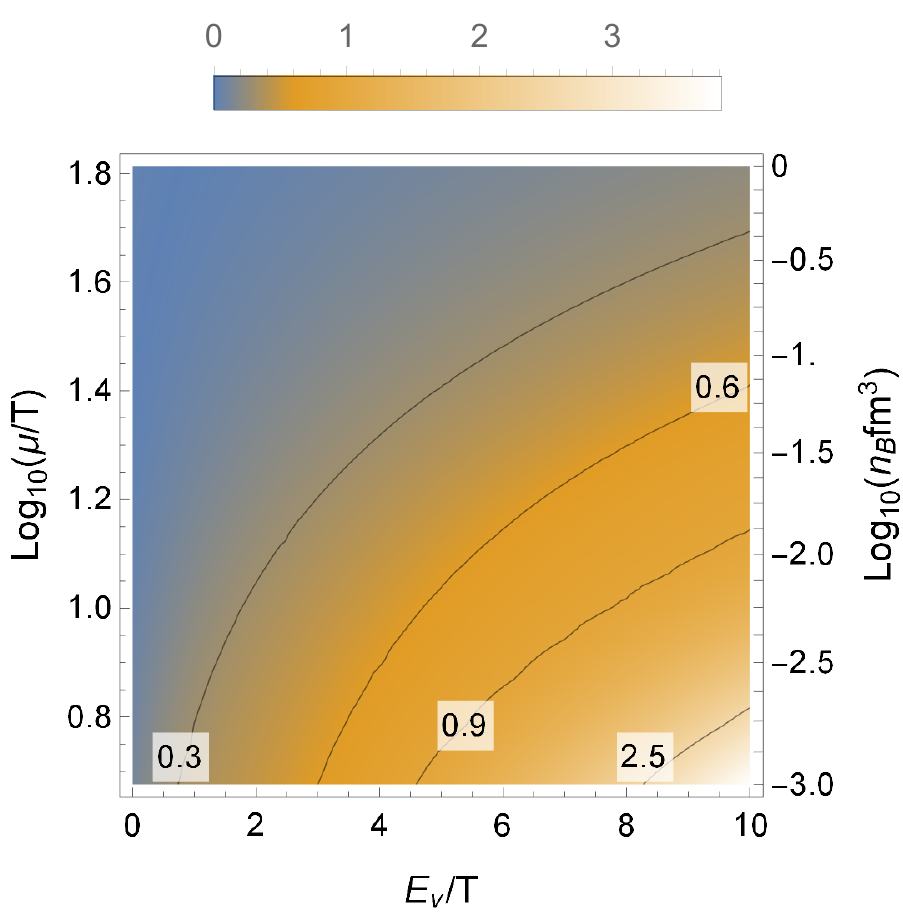}
\includegraphics[width=0.49\textwidth]{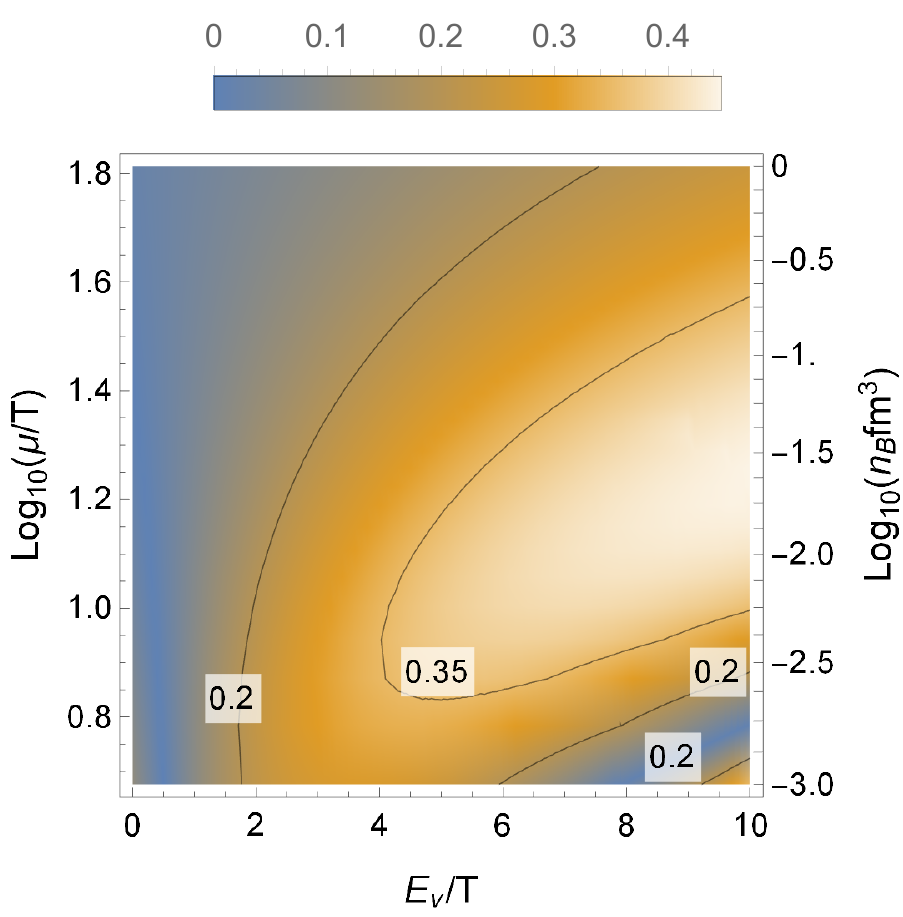}
\caption{Relative difference between the hydrodynamic approximation and the exact opacities, for neutrinos (\textbf{Left}) and anti-neutrinos (\textbf{Right}). }
\label{Fig:RDhyd_sum}
\end{center}
\end{figure}

The accuracy of the various approximations introduced before is then analyzed in detail, over the full parameter space of density and neutrino energy. The corresponding 2-dimensional plots of the relative differences are shown in figures \ref{Fig:RDdegen} to \ref{Fig:RDdiffdegen}.  The main conclusion from this analysis is that the accuracy of the approximations are typically better at high baryonic density. In particular, for $n_B \gtrsim 10^{-1}\,\mathrm{fm^{-3}}\big(T/(10\,\text{MeV})\big)^3$, the hydrodynamic approximation is within $0$ to $50\%$ from the exact neutrino opacity, whereas the error is less than $30\%$ for anti-neutrinos (see figure \ref{Fig:RDhyd_sum}). That is, extracting only the leading order transport coefficients $\s$ and $D$ from the holographic calculation is a sufficient input to obtain a good estimate of the opacities at those densities.

At densities $n_B \lesssim 10^{-2}\,\mathrm{fm^{-3}}\big(T/(10\,\text{MeV})\big)^3$, the hydrodynamic approximation is much cruder, with errors exceeding $100\%$ at high neutrino energy. This means that higher order transport coefficients are required to produce a reasonable approximation. For even smaller densities, as $\m/T$ becomes of order 1, $r_H E_\n$ becomes larger than 1 for $E_\n \gtrsim T$. In this case, the hydrodynamic expansion breaks down, and the full holographic 2-point function is needed to compute the opacities.
The reason for this breakdown is that  the hydrodynamic expansion \eqref{PVtRNrH} and \eqref{PVpRN4} is an expansion in $(r_H \omega, r_H k)$, where $\omega$ and $k$ depend on the leptonic energies in the radiative integrals. In particular, large values of $r_H\omega$ and $r_H k$ are explored for large $r_H E_\n$. This is discussed in detail in section \ref{Sec:App}.

\subsubsection*{Comparison with the literature}

\begin{figure}[h]
\begin{center}
\includegraphics[scale=1.]{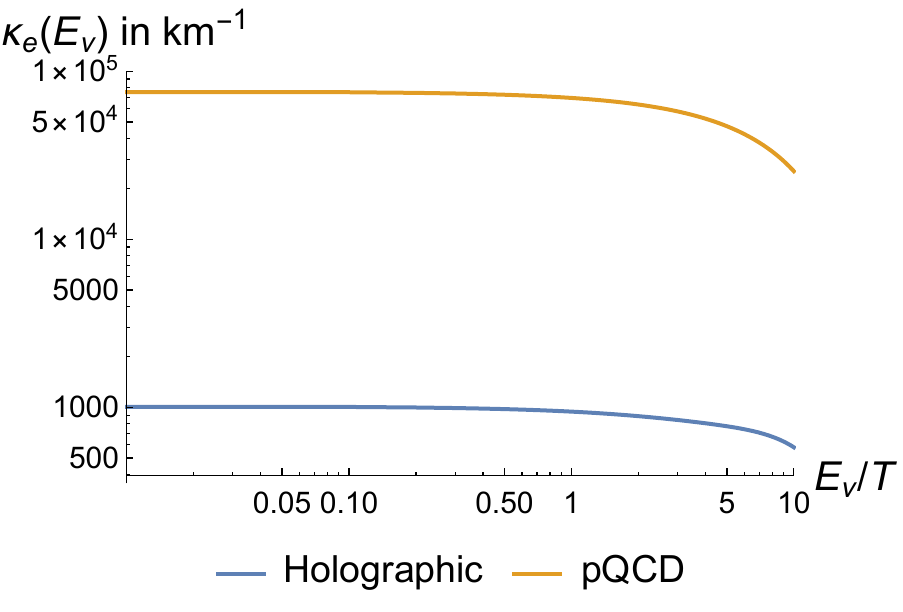}
\caption{Neutrino opacity from our holographic result (blue) compared with the perturbative QCD result \cite{Iwamoto:1982zz} (orange), at $n_B=0.11\,\mathrm{fm^{-3}}$ and  $T=10\,\text{MeV}$. The opacity is expressed in $\mathrm{km^{-1}}$.}
\label{Fig:kpQCD_sum}
\end{center}
\end{figure}

In the last section of this work, our results are compared with other calculations of the neutrino radiative coefficients from the literature. We focus on the recent results in non-relativistic nuclear matter from \cite{Oertel:2020pcg}, and the calculations in weakly-coupled quark matter from \cite{Iwamoto:1982zz}. The opacities computed from the holographic model
\be
\nn \kappa_{e^-}(0) \simeq 6.2\times 10^2\, \mathrm{km^{-1}}\, \left(\frac{n_B}{0.1\,\mathrm{fm^{-3}}}\right)^{\frac{5}{3}} \,   \left(\frac{(M\ell)^3}{(M\ell)^3_{\text{free}}}\right)^{-\frac{1}{2}}\left(\frac{w_0^2(M\ell)^3}{(w_0^2(M\ell)^3)_{\text{free}}}\right)^{\frac{5}{6}} \, ,
\ee
are found to be about an order of magnitude larger than the results from approximate calculations in nuclear matter,
which is about two orders of magnitude smaller than the perturbative result in quark matter (see figure \ref{Fig:kpQCD_sum}). This indicates that, although the holographic matter is deconfined, the strong coupling implies that the neutrino opacity is highly suppressed compared with the perturbative estimate. We caution however that the model we used is not very close  to the real theory and more effort is needed to corroborate the results.

\subsection{Outlook}

The analysis presented in this article can be extended and improved in several ways. As we restricted our analysis to the charged current correlators in this work, an obvious extension is the holographic analysis of the neutral current correlators and their impact on neutrino transport, simply by using the setup described here. As we already mentioned above, this will be the topic of a future publication.

As for the setup, perhaps the most natural place to search for improvements is the holographic model. We shall now provide a summarized description of the physics of the toy model, in order to sketch possible improvements.

The model describes a strongly coupled large-N$_c$ plasma, and N$_f$ quarks with $N_f\sim N_c$ so that they have non-trivial backreaction on the glue dynamics. There is a $U(N_f)\times U(N_f)$ chiral symmetry that is unbroken (no pions here). The theory is conformal like $N=4$ super-Yang-Mills, i.e. at zero temperature and density, the mesons are vectorial and  massless, and the spectrum continuous, as is the case in conformal theories. The spectral density is fixed by conformal invariance. This spectrum is quite similar in many respects (but not all) to what is expected in quark-gluon plasma phases.
The axial U(1)$_A$ is not anomalous here, but also does not enter in the dynamics.

This model is the simplest holographic model
in which the calculation of chiral current 2-point functions at finite baryon density can be performed. There are several directions for its improvement

\begin{itemize}

\item As already mentioned, a next step is to add an isospin chemical potential together with the baryon chemical potential. It is well known that in real world QCD, two different extra phases are possible in such a case. The first, \cite{IsoSon}, is pion condensation, that can be established from the chiral Lagrangian, while the other, \cite{Aha}, is $\rho$-condensation that also breaks the rotational symmetry. In the model we use, there are no pions, but one in principle could have vector meson condensation. This possibility was already found to be realized in \cite{Gubser08a,Gubser08b},  which discussed a similar holographic model but for a three-dimensional (ABJM) theory.
    A mapping of the phases and the determination of the (expected second order) phase transition of the four-dimensional theory is necessary before the calculations at finite baryon and isospin chemical potential is done.

\item A Chern-Simons (CS) term can be added. In the absence of a tachyon, such a term is unique and is the same as in $N=4$ super-Yang-Mills, \cite{book}.
It controls the P-odd structure of the correlators, it generates the chiral anomalies, and may have an interesting impact in the associated neutrino diffusion problem\footnote{The effects of the Chern-Simons term on transport in holographic theories has been discussed for example in \cite{Erdmenger08,Son09,Landsteiner11}.}.
In the vacuum of the holographic theory, the CS term affects correlators of currents starting from the three point functions, while it is explicitly independent of the string coupling constant (dilaton). However, at finite baryon density, it affects also the two-point functions of currents.

\item As flavor is added by adding flavour branes to the glue sector,  \cite{Karch02}, one could also upgrade the five-dimensional flavour action to the DBI action  which include a class of long distance non-linearities of the flavour gauge fields, \cite{book}.

\item  The model used is relatively close to $\mathcal{N}=4$ SU($N_c$) SYM coupled to fundamental flavor fields ($N=2$ hypermultiplets), \cite{Karch02}, which is an interesting model to test our formalism.
This is a top-down model  with $N_f$ D7 branes embedded non-trivially in the ten-dimensional background space-time, $AdS_5\times S^5$. It is conformal to leading order, as the
running of the gauge coupling due to the presence of the hypermultiplets is subleading in $1/N_c$ for $N_f\sim {\cal O}(1)$.
Its associated holographic physics has been analysed in detail, \cite{Kruczenski03,review,un}. Because  the embedding of the flavor branes inside $AdS_5\times S^5$ is non-trivial the flavour gauge fields are subjected to a different open string metric than in the previous items.
In this case, the full DBI action is used as  well as the CS term.

\item A further improvement, but keeping to the top-down nature of the holographic theory, is to use the Sakai-Sugimoto model, \cite{SS}.
In this case the glue sector is confining and non-scale invariant while we have quarks and antiquarks with a chiral symmetry that is similar to QCD with chiral group $U(N_f)_L\times SU(N_f)_R$. This setup is close to QCD, with the only exception that the relevant field that is important for giving mass to the quarks, the open string tachyon is missing.\footnote{Of course, it is part of that theory, but for the relevant configuration, the tachyon string is non-local, \cite{ak}.}

\item The bifundamental open string tachyon, {which in the present setup is not included,} is important in the holographic setup, as it is the order parameters for chiral symmetry breaking in QCD, \cite{ckp},
as well as a way of adding a mass to the quarks. A simple holographic model that   includes the bifundamental open string tachyon, using Sen's string theory action, \cite{Sen03},  was proposed and analyzed in \cite{tachyon1}-\cite{tachyon3}.
The model does extremely well in describing chiral symmetry breaking and meson spectra. It is therefore a good laboratory for testing the calculations of the present paper.

\item The last target calculation involves the holographic theory of V-QCD, \cite{V1}-\cite{V7}, the most complete holographic  model so far to address QCD dynamics in a variety of arenas. It is a semi-phenomenological model for the Veneziano limit of QCD,  \eqref{V1}. It includes all players in the dynamics and for this it is also computationally challenging, although a lot of progress has been seen recently and the model can describe reasonably well, a host of different QCD data, \cite{V7}.

\item So far in all the setups mentioned above, either quarks are all massless or all quarks have the same mass.
In the case studied in this paper, the flavor sector is assumed to contain two massless flavors. In QCD, this would correspond to including only the two lightest flavors up and down, and neglecting their masses. At the densities relevant for neutron stars, it is expected that neglecting the up and down masses is a good approximation. On the other hand, it cannot be excluded that neutron star cores exhibit some degree of strangeness. In order to take into account strange quarks, the current model would need to be extended to include a third massive flavor. Working with an $SU(3)$ chiral group instead of $SU(2)$ is just a matter of algebra, but including quark masses actually requires the bulk theory to include the bifundamental tachyon field $T^{ij}$, with a non-trivial matrix structure. This involves the analogue of the non-abelian DBI action for the tachyon. An example of this was worked out in the appendix of reference \cite{tachyon2}. The formalism needs to be developed so that we can address masses of the strange quark substantially different from those of up or down quarks.

\end{itemize}

Apart from the simplicity of the holographic model, our approach included other approximations and simplifications of the general formalism which are typical in the literature on neutrino transport.
\begin{itemize}
 \item The derivation of the Boltzmann equation for neutrinos used the semiclassical gradient approximation (see~\eqref{gradap} below), which holds when the mean free paths of the neutrinos are much longer than their de Broglie wavelengths. A particular higher order correction includes the effects of the (maximal) breaking of parity by the neutrinos (which are left-handed), and results in the so-called chiral kinetic theory \cite{CKT}. Those corrections were found to be particularly relevant to neutrino transport \cite{YY}.
 \item We used the so-called quasi-particle approximation for neutrinos, so that their propagator has the same form as the free propagator but with generalized particle distribution functions (see~\eqref{G<qp}-\eqref{G>qp} below).
\item We assumed that the neutrinos are sufficiently close to equilibrium so that their chemical potential is well defined, and at $\b$-equilibrium with the medium.
 \item We also assumed that the medium composed of electrons and quark matter was at thermal equilibrium. This is expected to be a good approximation as the astrophysical times should always be much longer than the thermalization time for the medium.

\end{itemize}
While it is expected that these approximations work well in many cases relevant for neutron stars and supernovae, it is also clear that they will not apply to all
regimes, and eventually a description of neutrino transport which is valid for neutrinos fully out of equilibrium is desirable. This would require to solve the full Kadanoff-Baym equations instead of the Boltzmann equations, which is much more involved numerically.

In addition to going fully out of equilibrium, there are other extensions to our formalism related to the leptonic component, that are mentioned in section \ref{nSE}:
\begin{itemize}
 \item We did not include muons (or muon neutrinos). While muons are relatively massive, their effect may be significant at the highest densities reached in neutron stars or core-collapse supernovae \cite{Bollig17,Loffredo22}.

 \item We did not include the purely electroweak interactions between neutrinos and leptons. Since these interactions are weak, they may be analyzed separately, and their effect can be added on top of the results presented here.

\end{itemize}

We leave such extensions of our approach for future work.

The structure of this paper is as follows:
In section 2, we review the formalism used to describe neutrino transport. The holographic model that is used to compute the charged current 2-point functions is introduced in section 3, with the calculation of the correlators described in section 4. Section 5 is devoted to the analysis of the results obtained for the neutrino radiative coefficients.
Appendix A reviews some basic results in thermal field theory, involved in the general formalism for neutrino transport, and discusses in more details the quasi-particle approximation. We review in appendix B  the weak vertices that are involved in neutrino interactions. In Appendix C, we collect the details of the calculation of the charged black hole background solution. The procedure for fixing the parameters of the model is described in appendix D, whereas appendix E contains the expressions for the fluctuation equations in Eddington-Finkelstein coordinates. The latter are well adapted for finding a numerical solution. Finally, appendix F describes in detail the degenerate and hydrodynamic limit of the radiative coefficients, with the derivation of the approximate expressions shown in the main text.

\vskip 3cm

\section{Formalism for the transport of neutrinos}

In this section, we give a complete review of the elements of formalism that are used to describe the transport of neutrinos. The idea is to make clear the connection between neutrino transport and the retarded chiral-current two-point function, which is the quantity that we compute in this work using holographic methods. We start from the basic definitions of the real-time correlators in the closed-time-path formalism, before deriving the Boltzmann equation, obeyed by the neutrino distribution function. The collision term in the Boltzmann equation depends on the neutrino self-energy in the medium of propagation. The charged current contribution to the self-energy is then computed explicitly at quadratic order in the Fermi weak coupling constant $G_F$, in terms of the chiral current two-point function. The final form of the neutrino Boltzmann equation is presented at the end of this section.

\subsection{Definitions in the closed-time-path formalism}

\begin{figure}[h]
\begin{center}
\includegraphics[scale=0.45]{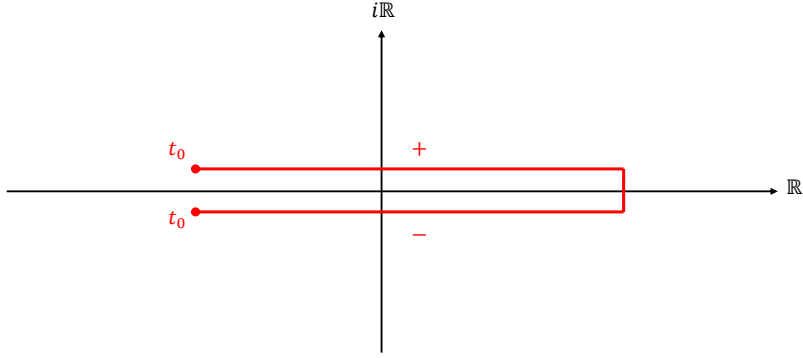}
\caption{The closed-time-path. The turning point of the contour on the right of the figure should thought of as being pushed to infinity.}
\label{Fig:CTP}
\end{center}
\end{figure}

The mathematical objects which contain the information about the transport of neutrinos in a given medium, are the (exact) real-time propagators of the neutrinos
\be
\label{F1} iG_{\a\b}(x_1,x_2) \equiv \left<\mathcal{T}\psi_\a(x_1) \bar{\psi}_\b(x_2)\right> \, ,
\ee
where $\psi$ the neutrino spinor field. $\mathcal{T}$ is a time-ordering operator, for which the possible choices are made explicit below (see \eqref{defGF}-\eqref{defG>}). The brackets refer to the expectation value in the medium at finite temperature. The convenient formalism to compute out-of-equilibrium real-time quantities such as \eqref{F1} is the so-called Schwinger-Keldysh, or closed-time-path (CTP) formalism. The latter relies on the fact that all real-time correlation functions can be written as correlation functions on a specific path in the complex time plane: the CTP, shown in figure \ref{Fig:CTP}.

In particular, the propagators can be expressed in terms of the two-point correlation function on the CTP, which is defined as
\be
\label{defGc} iG_{\a\b}(x_1,x_2) \equiv \left<T_C \psi_\a(x_1)\bar{\psi}_\b(x_2)\right> \equiv \text{Tr}\Big\{\rho(t_0)T_C\psi_\a(x_1)\bar{\psi}_\b(x_2)\Big\}\, ,
\ee
where $\rho(t_0)$ is the density matrix at the initial time on the CTP, $C$ denotes the CTP and $T_C$ is the time-ordering operator on the CTP. The CTP two-point function can be split into several pieces depending on the location of the points $x_1$ and $x_2$ on the path, which is written in matrix form as
\be
\label{matG} G(x_1,x_2) = \left(G^{\e\e'}(x_1,x_2)\right)_{\e,\e'=\pm } =
\begin{bmatrix}
G^{++}(x_1,x_2) & G^{+-}(x_1,x_2) \\
G^{-+}(x_1,x_2) & G^{--}(x_1,x_2)
\end{bmatrix} \, ,
\ee
where the indices $+$ and $-$ refer to the upper and lower branches of the path as indicated in figure~\ref{Fig:CTP} so that $\e$ ($\e'$) gives to the location of the point $x_1$ ($x_2$). The correlation functions $G^{\e\e'}$ can be defined in terms of regular propagators as
\be
\label{defGF} iG^{++}_{\a\b}(x_1,x_2)\equiv iG^F_{\a\b}(x_1,x_2) \equiv \left<T \psi_\a(x_1)\bar{\psi}_\b(x_2)\right>\, ,
\ee
\be
\label{defGFb} iG^{--}_{\a\b}(x_1,x_2)\equiv iG^{\bar{F}}_{\a\b}(x_1,x_2) \equiv \left<T_A \psi_\a(x_1)\bar{\psi}_\b(x_2)\right>\, ,
\ee
\be
\label{defG<} iG^{+-}_{\a\b}(x_1,x_2)\equiv iG^<_{\a\b}(x_1,x_2) \equiv -\left< \bar{\psi}_\b(x_2)\psi_\a(x_1)\right>\, ,
\ee
\be
\label{defG>} iG^{-+}_{\a\b}(x_1,x_2)\equiv iG^>_{\a\b}(x_1,x_2) \equiv \left< \psi_\a(x_1)\bar{\psi}_\b(x_2)\right>\, .
\ee
In the above expressions, $T$ and $T_A$ are respectively the real time-ordering and reverse time-ordering operators.

The retarded and advanced propagators are combinations of \eqref{defGF}-\eqref{defG>}
\be
\label{defGR} G^R_{\a\b}(x_1,x_2) \equiv G^F_{\a\b}(x_1,x_2) - G^<_{\a\b}(x_1,x_2) = -i\theta\big(x_1^0-x_2^0\big)\left<\{ \psi_\a(x_1),\bar{\psi}_\b(x_2) \}\right>\, ,
\ee
\be
\label{defGA} G^A_{\a\b}(x_1,x_2) \equiv G^F_{\a\b}(x_1,x_2) - G^>_{\a\b}(x_1,x_2) = i\theta\big(x_2^0-x_1^0\big)\left<\{ \psi_\a(x_1),\bar{\psi}_\b(x_2) \}\right>\, .
\ee
Note that the four propagators in \eqref{defGF}-\eqref{defG>} are not independent. In particular, the anti-commutation relations for the fermion field operators imply that
\be
\label{relG1} G^< + G^> = G^F + G^{\bar{F}} \, .
\ee
Also, from the definition of the time ordering
\be
\label{GF><} G^F(x_1,x_2) = \theta\big(x_1^0 - x_2^0\big) G^>(x_1,x_2) + \theta\big(x_2^0 - x_1^0\big) G^<(x_1,x_2) \, .
\ee
Therefore, all correlators in \eqref{defGF}-\eqref{defGA} can be expressed in terms of $G^<$ and $G^>$.

\subsubsection*{Relations at equilibrium for bosonic two-point functions}

Although only fermions were considered above, the CTP formalism is perfectly well adapted to describe bosonic real-time correlators as well. As we shall see in the next section, the transport of neutrinos is controlled by the chiral current real-time two-point functions in the medium. The latter is a bosonic correlator, which can be expressed in terms of the two-point function on the CTP
\be
\label{GJ} iG_{\m\n}(x_1,x_2) \equiv \left<T_C J_\m(x_1)J_\n(x_2)\right> \, ,
\ee
where $J$ refers to the chiral current and we omitted the flavor indices. When the medium is at equilibrium, the 2-point function \eqref{GJ} obeys further constraints that we present here. Only the results are given, but the derivations are standard and simple. They are reviewed in Appendix \ref{Sec:eq}.

The first useful property obeyed by the 2-point function at equilibrium is related to the time-translation invariance of the system. If we focus on the time dependence of the propagators, it implies that
\be
\label{GDt} G_{\m\n}(t_1,t_2) = G_{\m\n}(\D t,0) \equiv G_{\m\n}(\D t) \sp \D t \equiv t_1 - t_2 \, .
\ee
In particular, the retarded and advanced propagators are
\be
\label{GRGADt} iG_{\m\n}^R(\D t) = \theta(\D t) \left< \left[J_\m(\D t), J_\n(0) \right] \right> \sp iG_{\m\n}^A(\D t) = -\theta(-\D t) \left< \left[ J_\m(0), J_\n(-\D t) \right] \right> \, .
\ee
In momentum space, the expressions \eqref{GRGADt} imply that the behavior of the retarded 2-point function under a change of sign of $p^0$ is fixed
\be
\label{GRA3} \text{Im}\, G_{\m\n}^R(-p^0) = -\text{Im}\, G_{\m\n}^R(p^0) \sp \text{Re}\, G_{\m\n}^R(-p^0) = \text{Re}\, G_{\m\n}^R(p^0) \, .
\ee

The other equilibrium result that we  use is a consequence of the so-called Kubo-Martin-Schwinger (KMS) symmetry. The latter gives a relation between the forward and backward propagators
\be
\label{KMSG<p} G_B^<(p) = \ex^{-\b p^0} G_B^>(p) \, .
\ee
Using this result, $G_{\m\n}^<$ and $G_{\m\n}^>$ can be expressed in terms of the imaginary part of $G_{\m\n}^R$ only
\be
\label{J<ReJR} G_{\m\n}^<(p) = 2 i n_b(p^0) \text{Im}\,G_{\m\n}^R(p) \, ,
\ee
\be
\label{J>ReJR} G_{\m\n}^>(p) = 2 i \big(n_b(p^0)+1\big) \text{Im}\,G_{\m\n}^R(p) \, ,
\ee
where $n_b$ is the Bose-Einstein distribution
\be
\label{defnB} n_b(E) \equiv \frac{1}{\ex^{\b E} - 1}\, .
\ee

\subsection{Boltzmann equation for neutrinos}

\label{KBeq}

We introduce in this subsection the equation which controls the dynamics of neutrino transport. The fundamental equation obeyed by the neutrino propagator is an exact QFT result, called the Kadanoff-Baym equation \cite{KBbook}. Upon certain semi-classical limits, this equation results in the Boltzmann equation for the neutrino distribution function, which is what neutrino transport simulations aim at solving. We first review the derivation of the Kadanoff-Baym equation from the Schwinger-Dyson equation, and then explain how the Boltzmann equation arises.

Note that the curvature of the space-time has an influence on the transport of neutrinos inside a neutron star. The equations obeyed by the neutrino propagators should therefore be written in a generally covariant form. To keep the presentation of the formalism as simple as possible, we  consider the case of flat space-time. The covariant form can be inferred from the final form of the equations.

\subsubsection{Kadanoff-Baym equation}

To derive the Kadanoff-Baym equation, the starting point is the Schwinger-Dyson equation on the CTP contour. The latter relates the exact neutrino propagator $G$ to the free propagator $G^0$ and the neutrino self-energy $\Sigma$
\be
\label{SDeqa} G(x_1,x_2) = G^0(x_1,x_2) + \int_C\mathrm{d}^4u\mathrm{d}^4v\, G^0(x_1,u)\Sigma(u,v)G(v,x_2) \, ,
\ee
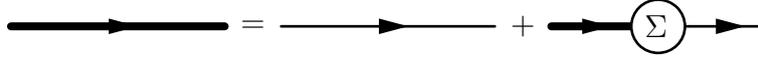
\begin{figure}[h]
\begin{center}
\begin{fmffile}{SD}
\be  \nn
\begin{gathered}
\begin{fmfgraph*}(80,50)
\fmfleft{l1}
\fmfright{r1}
\fmf{fermion,width=3}{l1,r1}
\fmfv{label=$=$}{r1}
\end{fmfgraph*}
\hphantom{\,\,\,=\,\,\,}
\end{gathered}
\begin{gathered}
\begin{fmfgraph*}(80,50)
\fmfleft{l1}
\fmfright{r1}
\fmf{fermion,width=1}{l1,r1}
\fmfv{label=$+$}{r1}
\end{fmfgraph*}
\hphantom{\,\,\,+\,\,\,}
\end{gathered}
\begin{gathered}
\begin{fmfgraph*}(80,50)
\fmfleft{l1}
\fmfright{r1}
\fmf{fermion,width=1}{l1,v1}
\fmfv{decor.shape=circle,decor.filled=empty,
decor.size=20,label=$\Sigma$,label.dist=-5}{v1}
\fmf{fermion,width=3}{v1,r1}
\end{fmfgraph*}
\end{gathered}
\ee
\end{fmffile}
\caption{Diagrammatic representation of the Schwinger-Dyson equation \eqref{SDeqa}. Thick lines correspond to exact fermion propagators and thin lines to free propagators.}
\label{Fig:SD}
\end{center}
\end{figure}
which is written diagrammatically in figure \ref{Fig:SD}. The area of integration $C$ contains integrations over spatial coordinates and over the CTP contour. The self-energy is equal to the interacting part of the 1PI 2-point function
\be
\label{defSE} -i\Sigma \equiv \left<\psi\bar{\psi}\right>_{\text{1PI}} - \left<\psi\bar{\psi}\right>_{\text{1PI, free}}  \, .
\ee
\eqref{SDeqa} is valid for any $x_1,x_2$ on the CTP and can be understood as a matrix equation, if we write the self-energy $\Sigma$ in matrix form as in \eqref{matG}
\be
\label{matS} \Sigma(x_1,x_2) = \left(\Sigma^{\e\e'}(x_1,x_2)\right)_{\e,\e'=\pm } =
\begin{bmatrix}
\Sigma^{++}(x_1,x_2) & \Sigma^{+-}(x_1,x_2) \\
\Sigma^{-+}(x_1,x_2) & \Sigma^{--}(x_1,x_2)
\end{bmatrix} \, .
\ee
In particular, the $+-$ component of \eqref{SDeqa} reads
\begin{align}
\nn G^<(x_1,x_2) \equiv G^{+-}(x_1,x_2) &= G^{0,+-}(x_1,x_2) + \\
\nn +& \int_{t_0}^\infty\mathrm{d}^3u\mathrm{d}^3v\mathrm{d}u_{0,+}\mathrm{d}v_{0,+} G^{0,++}(x_1,u)\Sigma^{++}(u,v)G^{+-}(v,x_2) -\\
\nn -& \int_{t_0}^\infty\mathrm{d}^3u\mathrm{d}^3v\mathrm{d}u_{0,+}\mathrm{d}v_{0,-} G^{0,++}(x_1,u)\Sigma^{+-}(u,v)G^{--}(v,x_2) -\\
\nn -& \int_{t_0}^\infty\mathrm{d}^3u\mathrm{d}^3v\mathrm{d}u_{0,-}\mathrm{d}v_{0,+} G^{0,+-}(x_1,u)\Sigma^{-+}(u,v)G^{+-}(v,x_2) +\\
\label{SDeq<} +& \int_{t_0}^\infty\mathrm{d}^3u\mathrm{d}^3v\mathrm{d}u_{0,-}\mathrm{d}v_{0,-} G^{0,+-}(x_1,u)\Sigma^{--}(u,v)G^{--}(v,x_2) \, ,
\end{align}
where the subscripts $+$ and $-$ on the real times $u_0$ and $v_0$ indicate on which branch of the CTP the integral is performed. We then use the fact that the free propagator $G^0$ is the inverse of the Dirac operator $i\slashed{\partial} - m$, which implies in particular that
\be
\label{DG0} (i\slashed{\partial}_x - m)G^{0,+-}(x,y) = 0 \sp (i\slashed{\partial}_x - m)G^{0,++}(x,y) = \d(x-y) \, .
\ee
Applying the Dirac operator to \eqref{SDeq<} therefore results in the following equation for $G^<$
\be
\label{dsSDeq<} (i\slashed{\partial}_{x_1}-m)G^<(x_1,x_2) = \int \mathrm{d}^4v\left[ \Sigma^{++}(x_1,v)G^<(v,x_2) - \Sigma^<(x_1,v)G^{--}(v,x_2) \right] \, .
\ee
\eqref{dsSDeq<} is called the Kadanoff-Baym equation for the propagator $G^<(x_1,x_2)$. In analogy to the neutrino propagator, we define $\Sigma^< = \Sigma^{+-}$ and $\Sigma^> = \Sigma^{-+}$.

The first step towards the Boltzmann equation, is to go from \eqref{dsSDeq<} to an equation for $G^<$ in momentum space. The appropriate way of doing so for correlators which generically are not translation-invariant, is via a \emph{Wigner transform}, that is a Fourier transform with respect to the separation between the two points
\be
\label{WF} F(X,k) \equiv \int \mathrm{d}^4y\, \ex^{ik\cdot y} F\left(X+\frac{y}{2},X-\frac{y}{2}\right) \, .
\ee
Note that the Wigner transform of a convolution is not the product of the Wigner transforms. We shall however consider the semiclassical \emph{gradient approximation}
\be
\label{gradap} k \gg \partial_X \, ,
\ee
which corresponds to requiring that the system is sufficiently dilute for the mean free path of a neutrino to be much larger than its de Broglie wavelength. In this approximation, the Wigner transform of a convolution is simply the product of the Wigner transforms. Then, assuming \eqref{gradap} and Wigner transforming \eqref{dsSDeq<} gives
\be
\label{KB<} (\slashed{K}-m)G^<(X,k) = \Sigma^{++}(X,k)G^<(X,k) - \Sigma^<(X,k)G^{--}(X,k) \, ,
\ee
where we defined
\be
\label{defK} K^\mu \equiv -k^\mu + \frac{i}{2}\partial^\mu_X \, .
\ee
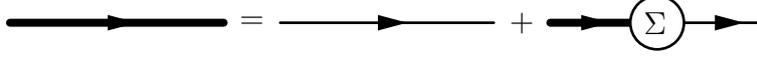
\begin{figure}[h]
\begin{center}
\begin{fmffile}{SD}
\be  \nn
\begin{gathered}
\begin{fmfgraph*}(80,50)
\fmfleft{l1}
\fmfright{r1}
\fmf{fermion,width=3}{l1,r1}
\fmfv{label=$=$}{r1}
\end{fmfgraph*}
\hphantom{\,\,\,=\,\,\,}
\end{gathered}
\begin{gathered}
\begin{fmfgraph*}(80,50)
\fmfleft{l1}
\fmfright{r1}
\fmf{fermion,width=1}{l1,r1}
\fmfv{label=$+$}{r1}
\end{fmfgraph*}
\hphantom{\,\,\,+\,\,\,}
\end{gathered}
\begin{gathered}
\begin{fmfgraph*}(80,50)
\fmfleft{l1}
\fmfright{r1}
\fmf{fermion,width=3}{l1,v1}
\fmfv{decor.shape=circle,decor.filled=empty,
decor.size=20,label=$\Sigma$,label.dist=-5}{v1}
\fmf{fermion,width=1}{v1,r1}
\end{fmfgraph*}
\end{gathered}
\ee
\end{fmffile}
\caption{Alternative writing of the Schwinger-Dyson equation. The difference with figure \ref{Fig:SD} is whether the exact propagator comes before or after the self energy in the right-hand side.}
\label{Fig:SD2}
\end{center}
\end{figure}
There is another way of writing the Schwinger-Dyson equation, which is shown in figure \ref{Fig:SD2}. Starting from this alternative writing, the adjoint Kadanoff-Baym equation can be shown to be
\be
\label{KB<s} G^<(X,k)\left(\overleftarrow{\slashed{K}^*}-m\right) =  G^{++}(X,k)\Sigma^<(X,k) - G^<(X,k)\Sigma^{--}(X,k) \, .
\ee
Taking the trace of the difference of \eqref{KB<} and \eqref{KB<s} results in an equation that depends only on the $+-$ and $-+$ correlators
\be
\label{KB<b} i\partial^{X}_\mu\text{Tr}\left\{\g^\mu G^<(X,k)\right\} =  -\text{Tr}\left\{G^>(X,k)\Sigma^<(X,k) - \Sigma^>(X,k)G^<(X,k)\right\} \, ,
\ee
where we used the fact that for every two-point function F
\be
\label{Fto<} F^{++} + F^{--} = F^< + F^> \, .
\ee

\subsubsection{The Boltzmann equation}

\label{QPA}

We now explain how the Boltzmann equation for the neutrino distribution function is derived from \eqref{KB<b}. This requires considering the so-called quasi-particle approximation.

The quasi-particle approximation consists in assuming that the propagator for the system out of equilibrium can be written in the same form as the free propagator at equilibrium\footnote{See appendix \ref{Sec:G0} for a review of the derivation of \eqref{G0f<}-\eqref{G0f>}, and appendix \ref{Sec:qp} for a more detailed discussion of the quasi-particle approximation.}
\begin{align}
\nn iG^{0,<}(k) = -(\slashed{k}+m+\mu\g_0)\frac{\pi}{E_p}&\Big[ n_f(E_p-\mu)\d(E_p-k^0-\mu)- \\
\label{G0f<} &- (1-n_f(E_p+\mu) )\d(E_p+k^0+\mu) \Big] \, ,
\end{align}
\begin{align}
\nn iG^{0,>}(k) = (\slashed{k}+m+\mu\g_0)\frac{\pi}{E_p}&\Big[ (1-n_f(E_p-\mu))\d(E_p-k^0-\mu) -\\
\label{G0f>} &- n_f(E_p+\mu) \d(E_p+k^0+\mu) \Big] \, ,
\end{align}
but replacing the Fermi-Dirac distribution by space-time dependent particle and anti-particle distributions, $f_{\nu}(X;\vec{k})$ and $f_{\bar{\nu}}(X;\vec{k})$
\begin{align}
\nn iG_\nu^{<}(X;k_\nu) = -(\slashed{k}_\nu+\mu_\nu\g_0)\frac{1-\g^5}{2}\frac{\pi}{E_\nu}&\Big[ f_{\nu}(X;\vec{k}_\nu)\d(E_\nu-k_\nu^0-\mu_\nu) -\\
\label{G<qp} &- (1-f_{\bar{\nu}}(X;-\vec{k}_\nu) )\d(E_\nu+k_\nu^0+\mu_\nu) \Big] \, ,
\end{align}
\begin{align}
\nn iG_\nu^{>}(X;k_\nu) = \frac{1-\g^5}{2}(\slashed{k}_\nu+\mu_\nu\g_0)\frac{\pi}{E_\nu}&\Big[ (1-f_{\nu}(X;\vec{k}_\nu))\d(E_\nu-k_\nu^0-\mu_\nu)- \\
\label{G>qp} &- f_{\bar{\nu}}(X;-\vec{k}_\nu) \d(E_\nu+k_\nu^0+\mu_\nu) \Big] \, .
\end{align}
In the above expressions, we neglected the neutrino mass, $E_\n$ is the on-shell neutrino energy
\be
\label{En} E_\n = \sqrt{\vec{k}_\n^2} \, ,
\ee
and $\m_\n$ is the chemical potential of neutrinos at $\b-$equilibrium with the medium, which is related to the electron ($\m_e$) and isospin ($\m_3$) chemical potentials via
\be
\label{mnbeq} \mu_\nu = \mu_e + \mu_3 \, .
\ee
Note also the presence of the left-handed projectors $(1-\g^5)/2$, which account for the left-handed nature of the neutrinos in the Standard Model.

The Boltzmann equations for neutrinos and anti-neutrinos are then obtained by substituting \eqref{G<qp}-\eqref{G>qp} into the Kadanoff-Baym equation \eqref{KB<b}
\be
\label{BEn} (K_\n\cdot\partial)f_\n(X;\vec{k}_\nu) = -\frac{i}{4} \text{Tr}\Big[\slashed{K}_\n \Sigma^<(K_\n)(1-f_\n) + \Sigma^>(K_\n) \slashed{K}_\n f_\n \Big] \, ,
\ee
\be
\label{BEnb} (K_\n\cdot\partial)f_{\bar{\n}}(X;\vec{k}_\nu) = \frac{i}{4} \text{Tr}\Big[\slashed{K}_\n \Sigma^<(-K_\n)f_{\bar{\n}} + \Sigma^>(-K_\n) \slashed{K}_\n (1-f_{\bar{\n}}) \Big] \, ,
\ee
where $K_\n$ is the on-shell neutrino momentum
\be
\label{Kn} K_\n^\l \equiv \big(E_\n, \vec{k}_\n\big) \, .
\ee

The quasi-particle approximation is exact in the limit of free particles {at equilibrium (both thermodynamic and $\b-$equilibrium). It is therefore justified when the neutrino mean free path is large compared to the typical neutrino wavelength, and the neutrinos are close to equilibrium. In a neutron star, the neutrinos are at equilibrium for layers of the star such that the optical depth to the surface is much larger than one for all neutrino energies. This is typically the case in the core of the star, but not near the neutrinosphere. The quasiparticle approximation is expected to be valid near the core, but it is not clear to what extent it is justified up to the neutrinosphere. In absence of alternative methods to the Boltzmann equation (solving directly the Kadanoff-Baym equation is too complicated), it is always assumed in astrophysical simulations that the quasiparticle approximation applies for neutrinos. We will therefore follow the same assumptions here.

As we shall see in the next subsection, the real time propagators for the electrons also appear in the neutrino Boltzmann equations, via the self-energies. The electrons are assumed to be at equilibrium with chemical potential $\m_e$, and in a regime where the quasi-particle approximation can be used. Within those assumptions, the electron propagators are given by
\begin{align}
\nn iG_e^{<}(t;p_e) = -(\slashed{p}_e+m_e+\mu_\e\g_0)\frac{\pi}{E_e}&\Big[ f_{e}(t;\vec{p}_e)\d(E_e-p_e^0-\mu_e) -\\
\label{Ge<qp} &- (1-f_{\bar{e}}(t;-\vec{p}_e) )\d(E_e+p_e^0+\mu_e) \Big] \, ,
\end{align}
\begin{align}
\nn iG_e^{>}(t;p_e) = (\slashed{p}_e+m_e+\mu_e\g_0)\frac{\pi}{E_e}&\Big[ (1-f_{e}(t;\vec{p}_e))\d(E_e-p_e^0-\mu_e) - \\
\label{Ge>qp} &-f_{\bar{e}}(t;-\vec{p}_e) \d(E_e+p_e^0+\mu_e) \Big] \, ,
\end{align}
where $f_e$ and $f_{\bar{e}}$ are the electron and positron distribution functions, and $E_e \equiv \sqrt{\vec{k}_e^2+m_e^2}$ is the on-shell electron energy.

\subsection{Charged current self-energy}

\label{nSE}

{As we have explained in the previous section, the  Boltzmann equation is determined by the neutrino self-energy.  In this section, we derive} the neutrino self-energy \eqref{defSE} at leading order in the weak coupling Fermi constant $G_F$. We  restrict our analysis in this work to the contribution from the charged current interaction of electronic neutrinos with the baryonic matter
\be
\label{JCint} u + e^- \rightleftharpoons d + \nu_e \, .
\ee
This means that several other components are not considered here:
\begin{itemize}

\item The neutrino self-energy receives a contribution from the \emph{neutral current} interactions
\be
\label{scatt} \n + q \rightleftharpoons \n + q \sp \bar{\n} + q \rightleftharpoons \bar{\nu} + q  \, ,
\ee
\be
\label{brem} q + q' \rightleftharpoons \nu + \bar{\nu} + q + q' \, ,
\ee
where $q$ and $q'$ are quarks. Those interactions are not negligible a priori and should be taken into account when addressing neutrino transport. The calculation of neutral current neutrino self-energies in holography will be the subject of a future work.

\item In general, the other charged leptons contribute if they are present in the medium. In particular, there may be a significant \emph{muon} component in the core of neutron stars~\cite{Potekhin13,Goriely10}. We assume in this work that the only charged leptons present in the medium are electrons. In presence of muons, the muonic neutrinos also couple to the medium, and their transport is described by a separate Boltzmann equation.

\item The propagating neutrinos do not interact only with the baryonic matter, but also with the leptons contained in the medium. Assuming the leptonic component to be composed of electrons, the corresponding charged current processes are given by
\be
\label{scattC} \n_e + e^- \rightleftharpoons \n_e + e^- \sp \bar{\n} + e^+ \rightleftharpoons \bar{\nu} + e^+  \, .
\ee
\be
\label{pprod} e^+ + e^- \rightleftharpoons \n_e + \bar{\n}_e \, .
\ee
The contribution to the neutrino self-energy from these leptonic processes can be derived from a weakly-coupled calculation, which can be added independently from the strongly coupled component considered here. The expressions can be found in \cite{Yueh77}.

\end{itemize}

The Feynman diagram for the $+-$ neutrino self-energy from the reaction \eqref{JCint} at order $\OO(G_F^2)$ is represented in figure \ref{Fig:SEW}.
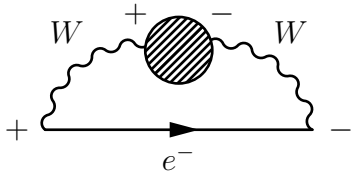
\begin{figure}[h]
\begin{center}
\vspace{1cm}
\begin{fmffile}{SEW}
\begin{fmfgraph*}(100,60)
\fmfleft{l1}
\fmfv{label=$+$,l.s=left}{l1}
\fmfright{r1}
\fmfv{label=$-$,l.s=right}{r1}
\fmftop{v1}
\fmffreeze
\fmftop{v2}
\fmffreeze
\fmf{fermion,width=1,label=$e^-$}{l1,r1}
\fmf{boson,width=1,label=$W$,left=0.5,tension=1}{l1,v1}
\fmf{boson,width=1,label=$W$,left=0.5,tension=1}{v1,r1}
\fmfv{label=$+$,l.d=15,l.a=140}{v1}
\fmfv{label=$-$,l.d=15,l.a=40}{v2}
\fmfblob{25}{v2}

\end{fmfgraph*}
\end{fmffile}
\caption{Diagram for the calculation of the charged current $+-$ neutrino self-energy at leading order in the electroweak interactions. Lines correspond to free propagators, whereas the hatched blob denotes the $W$ current-current correlator in the dense medium $\left<J^-_\mu J^+_\nu\right>$. This two-point function is the exact non-perturbative quantity, that is computed holographically in this work. For illustration, the one-loop contributions to this correlator are shown in figure \ref{Fig:JJ1L}, where only the quark component is considered here. The $+$ and $-$ are the Schwinger-Keldysh indices referring to the location of the operators on the CTP contour in figure \ref{Fig:CTP}. }
\label{Fig:SEW}
\end{center}
\end{figure}
Applying the Feynman rules in the limit where the neutrino momentum is much smaller than the W-boson mass\footnote{In which case the free W diagonal propagators reduce to $-iG_{W,\mu\nu}^{0;\pm\pm} = \frac{\d_{\mu\nu}}{M_W^2}$. See appendix \ref{Sec:nSE_corr} for a review of the weak vertices involved in neutrino interactions.} yields the following result for the self-energy
\be
\label{ASEW} \Sigma_c^{<}(X;k_\n) = -2iG_F^2\int\frac{\mathrm{d}^4k_e}{(2\pi)^4} \g^\l(1-\g^5)\big(iG_e^<(k_e)\big)\g^\s(1-\g^5)\big(iG_{c,\s\l}^>(k_e-k_\nu)\big) \, .
\ee
where we defined the W boson 2-point function
\be
\label{GRc} iG_{c,\s\l} \equiv \left<J^-_\s J^+_\l\right> \, ,
\ee
$J^{\pm}_\l$ being the W boson current. The expression for the backward self-energy $\Sigma_c^>$ is obtained from \eqref{ASEW} by exchanging the $<$ and $>$.
\begin{figure}[h]
\begin{center}
\begin{fmffile}{JJ1L}
\fmfcmd{
    path quadrant, q[], otimes;
    quadrant = (0, 0) -- (0.5, 0) & quartercircle & (0, 0.5) -- (0, 0);
    for i=1 upto 4: q[i] = quadrant rotated (45 + 90*i); endfor
    otimes = q[1] & q[2] & q[3] & q[4] -- cycle;
}
\fmfwizard

\be \nn
\begin{gathered}
\begin{fmfgraph*}(70,60)
\fmfleft{l1}
\fmfv{label=$+$,l.s=left,l.d=0.12w,d.sh=otimes,d.f=empty}{l1}
\fmfright{r1}
\fmfv{label=$-$,l.s=right,l.d=0.12w,d.sh=otimes,d.f=empty}{r1}
\fmf{fermion,width=1,label=$u$,right=1,tension=1}{l1,r1}
\fmf{fermion,width=1,label=$d$,right=1,tension=1}{r1,l1}
\end{fmfgraph*}
\end{gathered}
\hspace{3cm}
\begin{gathered}
\begin{fmfgraph*}(70,60)
\fmfleft{l1}
\fmfv{label=$+$,l.s=left,l.d=0.12w,d.sh=otimes,d.f=empty}{l1}
\fmfright{r1}
\fmfv{label=$-$,l.s=right,l.d=0.12w,d.sh=otimes,d.f=empty}{r1}
\fmf{fermion,width=1,label=$\nu$,right=1,tension=1}{l1,r1}
\fmf{fermion,width=1,label=$e^-$,right=1,tension=1}{r1,l1}
\end{fmfgraph*}
\end{gathered}
\ee
\end{fmffile}
\vspace{1cm}
\caption{1-loop diagram for the W $-+$ current-current correlator, from the quark (left) and lepton (right) contributions. Only the quark component is included in this work.}
\label{Fig:JJ1L}
\end{center}
\end{figure}
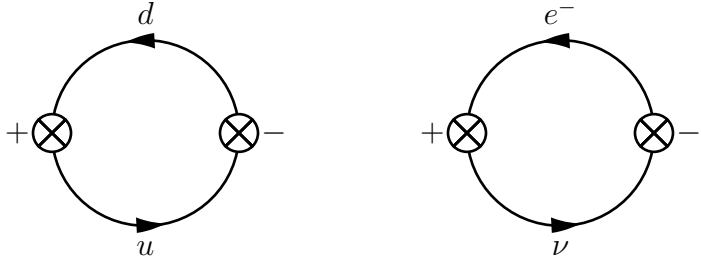

We now proceed to write the quark contribution to the W current $J_\l^+$ in terms of the chiral currents $J_\l^{(L/R)}$. As we shall see in the next section, the latter are the duals of the bulk gauge fields in the holographic set-up.
We assume here for simplicity that $N_f$ is even and the quarks are divided into an equal number of up and down type quarks. Later we will set $N_f=2$.
The quark W current is
\be
\label{JWq} J^{+,\l} = - M_{ij}^{\text{CKM}} \bar{u}^i \g^\l\frac{1-\g^5}{2} d^j \, ,
\ee
where $u$ is a vector that gathers the $N_f/2$ up flavors of quarks (of weak isospin $I^3=1/2$) and $d$ gathers the $N_f/2$ down flavors of quarks (of weak isospin $I^3=-1/2$), $N_f$ being the number of flavors. $M_{ij}^{\text{CKM}}$ is the CKM matrix that determines the mixing between mass and weak eigenstates of the quarks. As for the chiral currents, they are expressed as
\be
\label{JLRq} J_{ij}^{(L/R),\l} = \bar{q}^i \g^\l\frac{1\mp\g^5}{2} q^j \, ,
\ee
where the minus sign is for the left-handed current. The vector $q$ contains all the $N_f$ flavors of quarks
\be
\label{defq} q =
\begin{pmatrix}
u \\
d
\end{pmatrix}
\, .
\ee
In order to write the W current \eqref{JWq} in terms of the chiral currents \eqref{JLRq}, we introduce the enlarged CKM matrix
\newsavebox{\zbox}
\newlength{\wdz}
\sbox{\zbox}{$\begin{matrix}x&x&x&x&x\\x&x&x&x&x\\x&x&x&x&x\end{matrix}$}
\settowidth{\wdz}{\usebox{\zbox}}
\be
\label{defMCKMt}
\td{M}^{\text{CKM}} \equiv \left(
\begin{array}{c|c}
\vphantom{\usebox{\zbox}}\makebox[\wdz]{\bigzero}&\makebox[\wdz]{$M^{\text{CKM}}$}\\
\hline
\vphantom{\usebox{\zbox}}\makebox[\wdz]{\bigzero}&\makebox[\wdz]{\bigzero}
\end{array}
\right) \, ,
\ee
of size $N_f\times N_f$. \eqref{JWq} can then be written as
\be
\label{JWq2} J^{+,\l} = - \td{M}_{ij}^{\text{CKM}} J^{(L),\l}_{ij} \, .
\ee

\subsection{Emissivity and absorption}

In this subsection, the results from Sections \ref{KBeq} and \ref{nSE} are combined to obtain the final form of the kinetic equation obeyed by the neutrino distributions. From there, the radiative coefficients that control the neutrino transport are identified, and classified according to the radiative process they correspond to.

Substituting the expression for the charged current self-energy \eqref{ASEW} into the Boltzmann equations \eqref{BEn} and \eqref{BEnb} results in the following kinetic equations for the neutrino and anti-neutrino distribution functions
\begin{align}
\nn \frac{K_\n\cdot\partial}{E_\n} f_\nu = \frac{iG_F^2}{4}&\int\frac{\mathrm{d}^3k_e}{(2\pi)^3}\frac{1}{E_eE_\nu}\times\\
\nn &\times\left[L_e^{\l\s} \left( (1-f_\nu)f_e G_{c,\s\l}^{>}(q_{e\nu}) - f_\nu(1-f_e)G_{c,\s\l}^{<}(q_{e\nu}) \right)+ \right. \\
\label{dtfn}  & \left. \hphantom{\times} + L_{\bar{e}}^{\l\s} \left( (1-f_\nu)(1-f_{\bar{e}})G_{c,\s\l}^{>}(q_{\bar{e}\nu}) - f_\nu f_{\bar{e}}G_{c,\s\l}^{<}(q_{\bar{e}\nu}) \right) \right]  \! ,
\end{align}
\begin{align}
\nn \frac{K_\n\cdot\partial}{E_\n} f_{\bar{\nu}} = -\frac{iG_F^2}{4}&\int\frac{\mathrm{d}^3k_e}{(2\pi)^3}\frac{1}{E_eE_\nu}\times \\
\nn &\times\left[L_e^{\l\s} \left( f_{\bar{\nu}}f_e G_{c,\s\l}^{>}(q_{e\bar{\nu}}) - (1-f_{\bar{\nu}})(1-f_e)G_{c,\s\l}^{<}(q_{e\bar{\nu}}) \right)+ \right. \\
\label{dtfnb}  & \left. \hphantom{\times} + L_{\bar{e}}^{\l\s} \left( f_{\bar{\nu}}(1-f_{\bar{e}})G_{c,\s\l}^{>}(q_{\bar{e}\bar{\nu}}) - (1-f_{\bar{\nu}}) f_{\bar{e}}G_{c,\s\l}^{<}(q_{\bar{e}\bar{\nu}}) \right) \right] \! ,
\end{align}
where $f_\nu$ and $f_{\bar{\nu}}$ both have argument $(X,\vec{k}_\nu)$, and $f_e$ and $f_{\bar{e}}$ argument $(X,\vec{k}_e)$. We defined several condensed notations for the momenta $k_e-k_\n$ in the different leptonic sectors
\be
\label{qe} q_{e\nu} \equiv (E_e-E_\nu-\mu_e+\mu_\nu,\vec{k}_e-\vec{k}_\nu) \sp q_{e\bar{\nu}} \equiv (E_e+E_\nu-\mu_e+\mu_\nu,\vec{k}_e+\vec{k}_\nu) \, ,
\ee
\be
\label{qeb} q_{\bar{e}\nu} \equiv (-E_e-E_\nu-\mu_e+\mu_\nu,-\vec{k}_e-\vec{k}_\nu) \sp q_{\bar{e}\bar{\nu}} \equiv (-E_e+E_\nu-\mu_e+\mu_\nu,-\vec{k}_e+\vec{k}_\nu) \, ,
\ee
and the lepton tensors
\be
\label{Le} L_{(e/\bar{e})}^{\l\s} \equiv \text{Tr}\left[(\slashed{K}_e\pm m_e)\g^\s(1-\g_5)\slashed{K}_\nu\g^\l(1-\g^5)\right] \, .
\ee
In \eqref{Le} the $+$ is for $e$ and the $-$ for $\bar{e}$, and the $K$'s refer to the on-shell momenta
\be
\label{Kenu} K_e \equiv (E_e,\vec{k}_e) \sp K_\nu \equiv (E_\nu,\vec{k}_\n) \, .
\ee
Computing explicitly the trace in \eqref{Le} gives
\be
\label{Le2} L_{\bar{e}}^{\l\s} = L_e^{\l\s} = 8\left( K_e^\l K_\nu^\s + K_e^\s K_\nu^\l - (K_e.K_\nu)\eta^{\l\s} + i\e^{\l\s\a\b}K_{e,\a}K_{\nu,\b} \right) \, .
\ee
Note that the antisymmetric part will vanish in the contraction with the current-current correlator if the medium is assumed to be isotropic and the interactions preserve parity.

The transport equation for neutrinos can be further simplified by assuming that the medium in which they scatter is at equilibrium. This implies that the electrons follow the Fermi-Dirac distribution
\be
\label{fenF} f_e(\vec{k}_e) = n_f(E_e - \mu_e)\equiv n_e \sp f_{\bar{e}}(\vec{k}_e) = n_f(E_e + \mu_e)\equiv n_{\bar{e}} \, ,
\ee
and the chiral current 2-point functions can be expressed in terms of the retarded correlators according to \eqref{J<ReJR} and \eqref{J>ReJR}. Then, \eqref{dtfn} and \eqref{dtfnb} become
\begin{align}
\nn\frac{K_\n\cdot\partial}{E_\n} f_\nu = -\frac{G_F^2}{2}&\int\frac{\mathrm{d}^3k_e}{(2\pi)^3}\frac{1}{E_eE_\nu}L_e^{\l\s}\times \\
\nn & \times\left[ \text{Im}\,G^R_{c,\s\l}(q_{e\n}) \left( (1-f_\nu)n_e (1+n_b(q_{e\nu}^0)) - f_\nu(1-n_e) n_b(q_{e\nu}^0)\right) +\right. \\
\label{dtfneq}   &\hphantom{\times}\left. + \text{Im}\,G^R_{c,\s\l}(q_{\bar{e}\n})\left( (1-f_\nu)(1-n_{\bar{e}})(1+n_b(q_{\bar{e}\nu}^0)) - f_\nu n_{\bar{e}} n_b(q_{\bar{e}\nu}^0) \right) \right]  \! ,
\end{align}
\begin{align}
\nn \frac{K_\n\cdot\partial}{E_\n} f_{\bar{\nu}} = \frac{G_F^2}{2}&\int\frac{\mathrm{d}^3k_e}{(2\pi)^3}\frac{1}{E_eE_\nu}L_e^{\l\s}\times \\
\nn &\times\left[\text{Im}\,G^R_{c,\s\l}(q_{e\bar{\n}})\left( f_{\bar{\nu}}n_e (1+n_b(q^0_{e\bar{\nu}})) - (1-f_{\bar{\nu}})(1-n_e)n_b(q^0_{e\bar{\nu}}) \right)+ \right. \\
\label{dtfnbeq}  & \left. \hphantom{\times} + \text{Im}\,G^R_{c,\s\l}(q_{\bar{e}\bar{\n}})\left( f_{\bar{\nu}}(1-n_{\bar{e}})(1+n_b(q^0_{\bar{e}\bar{\nu}})) - (1-f_{\bar{\nu}}) n_{\bar{e}}n_b(q^0_{\bar{e}\bar{\nu}}) \right) \right]  \! .
\end{align}
The emissivities $j(E_{\nu}),\bar{j}(E_{\nu})$ and mean free paths $\l(E_{\nu}),\bar{\l}(E_{\nu})$ are defined such that
\be
\label{defjl}\frac{K_\n\cdot\partial}{E_\n} f_\nu \equiv j(E_\nu)(1-f_\nu) - \frac{1}{\l(E_\nu)}f_\nu \, ,
\ee
\be
\label{defjlb} \frac{K_\n\cdot\partial}{E_\n} f_{\bar{\nu}} \equiv \bar{j}(E_\nu)(1-f_{\bar{\nu}}) - \frac{1}{\bar{\l}(E_\nu)}f_{\bar{\nu}} \, .
\ee
The two radiative coefficients are themselves the sum of two terms, corresponding to the contributions from electrons and positrons
\be
\label{jnC0} j(E_\nu) = j_{e^-}(E_\nu) + j_{e^+}(E_\nu) \sp \bar{j}(E_\nu) = \bar{j}_{e^-}(E_\nu) + \bar{j}_{e^+}(E_\nu)\, ,
\ee
\be
\label{lnC0} \frac{1}{\l(E_\nu)} = \frac{1}{\l_{e^-}(E_\nu)} + \frac{1}{\l_{e^+}(E_\nu)}\sp \frac{1}{\bar{\l}(E_\nu)} = \frac{1}{\bar{\l}_{e^-}(E_\nu)} + \frac{1}{\bar{\l}_{e^+}(E_\nu)} \, .
\ee
In the quasi-particle picture, each of these coefficients can be associated with a given weak interaction process between the neutrinos and the baryonic matter
\be
\label{beta1} e^- + u \rightleftharpoons \nu + d \, : \, \,\, \big(\,j_{e^-} \, , \, \l_{e^-}\big) \sp d + e^+ \rightleftharpoons \bar{\nu} + d \, : \, \,\, \big(\,\bar{j}_{e^+} \, , \, \bar{\l}_{e^+}\big) \, ,
\ee
\be
\label{beta2} u \rightleftharpoons \nu + d + e^+ \, : \, \,\, \big(\,j_{e^+} \, , \, \l_{e^+}\big) \sp d \rightleftharpoons \bar{\nu} + u + e^- \, : \, \,\, \big(\,\bar{j}_{e^-} \, , \, \bar{\l}_{e^-}\big) \, .
\ee
These are identified as the various versions of the $\beta$ reaction \eqref{JCint}. Note that, as mentioned before, the purely leptonic processes are not included. Also, as far as the baryonic component of the medium is concerned, the quasi-particle picture is not expected to give a good description of the interaction of the leptons with the strongly-coupled QCD matter. Writing the weak reactions as in \eqref{beta1}-\eqref{beta2} corresponds to approximating the chiral current two-point functions with the 1-loop contribution, which comes from the diagrams in figure \ref{Fig:JJ1L}. However, at strong coupling, the exact 2-point functions receive contributions from all numbers of loops, which means that the weak processes that occur typically involve many quarks and gluons. It is still useful to make the identification as in \eqref{beta1}-\eqref{beta2} because it summarizes the exchange of flavor charges that occurs in each reaction. It also makes it clear what is the relation between the various coefficients that we defined and the weak processes that are usually considered in the literature, from a weakly-coupled perspective.

From \eqref{dtfneq} and \eqref{dtfnbeq}, the expressions for the various contributions to the emissivity and absorption can be identified to be
\be
\nn j_{e^-}(E_\nu) = -\frac{G_F^2}{2}\int\frac{\mathrm{d}^3k_e}{(2\pi)^3}\frac{1}{E_eE_\nu}L_e^{\l\s} \text{Im}\,G^R_{c,\s\l}(q_{e\n}) n_e\big(1+n_b(q_{e\nu}^0)\big) \, ,
\ee
\be
\label{j} j_{e^+}(E_\nu) = -\frac{G_F^2}{2}\int\frac{\mathrm{d}^3k_e}{(2\pi)^3}\frac{1}{E_eE_\nu}L_e^{\l\s} \text{Im}\,G^R_{c,\s\l}(q_{\bar{e}\n}) (1-n_{\bar{e}})\big(1+n_b(q_{\bar{e}\nu}^0)\big)  \, ,
\ee

\be
\nn \frac{1}{\l_{e^-}(E_\nu)} = -\frac{G_F^2}{2}\int\frac{\mathrm{d}^3k_e}{(2\pi)^3}\frac{1}{E_eE_\nu}L_e^{\l\s}\text{Im}\,G^R_{c,\s\l}(q_{e\n})(1-n_e)n_b(q_{e\nu}^0) \, ,
\ee
\be
\label{l} \frac{1}{\l_{e^+}(E_\n)}= -\frac{G_F^2}{2}\int\frac{\mathrm{d}^3k_e}{(2\pi)^3}\frac{1}{E_eE_\nu}L_e^{\l\s}  \text{Im}\,G^R_{c,\s\l}(q_{\bar{e}\n}) n_{\bar{e}}n_b(q_{\bar{e}\nu}^0) \, ,
\ee

\be
\nn \bar{j}_{e^-}(E_\nu) = -\frac{G_F^2}{2}\int\frac{\mathrm{d}^3k_e}{(2\pi)^3}\frac{1}{E_eE_\nu}L_e^{\l\s} \text{Im}\,G^R_{c,\s\l}(q_{e\bar{\n}}) (1-n_e)n_b(q_{e \bar{\nu}}^0) \, ,
\ee
\be
\label{jb}  \bar{j}_{e^+} = -\frac{G_F^2}{2}\int\frac{\mathrm{d}^3k_e}{(2\pi)^3}\frac{1}{E_eE_\nu}L_e^{\l\s} \text{Im}\,G^R_{c,\s\l}(q_{\bar{e}\bar{\n}}) n_{\bar{e}} n_b(q_{\bar{e}\bar{\nu}}^0)  \, ,
\ee

\be
\nn \frac{1}{\bar{\l}_{e^-}(E_\nu)} = -\frac{G_F^2}{2}\int\frac{\mathrm{d}^3k_e}{(2\pi)^3}\frac{1}{E_eE_\nu}L_e^{\l\s} \text{Im}\,G^R_{c,\s\l}(q_{e\bar{\n}})n_e\big(1+n_b(q_{e\bar{\nu}}^0)\big) \, ,
\ee
\be
\label{lb} \frac{1}{\bar{\l}_{e^+}(E_\nu)} = -\frac{G_F^2}{2}\int\frac{\mathrm{d}^3k_e}{(2\pi)^3}\frac{1}{E_eE_\nu}L_e^{\l\s}  \text{Im}\,G^R_{c,\s\l}(q_{\bar{e}\bar{\n}}) (1-n_{\bar{e}})\big(1+n_b(q_{\bar{e}\bar{\nu}}^0)\big) \, .
\ee
\vskip 1cm

\paragraph*{Detailed balance} Because the medium in which the neutrinos scatter is assumed to be at equilibrium, the charged current emissivities and mean free paths \eqref{j}-\eqref{lb} are actually related by a detailed balance condition. In terms of the fermionic and bosonic equilibrium distribution functions, detailed balance refers to the equalities
\be
\label{dben} n_e\big(1+n_b(q_{e\nu}^0)\big) = (1-n_e)n_b(q_{e\nu}^0)\ex^{-\b(E_\nu-\mu_\nu)} \, ,
\ee
\be
\label{dbebn} (1-n_{\bar{e}})\big(1+n_b(q_{\bar{e}\nu}^0)\big) = n_{\bar{e}}n_b(q_{\bar{e}\nu}^0)\ex^{-\b(E_\nu-\mu_\nu)} \, ,
\ee
\be
\label{dbenb} n_e \big(1+n_b(q_{e\bar{\nu}}^0)\big) = (1-n_e)n_b(q_{e\bar{\nu}}^0)\ex^{-\b(-E_\nu-\mu_\nu)} \, ,
\ee
\be
\label{dbebnb} (1-n_{\bar{e}}) \big(1+n_b(q_{\bar{e}\bar{\nu}}^0)\big) = n_{\bar{e}} n_b(q_{\bar{e}\bar{\nu}}^0)\ex^{-\b(-E_\nu-\mu_\nu)} \, ,
\ee
which imply that
\be
\label{dbjln} j_{e^-,e^+}(E_\nu) = \frac{\ex^{-\b(E_\nu-\mu_\nu)}}{\l_{e^-,e^+}(E_\nu)} \, ,
\ee
\be
\label{dbjlnb} \bar{j}_{e^-,e^+}(E_\nu) = \frac{\ex^{-\b(E_\nu+\mu_\nu)}}{\bar{\l}_{e^-,e^+}(E_\nu)} \, .
\ee

Due to the detailed balance relations, the emissivity and absorption are not independent quantities. It is therefore sufficient to study one of the two quantities, or a linear combination. The usual quantity that is considered is the opacity corrected for stimulated absorption \cite{Oertel:2020pcg},
\be
\label{N1} \kappa(E_\n) \equiv j(E_\n) + \frac{1}{\l(E_\n)} \, ,
\ee
which determines the ($E_\n$-dependent) location of the neutrinosphere.

We conclude this section by commenting on the expressions obtained for the neutrino radiative coefficients \eqref{j}-\eqref{lb}. Up to kinematic and statistical factors that are straightforwardly determined from the quasi-particle approximation \eqref{Ge<qp}-\eqref{Ge>qp}, all the contributions are expressed in terms of only one function : {\em the imaginary part of the retarded 2-point function for the charged chiral currents}. All the processes in \eqref{beta1}-\eqref{beta2} are captured by this single correlator. Computing this correlator in neutron-star matter is a strongly-coupled problem, which is why the transport of neutrinos in neutron stars remains an unsettled issue. In this work, we consider a simple holographic model where the strongly-coupled computation of the chiral current 2-point function can be done exactly. The holographic model and the computation of the retarded correlator are described in the next sections.

\section{The holographic model}

We introduce in this section the holographic model that is used to compute the charged current retarded correlator.  It is the simplest bottom-up model describing the dynamics of chiral current operators.
We assume therefore the strongly interacting medium is described by a strongly interacting quantum  theory with $N_f$ quarks and $U(N_f)_L\times U(N_f)_R$ chiral symmetry.
According to holographic duality, this theory is dual to a five dimensional gravitational theory that lives on five dimensional Anti-de Sitter space AdS$_5$, which is a constant negative curvature space with a  four-dimensional boundary. It is this form of the theory that we solve using gravitational methods in order to compute the two-point current-current correlator.

The background solution of this model at finite temperature and density will be then  reviewed, and the expressions of the particle densities will be  determined as a function of the chemical potentials.

\subsection{Action}

We consider  a five-dimensional asymptotically AdS bulk theory, whose field content is dictated by the types of operators that we want the dual (boundary) quantum field theory to include. In the present case, the operators of interest are the chiral currents $J^{(L/R)}_\m$, which are dual to chiral gauge fields in the five dimensional bulk $\mathbf{L}_M$ and $\mathbf{R}_M$. The latter, are elements of the Lie algebra of the chiral group U($N_f$)$_L\times$ U($N_f$)$_R$. The bulk gravitational action is constructed as the sum of a color and a flavor part
\be
\label{Sb} S = S_c + S_f \, .
\ee
The action for the color sector is the 5-dimensional Einstein-Hilbert action
\be
\label{SEH} S_{\text{c}} = M^3N_c^2 \int\mathrm{d}^5x\sqrt{-g}\, \left(R + \frac{12}{\ell^2}\right) \, ,
\ee
where $R$ is the 5D Ricci scalar, $M$ the 5D Planck mass, $\ell$ the AdS radius and $N_c$ the number of colors.  For the flavor sector, we make the simplest choice of a quadratic Yang-Mills action for the chiral gauge fields
\be
\label{SYM} S_f = - \frac{1}{8\ell}(M\ell)^3w_0^2 N_c \int\mathrm{d}^5x\sqrt{-g} \left(\text{Tr}\,\mathbf{F}_{MN}^{(L)}\mathbf{F}^{MN,(L)} + \text{Tr}\,\mathbf{F}_{MN}^{(R)}\mathbf{F}^{MN,(R)} \right) \, ,
\ee
where $w_0$ is the flavor Yang-Mills coupling, and $\mathbf{F}^{(L/R)}$ are the field strengths of the gauge fields $\mathbf{L}$ and $\mathbf{R}$
\be
\label{defF} \mathbf{F}^{(L)} \equiv \mathrm{d}\mathbf{L} - i\mathbf{L}\wedge \mathbf{L} \sp \mathbf{F}^{(R)} \equiv \mathrm{d}\mathbf{R} - i\mathbf{R}\wedge \mathbf{R} \, .
\ee

As usual in holographic theories, the number of colors $N_c$ is assumed to be large in order for the semi-classical treatment of the bulk theory to be valid. Since we are interested in describing dense baryonic matter, the back-reaction of the flavor sector on the glue sector will play an important role. In order for this back-reaction to be finite, we consider the so-called Veneziano large $N$ limit
\be
\label{V1} N_c \to \infty \sp N_f \to \infty \sp \frac{N_f}{N_c} \text{ fixed} \, .
\ee
Although $N_c$ and $N_f$ are assumed to be large, finite values of $N_c$ and $N_f$ will eventually be substituted in the large $N$ result for phenomenological applications. Specifically, $N_c$ will be set to 3, and from now on we fix the flavor sector to be composed of $N_f = 2$ massless flavors.  When the chiral group is U(2)$_L\times$U(2)$_R$, the chiral currents and their dual gauge fields can be decomposed in the Pauli basis $\{\s_a\}$
\be
\nn J^{(L)}_\mu = \frac{1}{2}\hat{J}^{(L)}_\mu\mathbb{I}_2 + \frac{1}{2}\sum_{a=1}^3 J^{a,(L)}_\mu \s_a \sp \mathbf{L}_M = \frac{1}{2}\hat{L}_M\mathbb{I}_2 + \frac{1}{2}\sum_{a=1}^3 L^{a}_M \s_a \, ,
\ee
\be
\label{JsP} J^{(R)}_\mu = \frac{1}{2}\hat{J}^{(R)}_\mu\mathbb{I}_2 + \frac{1}{2}\sum_{a=1}^3 J^{a,(R)}_\mu \s_a \sp \mathbf{R}_M = \frac{1}{2}\hat{R}_M\mathbb{I}_2 + \frac{1}{2}\sum_{a=1}^3 R^{a}_M \s_a \, ,
\ee
and the CKM matrix \eqref{defMCKMt} takes the form
\be
\label{MIQ2} \td{M}^{\text{CKM}} =
\begin{pmatrix}
0 & M_{ud} \\
0 & 0
\end{pmatrix} \, .
\ee
Then, substituting the decomposition \eqref{JsP} into the definition of the charged current \eqref{JWq2} gives
\be
\label{JWqsP} J^{+,\l} = -\frac{1}{2} M_{ud}\Big(J_1^{(L),\l} - iJ_2^{(L),\l}\Big) \, .
\ee
Among the bulk gauge fields, the charged currents will therefore be dual to $L_1$ and $L_2$, that is the non-abelian left-handed gauge fields orthogonal to the isospin direction.

\subsection{Background solution}

\label{Sec:bsT}

We now present the background solution for the bulk action \eqref{SEH}, at finite temperature and density. The dual state of matter that it describes in the dual boundary theory corresponds to a plasma of deconfined (generalized) quarks and gluons. Introducing a finite density of deconfined baryonic matter is equivalent to sourcing the bulk baryon number gauge field with a chemical potential
\be
\label{smq} \hat{V}_0\Big|_{\text{boundary}} = 2\m \, ,
\ee
where we defined the vector gauge field
\be
\label{dV} \mathbf{V}_M \equiv \mathbf{L}_M + \mathbf{R}_M = \frac{1}{2}\hat{V}_M\mathbb{I}_2 + \frac{1}{2}\sum_{a=1}^3 V^{a}_M \s_a \, .
\ee
$\m$ is the quark number chemical potential, related to the baryon number chemical potential by $\mu_B = N_c\m$. Then, the background solution is given by the solution of the Einstein-Maxwell equations obeying the boundary condition \eqref{smq}, together with appropriate regularity conditions in the IR. The derivation of the solution is reviewed in appendix \ref{Sec:bgd}. It corresponds to an asymptotically AdS$_5$ Reissner-Nordström (RN) black-hole, with metric

\be
\label{ds2RN} \mathrm{ds}^2 = \ex^{2A(r)}\left( -f(r)\mathrm{dt}^2 + f(r)^{-1}\mathrm{dr}^2 + \vec{\mathrm{d x}}^2 \right) \, ,
\ee
where
\be
\label{afRNBH} \ex^{A(r)} = \frac{\ell}{r} \sp f(r) = 1 - \left(\frac{r}{r_H}\right)^4\left(1 + 2\left(1 - \pi T r_H\right)\right) + 2\left(1 - \pi T r_H\right)\left(\frac{r}{r_H}\right)^6  \, ,
\ee
\be
\label{rHRN} r_H = \frac{2}{\pi T} \left( 1 + \sqrt{1 + \frac{w_0^2}{3N_c}\frac{\m^2}{\pi^2T^2}} \right)^{-1} \, .
\ee
The background solution for the gauge field is given by
\be
\label{AhA3s2} \hat{V}_0 = 2\m\left(1 - \left(\frac{r}{r_H}\right)^2\right)  \, .
\ee
In \eqref{ds2RN}, the coordinate $r$ is the holographic coordinate, defined such that the AdS  boundary is located at $r=0$ and the horizon at $r = r_H$. Note that the background that we consider is such that all non-abelian gauge fields vanish. In particular, the field dual to the isospin current is not sourced, meaning that the dual thermal state is isospin symmetric, with isospin chemical potential
\be
\label{mu30} \m_3 = 0 \, .
\ee 

In the following, we consider conditions relevant for neutron stars, where the baryon {chemical potential}  is much higher than the temperature, i.e.  $\mu \gg T$. In this limit, the charged black-hole is nearly extremal and the horizon radius is essentially controlled by the chemical potential
\be
\label{rH2} r_H = \sqrt{\frac{3N_c}{w_0^2}}\frac{2}{\m}\left( 1 + \OO\left(\frac{T}{\mu}\right)\right) \, .
\ee

\subsection{Particle densities}

In \eqref{AhA3s2}, the background gauge field is expressed in terms of the baryon number chemical potential $\mu_B = N_c\m$. Instead of the chemical potential, the relevant physical observable is given by the dual thermodynamic state variable, that is the baryon density. In this subsection, we explain how the chemical potential $\m$ is traded for the baryon density $n_B$. We also compute the chemical potentials for the leptons at equilibrium with the baryonic matter.

The baryon density is defined to be the vev of the 0 component of the baryon current
\be
\label{nJ} n_B \equiv \frac{1}{N_c}\left<\hat J_L^0+\hat J_R^0\right>  \, ,
\ee
and the current vev is obtained by differentiating the grand-canonical potential $\Omega$ with respect to $\mu_B$. The holographic correspondence states that $\Omega$ is equal to minus the Euclidean on-shell bulk action, \cite{RN1,RN2}
\be
\label{SosRN} \Omega = -S^E_{\text{on-shell}} = - (M\ell)^3 \left(N_c^2r_H^{-4} + N_c\frac{w_0^2}{6}\m^2 r_H^{-2} \right)V_3 \, ,
\ee
where $V_3$ is the volume of the boundary 3-dimensional Euclidean space. This gives the following expression for the equilibrium density
\be
\label{nYM1} n_B(\m,T) = -\frac{1}{V_3}\frac{\partial\Omega}{\partial \mu_B} = (M\ell)^3w_0^2 (r_H T)^{-2} T^2\m  \, ,
\ee
where the dimensionless quantity $r_H T$ is a function of the ratio $\m/T$ that we reproduce here for convenience
\be
\label{rHRNnew} r_H T = \frac{2}{\pi } \left( 1 + \sqrt{1 + \frac{w_0^2}{3N_c}\frac{\m^2}{\pi^2T^2}} \right)^{-1} \, .
\ee

\begin{figure}[h!]
\begin{center}
\includegraphics[scale=1.]{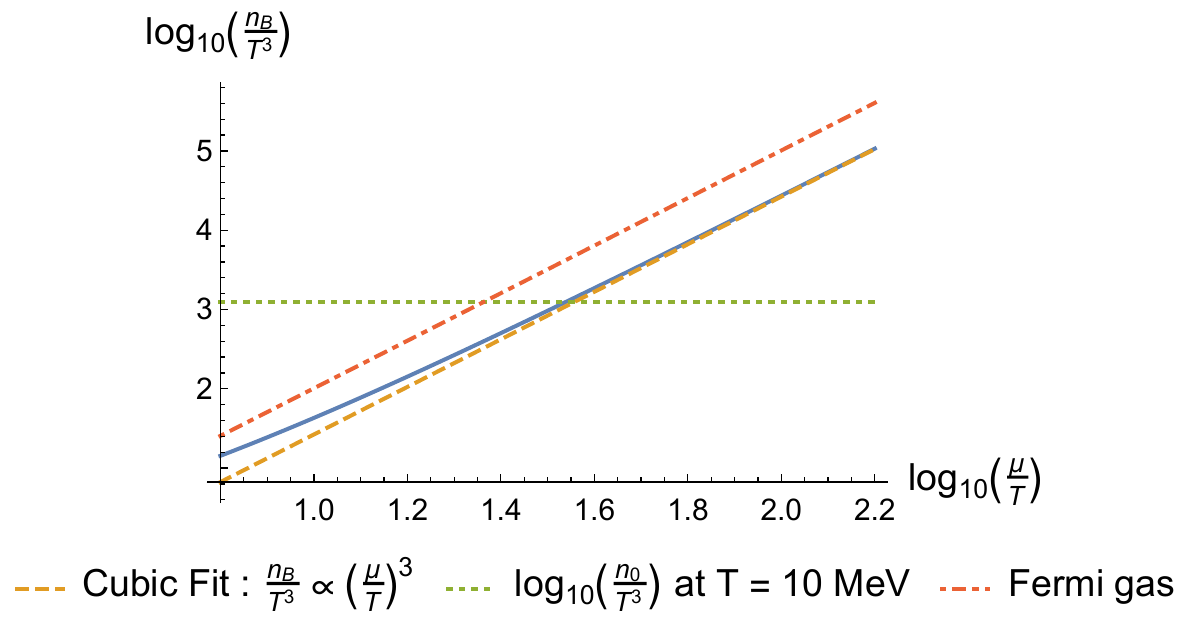}
\caption{The dependence of the baryon number density $n_B$ on the baryonic chemical potential $\m$. The precise quantity that is shown is the density in units of the temperature, $n_B/T^3$. The latter does not depend independently on $T$ and $\m$, but only on $\m/T$. The dashed orange line corresponds to the cubic fit at high density $\m \gg T$, \eqref{mueq2}. For comparison with the typical scales in a neutron star, the dotted green line indicates the value of the nuclear saturation density $n_0 \simeq 0.16\,\mathrm{fm^{-3}}$, in units of the temperature at $T = 10\,\mathrm{MeV}$. We also compare our result with the degenerate Fermi gas at $\b-$equilibrium, which is shown as the red dot-dashed line.}
\label{Fig:nB}
\end{center}
\end{figure}

At high density $\m \gg T$, the expression \eqref{nYM1} simplifies and $n_B$ is found to behave as $\m^3$
\be
\label{mueq2} n_B(\m) = \frac{(M\ell)^3w_0^4}{12 N_c}\m^3 \left(1 + \OO\left(\frac{T}{\mu}\right)\right) \, .
\ee
The profile for $n_B/T^3$ as a function of $\m/T$ is shown in figure \ref{Fig:nB}, where the parameters are those of Appendix \ref{Sec:param}. For comparison, figure \ref{Fig:nB} also shows the relation between $n_B$ and $\m$ in the case where all fermionic species are described by a degenerate Fermi gas. The quark matter described by our model is seen to have a harder equation of state than the degenerate Fermi gas, but the two are relatively close.

\begin{figure}[h!]
\begin{center}
\includegraphics[scale=1.]{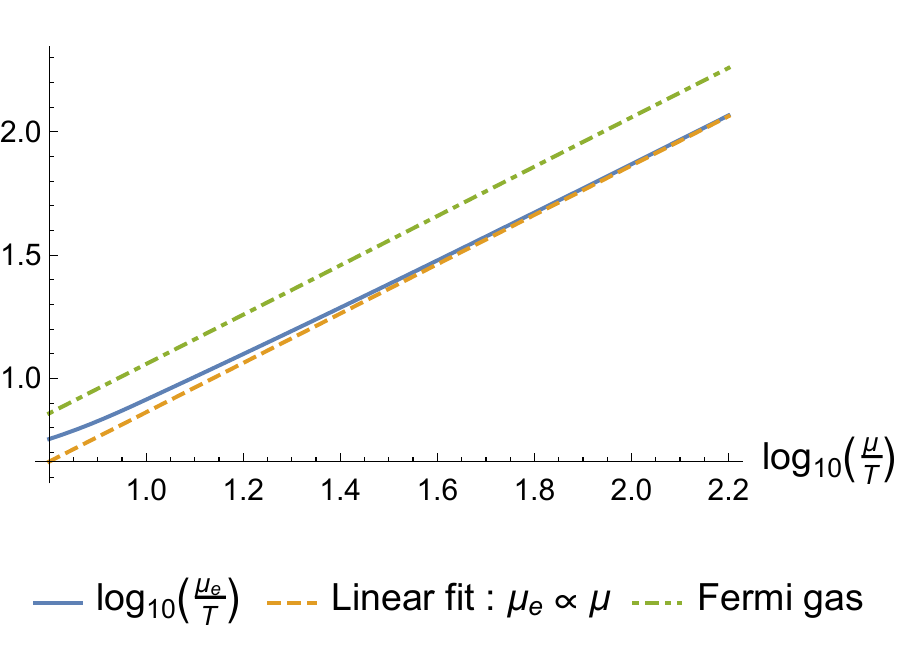}
\caption{The dependence of the leptonic chemical potentials $\mu_e=\mu_\n$ on the baryonic chemical potential $\m$. 
The ratio $\mu_e/T$   
does not depend independently on $T$ and $\m$, but only on $\m/T$.  
The exact numerical result is shown in blue  
and the dashed orange line corresponds to the linear fit at high density $\m \gg T$, \eqref{A1}. 
We compare our result with the degenerate Fermi gas at $\b-$equilibrium, which is shown as the green 
dot-dashed line.}
\label{Fig:muen}
\end{center}
\end{figure}

In addition to the particle densities, the calculations for the transport of neutrinos close to equilibrium also require the knowledge of the leptonic chemical potentials $\m_e$ and $\m_\n$ for given $\m$ and $T$. $\m_\n$ is related to the isospin chemical potential $\m_3$ and the electron chemical potential $\m_e$ via the condition of $\b-$equilibrium \eqref{mnbeq}. Since $\m_3$ is set to 0, $\m_\n$ and $\m_e$ are equal in the medium that we consider. 

The electrons are described by a relativistic Fermi liquid at equilibrium, so the relation between the electron density $n_{e^-}$ and the electron chemical potential $\m_e$ is known explicitly
\be
\label{nemue} \frac{n_{e^-}}{T^3} = \frac{1}{\pi^2} \int_{\frac{m_e}{T}}^{\infty} \mathrm{d}x \frac{x\sqrt{x^2-\left(\frac{m_e}{T}\right)^2}}{1 + \ex^{x-\frac{\mu_e}{T}}} \, .
\ee
Likewise, the positron density is given by
\be
\label{nebmue} \frac{n_{e^+}}{T^3} = \frac{1}{\pi^2} \int_{\frac{m_e}{T}}^{\infty} \mathrm{d}x \frac{x\sqrt{x^2-\left(\frac{m_e}{T}\right)^2}}{1 + \ex^{x+\frac{\mu_e}{T}}} \, .
\ee
Moreover, the electron fraction
\be
\label{defYe} Y_e \equiv \frac{n_{e^-} - n_{e^+}}{n_B} \, ,
\ee
is fixed by the condition of charge neutrality in the medium
\be
\label{YeC0} Y_e = \frac{N_c}{6} \, .
\ee 
Combining \eqref{nemue}-\eqref{YeC0} and \eqref{nYM1} then gives a relation that fixes $\m_e$ as a function of $\m$ and $T$.

At high density $\m \gg T$, the leptonic chemical potentials behave linearly in $\m$
\be
\label{A1} \m_\n = \mu_e = M\ell\, w_0^{\frac{4}{3}} \left(\frac{\pi^2}{24}\right)^{\frac{1}{3}} \m \left(1 + \OO\left(\frac{T}{\mu}\right)\right) \, .
\ee
In figure \ref{Fig:muen} we show the dependence of $\m_e$ on $\m/T$, for the parameters of Appendix \ref{Sec:param}. Figure \ref{Fig:muen} also shows the comparison with the chemical potentials in a degenerate Fermi gas at $\b-$equilibrium. $\m_e$ is observed to be of the same order as in the Fermi gas.

\section{Holographic calculation of the chiral current 2-point functions}

\label{Sec:cfT}

This section discusses the calculation of the retarded 2-point function for the charged current \eqref{JWq2}, in the holographic model presented above. 
We follow the now standard prescription of \cite{Son:2002sd}, and study the linearized field equations for small perturbations $\d L^{1,2}$ of the bulk gauge fields dual to the chiral currents $J_{(L)}^{1,2}$
\be
\label{EoML} \partial_M\left(\sqrt{-g}\,\left(\partial^M\d L^{N,a} - \partial^N\d L^{M,a} \right)  \right) = 0 \, .
\ee
We choose the axial gauge
\be
\label{Ag} \mathbf{L}^{1,2}_r = 0 \, ,
\ee
and define the 4-dimensional Fourier transform of the perturbation as
\be
\label{ansL} \d L^a_\mu(r;t,\vec{x}) = \int \frac{\intd k^4}{(2\pi)^4} \ex^{-i(\omega t - \vec{k}.\vec{x})} \mathcal{L}^a_{\mu,k}(r) \, .
\ee
To avoid clutter, the $k$ dependence of $\mathcal{L}^a_{\mu}$ will not be written explicitly in the following. We also omit the flavor index in most places, since the action is invariant under exchange of $\mathcal{L}^1$ and $\mathcal{L}^2$. We first present the general expression for the 2-point functions. We then  study its behavior in the hydrodynamic limit.

\subsection{General expression}

The general tensor structure of the chiral current 2-point function can be inferred from the symmetries of the background. The finite temperature plasma is invariant under SO(3) spatial rotations, as well as chiral transformations\footnote{Remember that this model does not account for neither explicit or spontaneous breaking of the chiral symmetry. Chiral symmetry imposes that the correlator should obey the Ward identity $\left< J_\l J_\s\right> k^\s = 0$.}. This implies that the 2-point function can be decomposed into a longitudinal and transverse part to the 3-momentum $\vec{k}$, according to
\be
\label{ltcR} \left<J^{(L),a}_\l J^{(L),b}_\s\right>^R(\omega,\vec{k}) = \d^{ab}\left(P_{\l\s}^{\perp}(\omega,\vec{k}) i\Pi_{(L)}^\perp(\omega,\vec{k}) + P_{\l\s}^{\parallel}(\omega,\vec{k}) i\Pi_{(L)}^\parallel(\omega,\vec{k}) \right)  \, ,
\ee
where $a,b \in \{1,2\}$ and the non-zero components of $P^\perp$ and $P^\parallel$ are
\be
\label{Po} P^\perp_{ij}(\omega,\vec{k}) = \d_{ij} - \frac{k_ik_j}{\vec{k}^2} \, ,
\ee
\be
\label{Pp} P^\parallel_{00} = \frac{\vec{k}^2}{\omega^2-\vec{k}^2} \,\, , \,\, P^\parallel_{0i}= P^\parallel_{i0} = -\frac{\omega k_i}{\omega^2-\vec{k}^2} \,\, , \,\, P^\parallel_{ij} = \frac{k_ik_j}{\vec{k}^2}\frac{\omega^2}{\omega^2-\vec{k}^2} \, .
\ee

The sum of the two projectors is the flat 4-dimensional projector transverse to $k^\mu$
\be
\label{P4} P_{\mu\nu}^\perp + P_{\mu\nu}^\parallel = \eta_{\mu\nu} - \frac{k_\mu k_\nu}{k^2} \equiv P_{\mu\nu} \sp k_\mu = \left(-\omega,\vec{k}\right) \, .
\ee
Note that we did not include any term involving the Levi-Civita tensor in \eqref{ltcR} because the bulk action is symmetric under parity $(\vec{x}\leftrightarrow -\vec{x})$. Also, the polarization functions have the following properties :
\begin{itemize}
\item At $\vec{k}=0$, the transverse and longitudinal directions cannot be distinguished anymore, so that the 2-point function should be written as
\be
\label{eqpt1} \left<J^{(L),a}_\l J^{(L),b}_\s\right>^R(\omega,0) = \d^{ab} P_{\l\s}\, i\Pi_{(L)}(\omega) \, ,
\ee
which implies that
\be
\label{eqpt2} \Pi_{(L)}^\parallel(\omega,0) = \Pi_{(L)}^\perp(\omega,0) \, .
\ee
\item Due to the shape of the longitudinal projector \eqref{Pp}, and since the retarded 2-point function should be regular at $\omega = k$, $\Pi^\parallel$ vanishes for light-like momenta as
\be
\label{Ppwk0} \left.\Pi^\parallel(\omega,\vec{k})\right|_{\omega^2\to\vec{k}^2} \sim  \omega^2 - \vec{k}^2 \, .
\ee
\end{itemize}

The 3-dimensional part of the gauge field perturbation can also be decomposed into transverse and longitudinal parts
\be
\label{Vt} \mathcal{L}_i^\perp = \mathcal{L}_i - \frac{k_i}{\vec{k}^2}\left(k_j\mathcal{L}_j\right)  \, ,
\ee
\be
\label{Vp} \mathcal{L}_i^\parallel = \frac{k_i}{\vec{k}^2}\left(k_j\mathcal{L}_j\right) \, .
\ee

We now write the equations of motion \eqref{EoML} component by component. In the axial gauge \eqref{Ag}, the $N=r$ component implies a constraint
\be
\label{EoMLr} \partial_r\mathcal{L}_0 + \frac{f(r)}{\omega}\partial_r\left(k_i\mathcal{L}_i\right) = 0 \, ,
\ee
and the other equations of motion are
\be
\label{EoMVh0} \partial_r\left(\frac{1}{r} \partial_r\mathcal{L}_0\right) - \frac{\sqrt{\vec{k}^2}}{rf(r)} E^\parallel = 0 \, ,
\ee
\be
\label{EoMVhpi} \partial_r\left(\frac{f(r)}{r}\partial_r\mathcal{L}_i^\parallel\right) + \frac{1}{rf(r)}\frac{\omega k_i}{\sqrt{\vec{k}^2}}E^\parallel = 0 \, ,
\ee
\be
\label{EoMVht} \partial_r\left(\frac{f(r)}{r}\partial_r\mathcal{L}_i^\perp\right) + \frac{1}{rf(r)}\left(\omega^2 - f(r)\vec{k}^2\right)\mathcal{L}_i^\perp = 0 \, .
\ee
We defined the longitudinal electric field
\be
\label{defEp} E^\parallel \equiv \sqrt{\vec{k}^2}\mathcal{L}_0 + \frac{\omega}{\sqrt{\vec{k}^2}}\left(k_j\mathcal{L}_j\right) \, .
\ee
Because of the constraint \eqref{EoMLr}, \eqref{EoMVh0} and \eqref{EoMVhpi} are actually the same equation, which can be written as a differential equation for $E^\parallel$
\be
\label{EoMEp}  \partial_r\left(\frac{f(r)}{r}\frac{\partial_r E^\parallel}{\omega^2-f(r)\vec{k}^2}\right) + \frac{1}{rf(r)} E^\parallel = 0 \, .
\ee

The charged current retarded 2-point function is then extracted from the solution to the equations of motion \eqref{EoMVht} and \eqref{EoMEp}, with infalling boundary conditions at the horizon \cite{Son:2002sd}. The on-shell action for the infalling solution reads
\begin{align}
\nn S_{\text{on-shell}} = \frac{1}{8\ell}(M\ell)^3w_0^2 N_c \int \frac{\intd^4k}{(2\pi)^4} \bigg[\frac{\ell}{r}f(r) \bigg(& \mathcal{L}_i^{\perp,a}(-k) \partial_r\mathcal{L}_i^{\perp,a}(k) - \\
\label{Sosb} & - E^{\parallel,a}(-k)\frac{\partial_rE^{\parallel,a}(k)}{\omega^2 - f(r)k^2} \bigg) \bigg]^{r=\e}_{r=r_H} \, ,
\end{align}
with $\e$ a UV cut-off. The AdS boundary contribution to \eqref{Sosb} can be rewritten as
\begin{align}
\nn S_{\text{on-shell}} = \frac{1}{8\ell}(M\ell)^3w_0^2 N_c \int_{r=\e} \frac{\intd^4k}{(2\pi)^4} \bigg[\frac{\ell}{r}\mathcal{L}^\l_a(-k)\d^{ab}\bigg(& P_{\l\s}^\perp(k) \frac{\partial_r\mathcal{L}_i^\perp(k)}{\mathcal{L}_i(k)}+ \\
\label{Sosb2} &+ P_{\l\s}^\parallel(k)\frac{\partial_rE^\parallel(k)}{E^\parallel(k)} \bigg)\mathcal{L}_b^\s(k)\bigg]^{r=\e} \, .
\end{align}
According to the prescription of \cite{Son:2002sd}, this implies that the polarization functions for the left-handed chiral currents are given by
\be
\label{PLt} \Pi_{(L)}^{\perp}(\omega,\vec{k}) = -\frac{1}{4}(M\ell)^3w_0^2 N_c \frac{1}{\e} \left.\frac{\partial_r\mathcal{L}_i^\perp}{\mathcal{L}_i^\perp} \right|_{r=\e} \, ,
\ee
\be
\label{PLp} \Pi_{(L)}^{\parallel}(\omega,\vec{k}) = -\frac{1}{4}(M\ell)^3w_0^2 N_c \frac{1}{\e} \left.\frac{\partial_rE^\parallel}{E^\parallel} \right|_{r=\e} \, .
\ee
To obtain the polarization functions for the charged current $\Pi^{\perp,\parallel}_c$, \eqref{JWqsP} implies that \eqref{PLt}-\eqref{PLp} should be multiplied by a factor $\frac{1}{2}|M_{ud}|^2$
\be
\label{PVt} \Pi_c^{\perp}(\omega,\vec{k}) = -\frac{1}{8}(M\ell)^3w_0^2 N_c |M_{ud}|^2 \frac{1}{\e} \left.\frac{\partial_r\mathcal{L}_i^\perp}{\mathcal{L}_i^\perp} \right|_{r=\e} \, ,
\ee
\be
\label{PVp} \Pi_c^{\parallel}(\omega,\vec{k}) = -\frac{1}{8}(M\ell)^3w_0^2 N_c |M_{ud}|^2 \frac{1}{\e} \left.\frac{\partial_rE^\parallel}{E^\parallel} \right|_{r=\e} \, .
\ee
Whether the expressions \eqref{PVt} and \eqref{PVp} give finite results or need to be regularized depends on the near-boundary behavior of the solutions. The latter is obtained by solving \eqref{EoMVht} and \eqref{EoMEp} at $r\to 0$
\be
\label{nbt} \mathcal{L}^\perp = \mathcal{L}^\perp_0 + r^2\left(\mathcal{L}^\perp_2 - \frac{1}{2}(\omega^2-\vec{k}^2)\mathcal{L}^\perp_0 \log{r}\right)\Big(1 + \OO\big(r^2\big)\Big) \, ,
\ee
\be
\label{nbp} E^\parallel = E^\parallel_0 + r^2\left(E^\parallel_2 - \frac{1}{2}(\omega^2-\vec{k}^2)E^\parallel_0 \log{r}\right)\Big(1 + \OO\big(r^2\big)\Big) \, ,
\ee
with the two independent integration constants given by the source $\mathcal{L}_0^\perp, E_0^\parallel$ and vev terms $\mathcal{L}_2^\perp, E_2^\parallel$. This behavior implies that the polarization functions are subject to a logarithmic UV divergence, which behaves as $(\omega^2 - k^2)\log{\e}$. The log term contributes only to the real part of the polarization functions, whereas the imaginary part does not need to be regularized. Since only the imaginary part enters in the expression for the neutrino radiative coefficients, no regularization is required for our purpose.

\subsection{Hydrodynamic limit}

\label{sec:hydro_charged}

We study in this section the hydrodynamic limit for the retarded 2-point function of the charged current, whose dual field is not sourced by a chemical potential. The expression for the correlators is given by \eqref{PVt} and \eqref{PVp}. The hydrodynamic limit corresponds to the limit of $\omega$ and $\vec{k}$ small compared with the temperature. In this regime, the correlators can be expressed in a systematic expansion in $\omega$ and $k$, whose coefficients correspond to the transport coefficients of the corresponding currents \cite{Kovtun12}. At leading order in the hydrodynamic expansion, the 00 component of the retarded 2-point function exhibits an imaginary diffusive pole of diffusivity $D$
\be
\label{H1} \left<J_0J_0\right>^R = \frac{\s \vec{k}^2}{\omega + iD\vec{k}^2}\left(1 + O\big(\omega, \vec{k}^2\big) \right)\, ,
\ee
where the transport coefficient $\s$ which controls the residue is the DC conductivity. The notation $O$ refers to a term that is at most of the indicated order, but can be much smaller depending on the relative values\footnote{The terms that appear in the expansion in the parentheses of \eqref{H1} do not correspond to a simple double Taylor expansion in $\omega$ and $\vec{k}^2$. Instead, corrections to the denominator will yield terms of the form $\frac{a \omega^2 + b \vec{k}^4}{\omega + iD\vec{k}^2}$. These are always small corrections for $\omega$ and $\vec{k}^2$ small and real, but they are not of a definite order in $\omega$ or $\vec{k}^2$.} of $\omega$ and $\vec{k}^2$. In particular, since the imaginary part of the retarded Green's function is odd in $\omega$ (see \eqref{GRA3}), the first correction to the numerator in \eqref{H1} should be of order $\OO(\omega^2)$ when $\vec{k}=0$. From \eqref{ltcR}, \eqref{H1} then implies that the leading order hydrodynamic approximation to the longitudinal polarization function is given by
\be
\label{H2} \Pi^\parallel(\omega,k) = \frac{-i\s(\omega^2-\vec{k}^2)}{\omega + iD\vec{k}^2}\left(1 + O\big(\omega, \vec{k}^2\big) \right) \, .
\ee
As for the transverse part of the correlator, it is not associated with any hydrodynamic mode, so that it is analytic in the hydrodynamic limit
\be
\label{H3} \Pi^\perp(\omega,k) = i c_0 \omega\left(1+ \OO\big(\omega^2,\vec{k}^2\big) \right) + \OO\big(\omega,\vec{k}^2\big) \, ,
\ee
where $c_0$ is a real constant, and we made explicit the decomposition into real and imaginary parts. Note that the corrections to the imaginary part start at order $\OO\big(\omega^3,\omega \vec{k}^2\big)$, since $\text{Im}\,\Pi^\perp$ is odd in $\omega$ according to \eqref{GRA3}. Using the fact that, when $\vec{k}=0$, the transverse and longitudinal polarization functions are equal, the coefficient at linear order in $\omega$ in (\ref{H3})  is shown to correspond to the conductivity
\be
\label{H4} \Pi^\perp(\omega,k) = -i\s\omega \left(1 + \OO\big(\omega^2,\vec{k}^2\big)\right) + \OO\big(\omega^2,\vec{k}^2\big) .
\ee
The shape of the correlators \eqref{H2} and \eqref{H4} is determined by hydrodynamics, but the transport coefficients $D$ and $\s$ are computed from the microscopic theory. In the present case, the holographic calculation makes it possible to extract analytic expressions for the leading order transport coefficients.

In this work, we are interested in conditions typical of neutron star matter, where the baryonic chemical potential $\mu$ is much larger than the temperature $T$. In this regime and for the bulk action \eqref{SYM}, it can be shown that the hydrodynamic approximation to the 2-point function \eqref{H2}-\eqref{H4} is valid not only at $\omega,k\ll T$, but extends to $T\ll \omega,k \ll \mu$ \cite{DP13}. In the following, we summarize the procedure for computing the hydrodynamic approximation to the 2-point function at $\mu\gg T$ from the equations of motion \eqref{EoMVht} and \eqref{EoMEp}, and give the expression for the transport coefficients $D$ and $\s$. We refer to \cite{DP13} for more details \footnote{The problem considered in \cite{DP13} is not exactly the same, as they consider perturbations of the gauge field in the group under which the black hole is charged (in our case, such a gauge field is dual to a current that enters the neutral current, but does not contribute to the charged current). In that case, the gauge field perturbation couples to the metric perturbation, and the linearized Maxwell equations have to be solved together with the linearized Einstein equations. The general method that they use still applies to the present case though, which is even simpler.}.

\subsubsection{Transverse correlator}

We start by analyzing the transverse polarization function \eqref{PVt}, whose expression at leading order in the hydrodynamic expansion is given by \eqref{H4}. To compute $\Pi^\perp$ in the hydrodynamic regime, we solve the equations of motion for the transverse fluctuations \eqref{EoMVht} at $\omega,k,T \ll \m$. To do this, a small parameter $\e\ll 1$ is introduced, and we consider the following scaling of the energy, momentum and temperature
\be
\label{HT1} r_H\omega \to \e\, r_H\omega \sp r_H k \to \e^a r_H k \sp r_H T \to \e^b r_H T \sp a,b>0 \, .
\ee
Since we are interested in the linear terms in the hydrodynamic expansion, we consider $(r_H k)^2\ll r_H\omega$, that is $a>1/2$. As far as the temperature exponent is concerned, $b<1$ corresponds to the usual hydrodynamic limit $\omega \ll T$, whereas $b\geq 1$ corresponds to the regime where $\omega > T$ but the hydrodynamic approximation remains valid as long as $\omega \ll \m$. The bulk is then divided into two regions where different approximations to the equation of motion \eqref{EoMVht} are valid
\begin{itemize}
\item \emph{The outer region}, where the holographic coordinate $r$ is sufficiently far from the horizon for
\be
\label{HT2} \frac{\omega^2}{f^2}\mathcal{L}^\perp \ll \partial_r^2 \mathcal{L}^\perp \sp \frac{k^2}{f}\mathcal{L}^\perp \ll \partial_r^2 \mathcal{L}^\perp \, ,
\ee
to be obeyed. For $a>1/2$, this region includes the boundary at $r=0$, and inside it the equation of motion \eqref{EoMVht} reduces to
\be
\label{HT3} \partial_r\left(\frac{f(r)}{r}\partial_r\mathcal{L}^\perp\right) = 0 \, .
\ee
The solution to \eqref{HT3} is given by
\be
\label{HT4} \mathcal{L}^\perp_{\text{out}} = A + B \int_0^r\intd r' \frac{r'}{f(r')} \, ,
\ee
with $A$ and $B$ two integration constants.

\item \emph{The inner region}, where the holographic coordinate $r$ is sufficiently close to the horizon for
\be
\label{HT5} \omega^2 \gg f(r)k^2 \, ,
\ee
to be obeyed\footnote{In general, the inner region is simply defined as a region where $r_H-r\ll r_H$. When considering $(r_H \vec{k})^2\ll 1$, the additional condition \eqref{HT5} can be added. This results in $\vec{k}^2$ disappearing from the inner equation of motion.}. For $b<1$, the outer region extends down to the horizon, and the solution in the inner region reduces to an infalling boundary condition for the outer solution at the horizon
\be
\label{HT5b} B = ir_H\omega A \, .
\ee
For $b\geq 1$, the solution in the inner region is better analyzed by zooming on the near-horizon geometry, which is done by defining
\be
\label{HT6} u \equiv \e\, \zeta \, ,
\ee
where $u = 1 - r/r_H$. $\zeta$ is the radial coordinate that describes the AdS$_2$-Schwarzschild factor of the near-horizon geometry (which has an additional $\mathbb{R}^3$ factor). For $b=1$, the equation of motion \eqref{EoMVht} reduces in the inner region to the equation for a massless scalar field in AdS$_2$-Schwarzschild
\be
\label{HT7} \partial_\zeta\left((4\pi r_H T \zeta + 12\zeta^2)\partial_\zeta\mathcal{L}^\perp\right) + \frac{(r_H \omega)^2}{4\pi r_H T \zeta + 12\zeta^2}\mathcal{L}^\perp = 0 \, .
\ee
The infalling solution in the inner region is then given by
\be
\label{HT8} \mathcal{L}^\perp_{\text{in}} = C \left( \frac{3\zeta}{3\zeta + \pi r_H T} \right)^{-\frac{i\omega}{4\pi T}} \, ,
\ee
with $C$ an integration constant.

For $b>1$, $\omega \gg T$ implies that the near-horizon region of the AdS$_2$-Schwarzschild space-time is not probed by the perturbation, and the equation of motion \eqref{EoMVht} reduces in the inner region to the equation for a massless scalar field in AdS$_2$
\be
\label{HT9} \partial_\zeta\Big(12\zeta^2\partial_\zeta\mathcal{L}^\perp\Big) + \frac{(r_H \omega)^2}{12\zeta^2}\mathcal{L}^\perp = 0 \, .
\ee
The infalling solution in the inner region is then given by
\be
\label{HT10} \mathcal{L}^\perp_{\text{in}} = C \exp{\left(\frac{ir_H\omega}{12\zeta}\right)} \, ,
\ee
with $C$ an integration constant.
\end{itemize}

The full solution at leading order in $\e$ is finally obtained by imposing that the outer and inner solutions \eqref{HT4} and \eqref{HT10} (or \eqref{HT8}) are equal in the region where they match. It can be shown that there exists such a matching region, where the outer and inner regions overlap. This region is reached by setting $u$ to be of order $\OO(\e^c)$, with $1-a<c<1$. In practice, this amounts to defining
\be
\label{HT11} u \equiv \e^c v \iff \zeta = \e^{c-1} v \, ,
\ee
and equating $\mathcal{L}_{\text{out}}^\perp(v)$ and $\mathcal{L}_{\text{in}}^\perp(v)$ for $v$ of order 1. Proceeding as such, the solution to \eqref{EoMVht} in the outer region is obtained as
\be
\label{HT12} \mathcal{L}_{\text{out}}^\perp(r) = A\left(1 + ir_H\omega \int_0^{\frac{r}{r_H}} \intd x\frac{x}{f(x)} + \mathcal{O}\big(\e^2,\e^{1+b},\e^{2a}\big) \right) \, .
\ee
Then, from \eqref{PVt}, the transverse polarization function is found to follow the hydrodynamic behavior \eqref{H4}
\begin{align}
\nn \Pi_c^\perp(\omega,\vec{k}) = -\frac{|M_{ud}|^2}{8}r_H^{-2} (M\ell)^3 w_0^2 N_c \bigg(-i&r_H\omega \left(1 + \OO\big((r_H\omega)^2,(r_H\vec{k})^2,r_H T\big)\right) +  \\
\label{PVtRNrH}  &  + \OO\big((r_H\omega)^2,(r_H\vec{k})^2,r_H^2 \omega T \big) \bigg) \, .
\end{align}
The DC conductivity is identified to be
\be
\label{sf} \s = \frac{|M_{ud}|^2}{8r_H}N_c (M\ell)^3w_0^2 \, ,
\ee
with $r_H$ the black-hole horizon radius, whose expression is given by \eqref{rHRN}. This result agrees with the universal result derived in \cite{LI08}. For $\mu\gg T$, the expression simplifies to
\be
\label{slT} \s = \frac{1}{16}|M_{ud}|^2 \sqrt{\frac{N_c}{3}} \left(M\ell\, w_0\right)^3  \m\left(1 + \OO\left(\frac{T}{\mu}\right)\right) \, .
\ee

\subsubsection{Longitudinal correlator}

We now turn to the longitudinal polarization function \eqref{PVp}, whose expression at leading order in the hydrodynamic expansion is given by \eqref{H2}. In this case, the equation of motion that has to be solved at $\omega,k,T \ll \mu$ is \eqref{EoMEp}. A small parameter $\e\ll 1$ is again introduced, and the same kind of scaling of the energy, momentum and temperature as in \eqref{HT1} is considered.
To describe the diffusive pole of the longitudinal correlator, we want to include in the calculation terms of order $(r_H \vec{k})^2$. Therefore, unlike the transverse case, we  now consider general positive values for the momentum exponent $a$.

The bulk is still divided into outer and inner regions, which are defined as in the previous section. The longitudinal equation of motion \eqref{EoMEp} is solved separately in each region as follows
\begin{itemize}
\item In the outer region,  \eqref{EoMEp} reduces to
\be
\label{HL1} \partial_r\left(\frac{f(r)}{r}\frac{\partial_r E^\parallel}{\omega^2 - f(r)k^2}\right) = 0 \, ,
\ee
with solution
\be
\label{HL2} E^\parallel_{\text{out}} = A + B \int_0^r\intd r' r'\frac{\omega^2 - f(r')\vec{k}^2}{f(r')} \, ,
\ee
with $A$ and $B$ two integration constants.

\item In the inner region, the shape of the equation of motion depends on the relative size of the energy $\omega$ and the temperature $T$.  For $b<1$, the outer region extends down to the horizon, and the solution in the inner region reduces to an infalling boundary condition for the outer solution at the horizon
\be
\label{HL3} B = \frac{iA}{(r_H\omega)^2+\frac{1}{2}i(r_H\vec{k})^2} \, .
\ee
For $b=1$, the equation of motion \eqref{EoMEp} in the inner region takes the same form \eqref{HT7} as in the transverse case (with $\mathcal{L}^\perp$ replaced by $E^\parallel$), and the infalling solution is
\be
 \label{HL4} E^\parallel_{\text{in}} = C \left( \frac{3\zeta}{3\zeta + \pi r_H T} \right)^{-\frac{i\omega}{4\pi T}} \, ,
\ee
with $C$ an integration constant.

For $b>1$ ($\omega \gg T$), \eqref{EoMEp} has the form \eqref{HT9} in the inner region and the infalling solution is given by
\be
\label{HL5} E^\parallel_{\text{in}} = C \exp{\left(\frac{ir_H\omega}{12\zeta}\right)} \, ,
\ee
with $C$ an integration constant.
\end{itemize}

By matching the two solutions, the solution to \eqref{EoMEp} in the outer region is obtained as
\begin{align}
\nn E^\parallel_{\text{out}}(r) = A\bigg(1 +  \frac{i}{(r_H\omega)^2+\frac{1}{2}i(r_H\vec{k})^2} \int_0^{\frac{r}{r_H}} \intd x\, x & \frac{(r_H\omega)^2 - f(x)(r_H\vec{k})^2}{f(x)} \times \\
\label{HL6} &\times \Big(1 + \mathcal{O}\big(\e,\e^b,\e^{2a}\big) \Big)\bigg)  \, .
\end{align}
Then, from \eqref{PVp}, the longitudinal polarization function is found to follow the hydrodynamic behavior \eqref{H2}
\be
\label{PVpRN4} \Pi_c^\parallel(\omega,\vec{k}) = r_H^{-2} \frac{-i r_H\s \big((\omega r_H)^2-(\vec{k}r_H)^2\big)}{\omega r_H + \frac{1}{2}i(\vec{k}r_H)^2} \Big( 1 + \OO\big(\omega r_H,  (\vec{k} r_H)^2, r_H T\big) \Big) \, .
\ee
The diffusivity is identified to be
\be
\label{D1} D = \frac{1}{2} r_H\, ,
\ee
which, for $\mu\gg T$, simplifies to
\be
\label{D2} D = \frac{\sqrt{3N_c}}{w_0} \m^{-1} \left(1 + \OO\left(\frac{T}{\mu}\right)\right) \, .
\ee

\section{Analysis of the radiative coefficients}

\label{Sec:res}

This section is dedicated to the analysis of the neutrino charged current radiative coefficients computed from the holographic model, which are the final target of this work. The coefficients are computed by performing the integrals over the loop electron momentum \eqref{j}-\eqref{lb}, where the charged current retarded 2-point function is computed holographically following Section \ref{Sec:cfT}. We first draw the consequences of the presence of the statistical factor to determine which coefficients dominate and which are suppressed.

We then introduce a set of approximations that help understand the behavior of the computed coefficients. These include the hydrodynamic approximation discussed in the previous section for the correlators. Finally, we present the numerical results for the radiative coefficients, and estimate the accuracy of the various approximations. We end the section by comparing the results of this work with some examples of radiative coefficients that are currently used to describe neutrino transport in simulations.

\subsection{Statistics at large baryonic density}

We assume in this subsection that the conditions in the medium where the neutrinos scatter are typical of neutron stars, so that the baryonic and electron densities are very high, $\mu/T, \mu_e/T \gg 1$. In these conditions, the medium at equilibrium is highly degenerate. We  investigate here the consequences of having such a highly degenerate medium for the neutrino radiative coefficients \eqref{j}-\eqref{lb}.
As far as the neutrino chemical potential $\m_\n$ is concerned, we recall that $\beta -$equilibrium \eqref{mnbeq} with $\m_3=0$ implies that it is equal to the electron chemical potential, $\m_e$.

The effect of a high density will appear via the statistical factors in \eqref{j}-\eqref{lb}, which contain the electron Fermi-Dirac distribution, and the Bose-Einstein distribution from the chiral currents correlator at equilibrium. Below, we review the degenerate limit of the statistical factors, which is valid at high density. These approximations depend on the fact that the distributions are evaluated against the imaginary part of the retarded 2-point functions for the chiral currents. By dimensional analysis, at $\mu \gg T$, the 2-point functions obey
\be
\label{JJhd} \text{Im}\,G^R_c(\omega,k) = \mu^2 F\left(\frac{\omega}{\mu},\frac{k}{\mu}\right)\left(1 + \OO\left(\frac{T}{\m}\right)\right) \, ,
\ee
where $F$ is some dimensionless function which vanishes linearly at $\omega = 0$. Given these properties, within the integrand of the radiative coefficients \eqref{j}-\eqref{lb}, the Bose-Einstein factors can be approximated by
\be
\label{nB_hD} n_b(\omega) = -\theta(-\omega) - \frac{\pi^2}{6}\left(\frac{T}{\mu}\right)^2\d'\left(\frac{\omega}{\mu}\right)+  \mathcal{O}\left(\frac{T}{\mu}\right)^{3} \, ,
\ee
where $\theta$ is the Heaviside distribution and $\d$ the Dirac distribution. When the integration path is such that $\omega \geq \omega_0 \gg T$, the Bose-Einstein distribution is exponentially suppressed
\be
\label{nBw0} n_b(\omega) = \ex^{-\b\omega_0}\left(\frac{T}{\mu}\d\left(\frac{\omega-\omega_0}{\mu}\right)+ \mathcal{O}\left(\frac{T}{\mu}\right)^2\right) + \mathcal{O}\left(\ex^{-2\b\omega_0} \right) \, .
\ee

At high electronic chemical potential $\mu_e\gg T$, the distributions for the electrons and positrons at equilibrium \eqref{fenF} are approximated by
\be
\label{nehD} n_e = \theta\big(\mu_e-E_e\big) - \frac{\pi^2}{6}\left(\frac{T}{\mu}\right)^2\d'\left(\frac{E_e-\mu_e}{\mu}\right) + \mathcal{O}\left(\frac{T}{\mu}\right)^3 + \mathcal{O}\big(\ex^{-\b\mu_e}\big) \, ,
\ee
\be
\label{nebhD} n_{\bar{e}} = \ex^{-\b(\mu_e+E_e)} + \mathcal{O}\big(\ex^{-2\b\mu_e}\big) \, .
\ee
In particular, \eqref{nebhD} implies that the following processes involving positrons
\be
\label{wpeb}  \nu + d + e^+ \rightarrow u \qquad \text{and} \qquad  d + e^+ \rightarrow \bar{\n} + u
\ee
are exponentially suppressed
\be
\nn \frac{1}{\l_{e^+}} \,:\,\, n_{\bar{e}}n_b(q^0_{\bar{e}\n}) = \mathcal{O}(\ex^{-\b(E_e+\mu_e)}) \, ,
\ee
\be
\label{jleb_hD_1}  \bar{j}_{e^+}\,:\,\, n_{\bar{e}}n_b(q^0_{\bar{e}\bar{\n}}) = \mathcal{O}(\ex^{-\b(E_e+\mu_e)}) \, ,
\ee
where the momenta $q_{\ell\ell'}$ were defined in \eqref{qeb}; their time components are
\be
\label{qe0} q_{e\nu}^0 \equiv E_e-E_\nu-\mu_e+\mu_\nu = E_e - E_\n \sp q_{e\bar{\nu}}^0 \equiv E_e+E_\nu-\mu_e+\mu_\nu = E_e + E_\n \, ,
\ee
\be
\label{qeb0} q_{\bar{e}\nu}^0 \equiv -E_e-E_\nu-\mu_e+\mu_\nu = -E_e-E_\n \sp q_{\bar{e}\bar{\nu}}^0 \equiv -E_e+E_\nu-\mu_e+\mu_\nu = -E_e+E_\n \, .
\ee
Moreover, \eqref{nB_hD} implies that the emission of a neutrino by the decay of an up quark is also suppressed
\be
\nn j_{e^+} \,:\,\, (1-n_{\bar{e}})\big(1+n_b(q^0_{\bar{e}\n})) = 1 - \theta(E_e + E_\n) + \OO\left(\frac{T}{\m}\right)^2 = \OO\left(\frac{T}{\m}\right)^2 \, .
\ee
The only positronic process that may play a significant role in the transport of neutrinos at high density, is the absorption of an anti-neutrino by the medium, resulting in the emission of a positron
\be
\label{wpeb2}  \bar{\n} + u \rightarrow d + e^+
\ee
The statistical factor for this process is approximated by
\be
\label{jlbeb_hD} \frac{1}{\bar{\l}_{e^+}}\,:\,\, (1-n_{\bar{e}})\big(1+n_b(q^0_{\bar{e}\bar{\n}})\big) = \theta(q^0_{\bar{e}\bar{\n}}) + \mathcal{O}\left(\frac{T}{\m}\right)^2 \, ,
\ee
which is of order 1 at electron energies $E_e \leq E_\n$.

The charged current sector also accounts for weak processes involving electrons
\be
\label{wpe} e^- + u \rightleftharpoons \n + d \qquad \text{and} \qquad d \rightleftharpoons \bar{\n} + u + e^-
\ee
For the processes involving neutrinos, the degenerate approximations are given by
\be
\label{je_hD_n} j_{e^-} \,:\,\, n_e\big(1 + n_b(q^0_{e\n})\big) = \theta(\mu_e-E_e)\theta(q^0_{e\nu}) + \mathcal{O}\left(\frac{T}{\mu}\right)^2 \, ,
\ee
\be
\label{le_hD}  \frac{1}{\l_{e^-}} \, : \,\, (1-n_e)n_b(q^0_{e\n}) = -\theta(E_e-\mu_e)\theta(-q^0_{e\nu}) + \mathcal{O}\left(\frac{T}{\mu}\right)^2 \, .
\ee
This shows that the emission of neutrinos  dominates for $E_\n < \m_\n$, whereas the absorption dominates for $E_\n > \m_\n$. From the detailed balance condition \eqref{dbjln}, the ratio of the subleading coefficient to the leading one is given by a Boltzmann factor $\ex^{-\b|E_\n-\m_\n|}$. When $|E_\n - \m_\n| \lesssim T^2/\m$, both terms are of the same order $\mathcal{O}(T/\mu)^2$.

 For the processes involving anti-neutrinos, the statistical factors are approximated by
\be
\label{jbe_hD} \bar{j}_{e^-} \, : \,\, (1-n_e)n_b(q^0_{e\bar{\n}}) = -\theta(E_e-\mu_e)\theta(-q^0_{e\bar{\n}}) + \mathcal{O}\left(\frac{T}{\mu}\right)^2 \, ,
\ee
\be
\label{lbe_hD}  \frac{1}{\bar{\l}_{e^-}} \, : \,\, n_e\big(1+n_b(q^0_{e\bar{\n}})\big) = \theta(\mu_e - E_e)\theta(q^0_{e\bar{\n}}) + \mathcal{O}\left(\frac{T}{\mu}\right)^2\, .
\ee
\eqref{jbe_hD} implies that the emission of anti-neutrinos is suppressed for $\mu_\n > 0$. From the detailed balance condition \eqref{dbjlnb}, the suppression of the emissivity with respect to the absorption is given by a Boltzmann factor $\ex^{-\b( E_\n + \m_\n)}$.

To summarize the contents of this section, we show in Table \ref{tab:cEn} the radiative processes that contribute to the transport of neutrinos for a given neutrino energy, as well as the associated radiative coefficients.
\begin{table}[h]
\centering
\begin{tabular}{|c|c|c|}
\hline
$E_\n$ & $E_\n < \m_\n$ & $\m_\n < E_\n $ \\
\hline
$\n$ processes & $e^- + u \rightarrow \n + d$ & $\n + d \rightarrow e^- + u$ \\
\hline
$\bar{\n}$ processes & \begin{tabular}{@{}c@{}} $\bar{\n} + u + e^- \rightarrow d $  \\ $\bar{\n} + u \rightarrow d + e^+ $ \end{tabular} & \begin{tabular}{@{}c@{}} $\bar{\n} + u + e^- \rightarrow d $  \\ $\bar{\n} + u \rightarrow d + e^+ $ \end{tabular}
\\
\hline
coefficients & $j_{e^-}, \bar{\l}_{e^-}, \bar{\l}_{e^+}$ & $\l_{e^-}, \bar{\l}_{e^-}, \bar{\l}_{e^+}$ \\
\hline
\end{tabular}
\caption{Radiative processes that contribute to the transport of neutrinos as a function of the neutrino energy $E_\n$, in the degenerate limit.}
\label{tab:cEn}
\end{table}

\subsection{Approximations to the neutrino radiative coefficients}

\label{Sec:App}

In this subsection, we present and analyze a set of approximations that result in simpler expressions for the radiative coefficients. Although not required for the calculation of the coefficients which can be done numerically, the approximations presented below provide some qualitative understanding of the numerical results. Here, we define and investigate the regime of validity of the approximations, leaving the numerical analysis of their accuracy for the next subsection. They are defined as follows:

\paragraph*{The degenerate approximation,} which corresponds to replacing the equilibrium statistical distributions $n_b$ and $n_f$ by their expression in the limit of $\mu \gg T$, \eqref{nB_hD} and \eqref{nehD}. This approximation is of course well known and purely related to the statistical factors whose general expression is analytically known. It is therefore not of much use for numerical calculations. However, it simplifies a lot the expression of the opacities, which helps a better conceptual understanding of the neutrino transport. The degenerate approximation is exact in the limit of $\mu/T \to \infty$.

\paragraph*{The hydrodynamic approximation,} where the charged current 2-point function is replaced by the leading order hydrodynamic expressions \eqref{H2} and \eqref{H4}, with the transport coefficients given by \eqref{sf} and \eqref{D1}. From Section \ref{sec:hydro_charged}, this approximation is expected to be exact when all energies $\mu_e, \mu_\n, E_\n$ are much smaller than the baryonic chemical potential $\m$, because in this case we will have $r_H\omega, r_H k \ll 1$. Below we discuss its validity in more detail.

At the level of the two-point functions, the criterion for the validity of the hydrodynamic approximation was shown in Section \ref{sec:hydro_charged} to correspond to $r_H\omega,r_H k \ll 1$.  The radiative coefficients \eqref{j}-\eqref{lb} are defined as integrals over the loop electron momentum containing the retarded charged current 2-point function, which is a function of two arguments $\omega$ and $k$. For the radiative coefficients, we therefore need to determine the region in the $(\omega,k)$ plane which contributes to the calculation of the integral.

\begin{figure}[h]
\begin{center}
\includegraphics[width=0.55\textwidth]{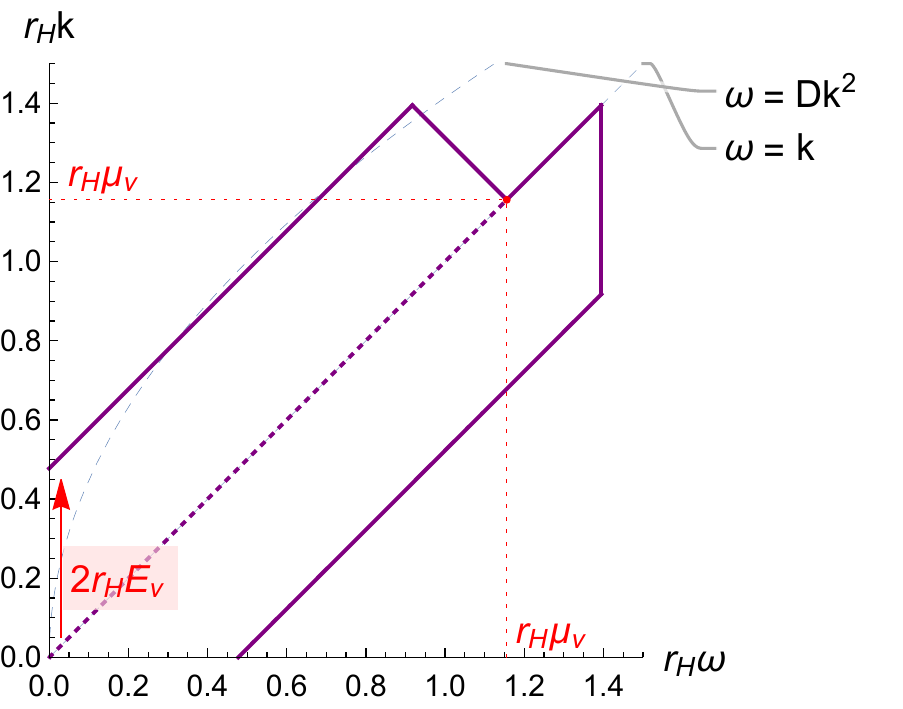}
\hspace{10mm}
\includegraphics[width=0.3\textwidth]{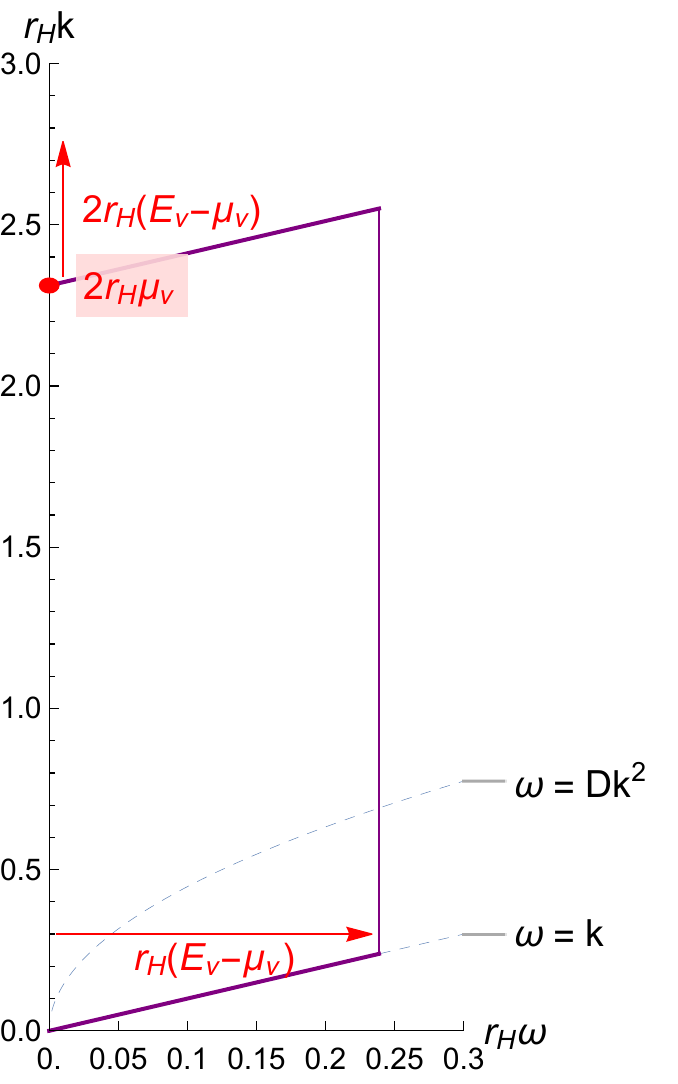}
\caption{Region of the $(\omega,k)$ plane which contributes to the calculation of the radiative coefficients, for $n_B = 1\,\text{fm}^{-3}$ and $T = 10\,\text{MeV}$. We consider neutrino energies between 0 and 100 MeV. The left plot shows the region for neutrino energies $E_\n < \mu_\n$, where the dotted line corresponds to $E_\n = 0$.  It should be combined with the right plot when $E_\n > \m_\n$.  }
\label{Fig:zoom_int}
\end{center}
\end{figure}

The corresponding region is shown in figure \ref{Fig:zoom_int}, where the baryon density $n_B$ is taken to be much larger than $T^3$, and the neutrino energy $E_\n$ to go from 0 to a few times $T$. For concreteness, the figure is constructed by setting $n_B = 1\,\mathrm{fm^{-3}}$, and $E_\n = 0-10T$, which is the range of energies investigated in the numerical analysis of the next subsection.
We use the anti-symmetry of the imaginary part of the retarded two-point function \eqref{GRA3} to restrict to the half space $\omega > 0$. Then, the boundaries of the integration area are determined from the Fermi-Dirac and Bose-Einstein distributions, which behave as step functions at high density, as well as the range of neutrino energies.

 All possible values of the angle $\theta$ between the electron and neutrino momentum are taken into account. In the left figure, the dotted line in the middle corresponds to $E_\n = 0$, and it separates the region relevant to the calculation of $1/\bar{\l}_{e^-}$ (below) from that relevant to $j_{e^-}$ (above). The output of this analysis is that the scales that control the size of the region are shown to be controlled by the leptonic energies\footnote{For the situation considered here where $n_B = 1\,\mathrm{fm^{-3}}$ and $E_\n < 100\,\text{MeV}$, $E_\n$ is always smaller than the chemical potential $\m_\n$. When $E_\n$ becomes significantly larger than $\m_\n$, the shape of the bounding curves deviates significantly from straight lines. However, the size of the region is still controlled by the scales indicated in figure \ref{Fig:zoom_int}, up to factors of order 1.}. This confirms that we enter the hydrodynamic regime when $\m_e,\m_\n$ and $E_\n$ are much smaller than $\m$.

Notice that, for $\mu \gg T$, $\mu_e$ (and $\mu_\n$) are actually proportional to $\mu$, according to \eqref{A1}. Whether or not the hydrodynamic approximation is relevant to the calculation of the radiative coefficients therefore depends on the parameters of the bulk action, $w_0$ and $M\ell$. For the values of the parameters derived from matching the ideal plasma thermodynamics \eqref{Pa3} and \eqref{Pa5}, the following number is found for the leptonic chemical potentials in units of the horizon radius
\be
\label{A3} r_H \m_\n = r_H \mu_e \simeq 1.17\,\left(\frac{(M\ell)^3}{(M\ell)^3_{\text{free}}}\right)^{\frac{1}{2}} \left(\frac{w_0^2(M\ell)^3}{\big(w_0^2(M\ell)^3\big)_{\text{free}}}\right)^{\frac{1}{6}} \left(1 + \OO\left(\frac{T}{\mu}\right)\right) \, ,
\ee
where we also set the number of colors to $N_c = 3$. This number is of order 1, which explains why the hydrodynamic approximation can produce sensible results. Notice that the dependence on the parameters of the bulk action is weak.

\paragraph*{The diffusive approximation,} where the leading hydrodynamic expression of the retarded 2-point function is used, and it is assumed that the time-time component of the retarded 2-point function dominates completely the integral in the calculation of the opacities \eqref{j}-\eqref{lb}. To show that this approximation is valid in the hydrodynamic regime, we evaluate the contribution of each component of the 2-point function to the integral, specifying to the case of the electronic neutrino emissivity $j_{e^-}$ (it is analogous for anti-neutrinos and/or positronic processes). The precise analysis is done in Appendix \ref{Sec:DiffApp}, and we reproduce here the main steps and results.

We consider the degenerate and hydrodynamic regime, where $r_H\m_e = \e$ is much smaller than 1, with all the leptonic energies of the same order and much larger than $r_H T$
\be
\label{A5}  r_H E_\n \sim r_H \m_\n = r_H\m_e = \e \sp r_H T  = \OO(\e^a) \sp \e \ll 1 ,\,\,\,\,\, a \gg 1 \, .
\ee
Since the temperature is assumed to be negligible, the degenerate expression of the statistical distributions can be used \eqref{nB_hD} and \eqref{nehD}. $j_{e^-}$ \eqref{j} can then be written as an integral over the first argument of the charged current 2-point function $\omega$
\be
\label{A6} j_{e^-}(E_\nu) = -\frac{G_F^2}{8\pi^2}\int_0^\pi\intd\theta\sin{\theta}\int_0^{\mu_\n-E_\n}\intd\omega\frac{\omega + E_\n}{E_\nu}L_e^{\l\s}\, \text{Im}\,G^R_{c,\s\l}\big(\omega,k(\omega,\theta)\big)\, ,
\ee
with
\be
\nn k(\omega,\theta) \equiv \sqrt{(\omega + E_\n)^2 + E_\n^2 - 2(\omega+E_\n)E_\n \cos{\theta}} \, .
\ee
From \eqref{H2}  and \eqref{H4}, the imaginary part of the hydrodynamic retarded correlator is given by
\be
\label{A6b} \text{Im}\,G^R_{c,\s\l}(\omega,k) =  -\s\omega\left(P_{\s\l}^{\perp}(\omega,k) + P_{\s\l}^{\parallel}(\omega,k) \frac{\omega^2-k^2}{\omega^2 + D^2k^4}\right) + \OO(\e^2) \, ,
\ee
where $P^\perp$ and $P^\parallel$ are the projectors defined in \eqref{Po}-\eqref{Pp}. The 2-point function reaches the maximum of the diffusion peak when $\omega$ is equal to $\omega^*(E_\n,\theta)$, whose expression is given in \eqref{DA10}. Since $\omega^*$ is of order $\OO(\e^2)$, which is parametrically smaller than the upper bound of the integral in the hydrodynamic limit, the integral can be split into two parts: the first part including the diffusion peak and the other being such that $\omega \gg k^2$. These two parts are respectively labeled by the subscripts ``diff" and ``lin". The following hydrodynamic scalings can then be derived for the contribution of each  component of the 2-point function to the emissivity
\be
\nn j^{(00)}_{e^-,\text{diff}} = \OO(\e^4) \sp j^{(00)}_{e^-,\text{lin}} = \OO(\e^4\log{\e})\, ,
\ee
\be
\label{A7} j^{(0i)}_{e^-,\text{diff}} = \OO(\e^6) \sp j^{(0i)}_{e^-,\text{lin}} = \OO(\e^4)\, ,
\ee
\be
\nn j^{\perp,(ij)}_{e^-,\text{diff}} = \OO(\e^6) \sp j^{\perp,(ij)}_{e^-,\text{lin}} = \OO(\e^4)\, ,
\ee
\be
\nn j^{\parallel,(ij)}_{e^-,\text{diff}} = \OO(\e^8) \sp j^{\parallel,(ij)}_{e^-,\text{lin}} = \OO(\e^4)\, ,
\ee
where the powers in $\e$ are determined by the form of the hydrodynamic 2-point function \eqref{A6b}.
As long as $\m_\n - E_\n$ is much larger than $\OO(\e^2)$, the term $j^{(00)}_{e^-,\text{lin}}$ dominates all the other contributions to the neutrino emissivity. When $E_\n$ is so close to the neutrino chemical potential that $\m_\n - E_\n$ is smaller than $\OO(\e^2)$, the integral includes only the diffusive part, and the leading term becomes $j^{(00)}_{e^-,\text{diff}}$. Since in both cases, the time-time component dominates, this shows that the diffusive approximation is valid in the hydrodynamic limit\footnote{With the exception of vanishing $E_\n$; see the discussion in appendix \ref{Sec:DiffApp}.}.

\begin{figure}[h]
\begin{center}
\includegraphics[width=0.49\textwidth]{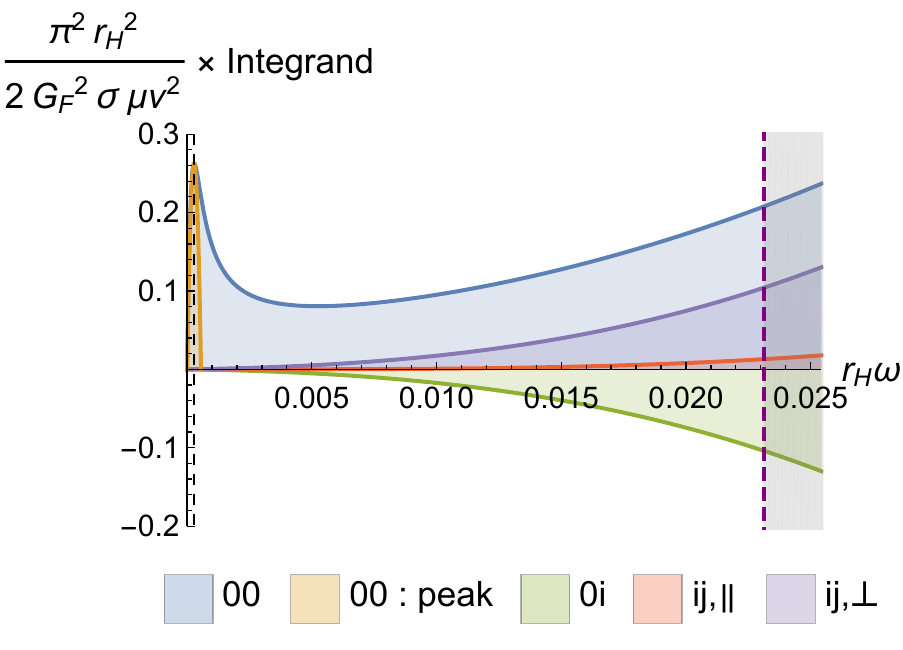}
\includegraphics[width=0.49\textwidth]{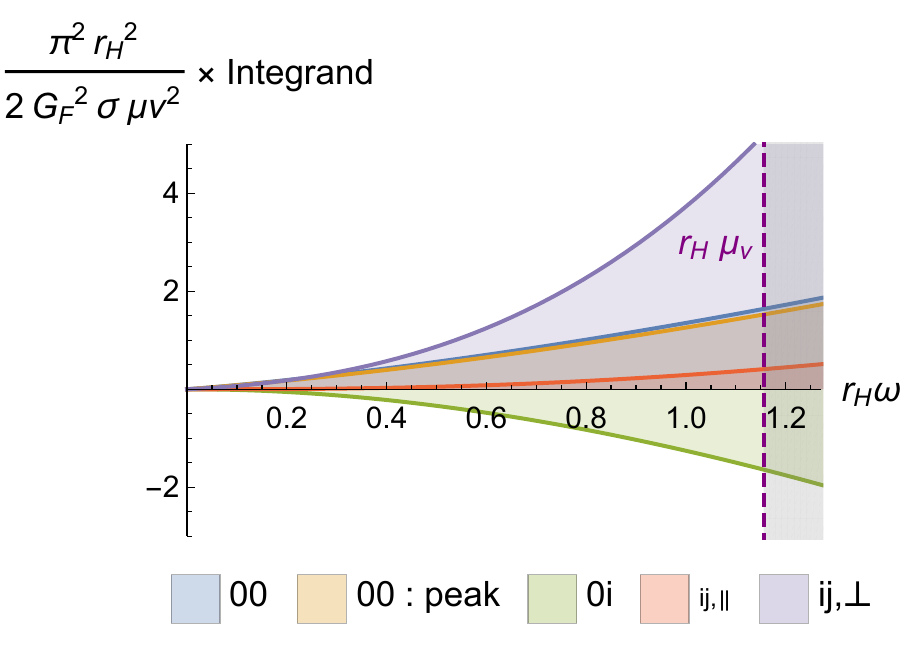}
\caption{Comparison between the contributions of the various components of the charged current 2-point function to the integrand of the neutrino emissivity, in the hydrodynamic approximation. We fix $n_B = 1\,\mathrm{fm^{-3}}$, $T=10\,\text{MeV}$ and $E_\n = \m_\n/2$. The right plot is for the actual values of the leptonic chemical potentials $\m_e$ and $\m_\n$ derived from equilibrium, whereas the left plot is for values fifty times smaller. The various contributions include the time-time (blue), time-space (green) and space-space (red) components. The orange line shows the contribution from the diffusive peak to the time-time component, which corresponds to the lower bound \eqref{DA11}. In the right plot the orange and blue lines are almost confounded. The areas under the curves give the corresponding contributions to the emissivity, whereas the grayed area does not contribute to the integral due to the statistical factor.}
\label{Fig:Diff}
\end{center}
\end{figure}

To illustrate the above discussion, figure \ref{Fig:Diff}  compares the contribution to the neutrino emissivity from the various components of the hydrodynamic 2-point function, at $E_\n = \m_\n/2$ and $n_B = 1\,\mathrm{fm^{-3}}$. Two cases are considered for the values of the leptonic chemical potentials. The right plot shows the result for the actual values of $\m_e$ and $\m_\n$ derived from thermodynamic \eqref{nemue} and $\beta-$equilibrium  \eqref{mnbeq}, whereas in the left plot we consider values fifty times smaller, which goes deeper into the hydrodynamic regime.

The left figure is completely consistent with the hydrodynamic scalings shown in \eqref{A7}. It confirms in particular that the time-time component of the 2-point function gives the largest contribution in the hydrodynamic limit. Also, it shows that the leading contribution to the time-time integral does not come from the diffusion peak itself, but rather from the region $D\vec{k}^2 \ll \omega < \m_\n - E_\n$.

For the actual equilibrium values of $\m_e$ and $\m_\n$ (on the right), the contribution from the time-time component is found to be of the same order as the other contributions. This indicates that the diffusive approximation to the neutrino radiative coefficients is not accurate to describe the actual result, and is only of the right order of magnitude. The latter is not surprising, since in this case $r_H \m_e$ and $r_H\m_\n$ are not much smaller than one.

\paragraph*{Approximate expressions} We now use the crudest diffusive (and degenerate) approximation to derive simple approximate expressions for the radiative coefficients. Details are again provided in Appendix \ref{Sec:DiffApp}. For generic values of the leptonic energies (much smaller than $\m$), the diffusive approximation is found to result in the following simplified expression for the neutrino opacity (defined in \eqref{N1})
\be
\label{A8} \kappa_{e^-}(E_\n) = \frac{8G_F^2\s}{3\pi^2}E_\n^4 \log{\left(\frac{D|\m_\n - E_\n|}{D^2 E_\n^2}\right)} + \OO(\e^4) \, ,
\ee
where $\e$ is the parameter of the hydrodynamic expansion \eqref{A5}. \eqref{A8} is valid as long as $|\m_\n-E_\n|$ is much larger than $\OO(\e^2)$.

 The point $E_\n = \m_\n$ is a peculiar point, since the opacity vanishes there in the degenerate limit. This translates in the presence of a dip at $E_\n = \m_\n$ in the logarithm of $\kappa_{e^-}$, which is clearly visible in the numerical analysis we carry out in section~\ref{sec:numres}. When taking into account the finite temperature, the finite value of the opacity at $E_\n = \m_\n$ can be calculated at leading order in $T/\m$ from the corrections to the equilibrium distribution functions \eqref{nB_hD} and \eqref{nehD}
\be
\label{A9} \kappa_{e^{-}}(\m_\n) = \frac{8G_F^2\s}{3}T^2 r_H^{-2} \left[4\log{\left(\frac{2\m_\n}{m_e}\right)} - 1\right](1+\OO(\e)+\OO(m_e/\m_\n)^2\big) \, .
\ee
This expression gives the depth of the dip in opacity in the hydrodynamic limit. In Appendix \ref{Sec:DiffApp}, the typical width of the dip is also estimated
\be
\label{A10} \D E_\n = \sqrt{\frac{3N_c}{w_0^2}}\frac{\m_\n^2}{\m} \, .
\ee
As for the anti-neutrino opacity, the degenerate and diffusive approximation is given by
\be
\label{A11} \bar{\kappa}_{e^-}(E_\n) = \frac{8G_F^2\s}{3\pi^2}E_\n^4 \log{\left(\frac{D(\m_\n + E_\n)}{D^2 E_\n^2}\right)} + \OO(\e^4) \, ,
\ee
\be
\label{A12} \bar{\ka}_{e^+}(E_\n) = \frac{8G_F^2\s}{3\pi^2}E_\n^4 \log{\left(D E_\n\right)} + \OO(\e^4) \, .
\ee

Note that the leading log term in \eqref{A9} and \eqref{A12} actually vanishes as the neutrino energy goes to zero. In this limit, the appropriate expression is given by the term of order $\OO(\e^4)$, which results in the following approximation
\be
\label{A13} \bar{\ka}_{e^-}(0) = \ka_{e^-}(0) \equiv \ka_{e,0} = \frac{G_F^2 \s}{\pi^2}\m_\n^4 + \OO(\e^5) \, .
\ee
This gives an estimate of the typical size of the opacity at a given baryon density.

To summarize, the analysis of Appendix \ref{Sec:DiffApp} shows that the approximate expressions given above \eqref{A8} and \eqref{A11}-\eqref{A13} are valid in the hydrodynamic limit. This occurs  when the leptonic energies are much smaller than the baryonic chemical potential. Figure \ref{Fig:zoom_int} and equation \eqref{A3} indicate that the conditions in the medium, where the neutrinos scatter, are  such that the leptonic and baryonic energies are of the same order. Therefore, the approximate expressions shown above are expected to give a rough estimate of the exact opacities.

\begin{figure}[h]
\begin{center}
\includegraphics[width=0.49\textwidth]{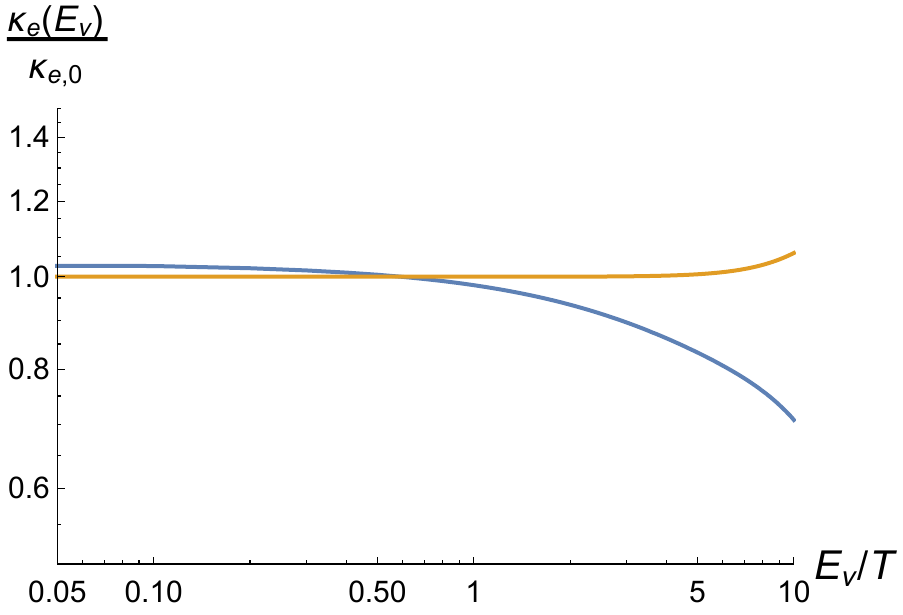}
\includegraphics[width=0.49\textwidth]{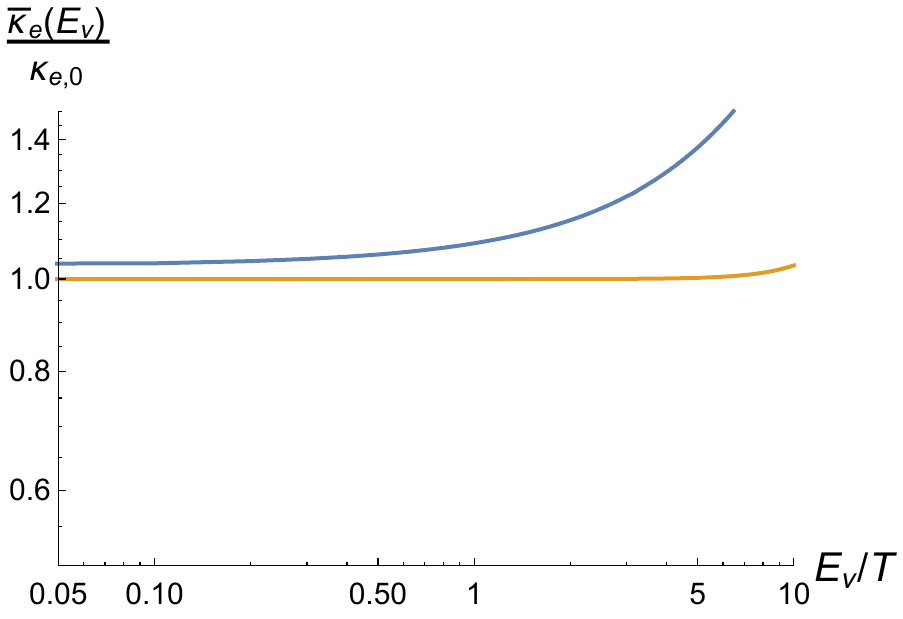}
\caption{Comparison of the hydrodynamic-degenerate approximations (orange), with the exact numerical opacities (blue), for neutrinos (\textbf{Left}) and anti-neutrinos (\textbf{Right}). The approximations are given by \eqref{A8} and \eqref{A11}-\eqref{A12}, summed with the typical value at $E_\n=0$, $\kappa_{e,0}$ in \eqref{A13}. The medium is characterized by $T=10\,\text{MeV}$ and $n_B = 0.31\,\mathrm{fm^{-3}}$, and the opacities are normalized by $\kappa_{e,0}$. }
\label{Fig:Appdd}
\end{center}
\end{figure}

To obtain  a more precise idea of the usefulness of those expressions in the present context, we would like to compare them with the exact opacities. In figure \ref{Fig:Appdd}, the plots of the approximate expressions as a function of neutrino energy are shown together with the numerical solution for the opacities, which is discussed in the next subsection. The state variables that characterize the medium are fixed to $T=10\,\text{MeV}$ and $n_B = 0.31\,\mathrm{fm^{-3}}$. The approximation to the neutrino opacity is given by \eqref{A8}, and for anti-neutrinos in~\eqref{A11}-\eqref{A12}, to which we add the expression at zero-energy \eqref{A13}. It is observed that $\ka_{e,0}$ as defined in \eqref{A13} gives a good approximation of the opacities at zero neutrino energy. However, the energy dependence given by \eqref{A8} and \eqref{A11}-\eqref{A12} is quite far from the actual result. This is particularly striking in the case of the neutrino opacity, where even the monotonicity is not correctly reproduced.

Therefore, the conclusion from figure \ref{Fig:Appdd} is that, although the expression \eqref{A13} gives a good estimate of the opacities at leading order in the hydrodynamic limit, the dependence on the neutrino energy derived from \eqref{A8} and \eqref{A11}-\eqref{A12} is not accurate. It is likely that more accurate expressions could be obtained by including the terms of order $\OO(\e^4)$, beyond the leading log term. However, those result in complicated expressions that are not very useful for a qualitative understanding.

\subsection{Numerical results}\label{sec:numres}

We present here the results of the numerical calculation of the neutrino transport coefficients \eqref{j}-\eqref{lb}. We first discuss the strongly-coupled component which is computed holographically. This  is the imaginary part of the charged current polarization functions. The latter are calculated according to the procedure described in Section \ref{Sec:cfT}. In particular, we are interested in estimating the accuracy of the hydrodynamic approximation \eqref{PVtRNrH} and \eqref{PVpRN4} to the 2-point function for the parameters of interest. Then, we analyze the radiative coefficients themselves, that are obtained by computing the integrals over the loop electron momentum \eqref{j}-\eqref{lb}, which include the charged current 2-point function. We estimate the accuracy of the approximations introduced in the previous subsection over a range of parameters relevant for neutron stars.

In the following, we  fix the temperature to a value that is typically relevant to neutrino transport calculations, for example in a cooling proto-neutron star
\be
\label{TYe} T = 10\,\text{MeV} \, .
\ee
We shall investigate the numerical results for the remaining 2-dimensional parameter space, spanned by the baryon number density $n_B$ and neutrino energy $E_\n$.

\subsubsection{Charged current polarization functions}

Figure \ref{Fig:ImP} shows the numerical result for the imaginary part of the charged current polarization functions $\Pi^\perp(\omega,k)$ and $\Pi^\parallel(\omega,k)$, for $n_B = 1\,\mathrm{fm^{-3}}$. In terms of chemical potentials, this corresponds to $\mu/T \simeq 65$. $r_H \omega$ and $r_H k$ 	are varied between $0$ and $2$, which includes the region over which the integral is performed to compute the radiative coefficients\footnote{In principle, the integrals go up to infinite electron momentum, but the contribution from high energies is exponentially suppressed by the statistical factors. In practice, computing the integral over a finite region as the one shown in figure \ref{Fig:RDP_Hyd} is sufficient, and the contribution from outside of this region is completely negligible.} (see figure \eqref{Fig:RDP_Hyd}).

For the smallest values of $\omega$ and $k$, the hydrodynamic approximations \eqref{PVtRNrH} and \eqref{PVpRN4} are expected to be relevant. This is confirmed at the qualitative level by comparing figure \ref{Fig:ImP} with the hydrodynamic result plotted in figure \ref{Fig:ImP_Hyd}: whereas the longitudinal polarization function $\text{Im}\,\Pi^\parallel$ shows a peak at 	a location set by the position of the diffusive pole $\omega = D k^2$, the transverse one $\text{Im}\,\Pi^\perp$ goes to zero near the origin, remaining relatively close to a linear behavior in $\omega$ up to $r_H\omega = 1$.

\begin{figure}[h]
\begin{center}
\includegraphics[width=0.45\textwidth]{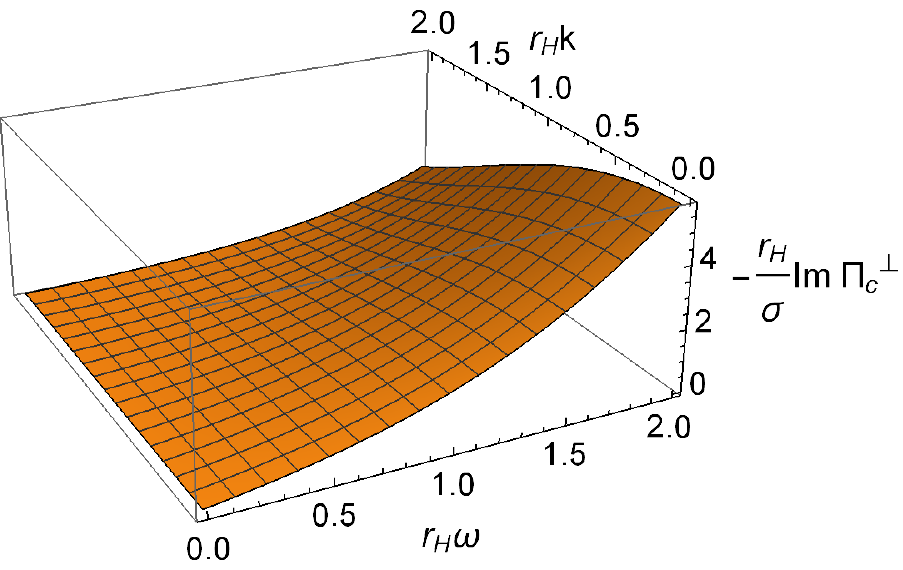}
\hspace{5mm}
\includegraphics[width=0.45\textwidth]{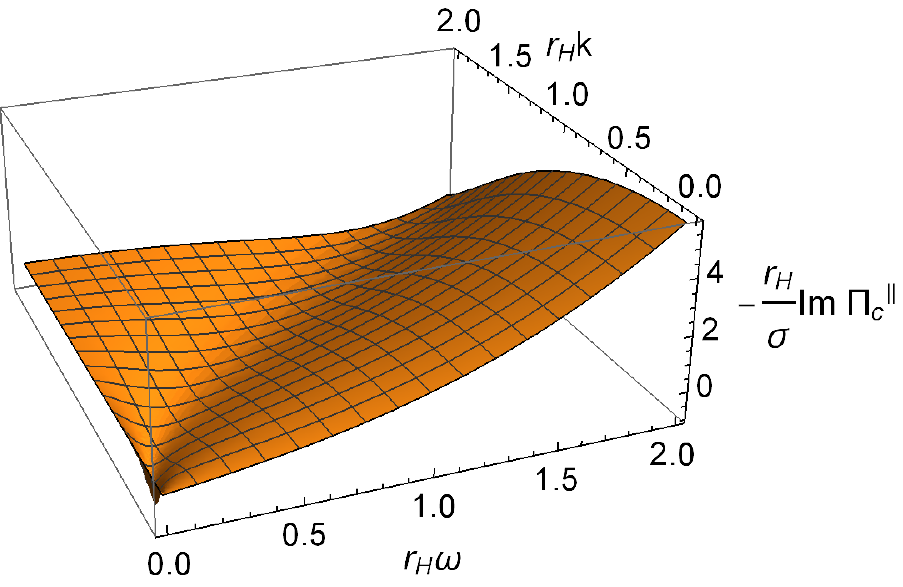}
\caption{Imaginary part of the transverse (\textbf{Left}) and longitudinal (\textbf{Right}) charged current retarded polarization functions. The energy and momentum are expressed in units of $r_H^{-1}$, and the polarization functions are normalized by $\s/r_H$. The medium is characterized by $n_B = 1\,\text{fm}^{-3}$ and $T = 10\,\text{MeV}$.}
\label{Fig:ImP}
\end{center}
\end{figure}

\begin{figure}[h]
\begin{center}
\includegraphics[width=0.45\textwidth]{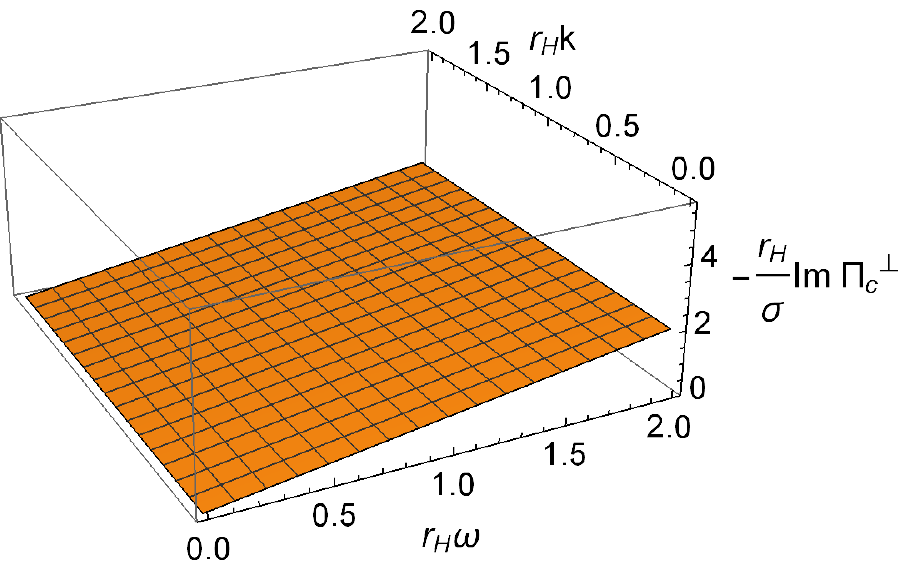}
\hspace{5mm}
\includegraphics[width=0.45\textwidth]{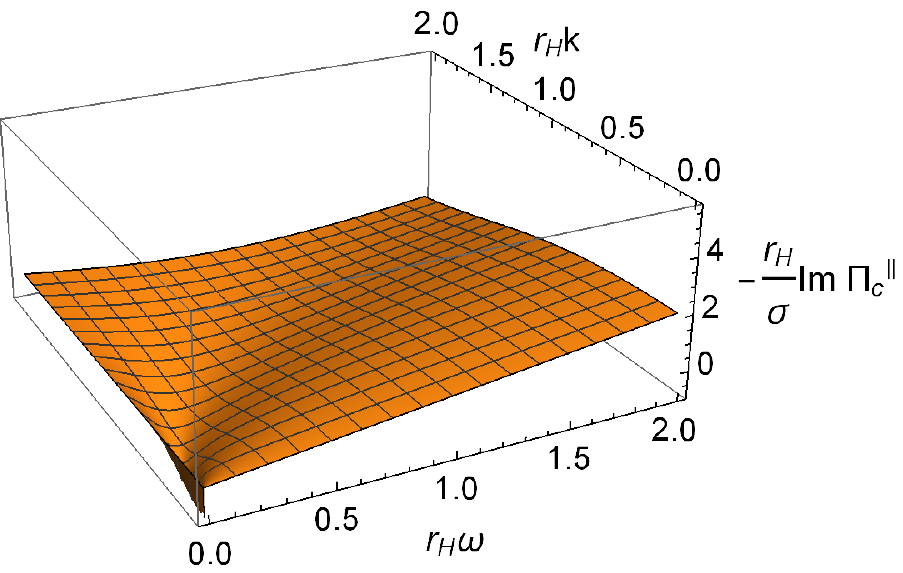}
\caption{Same as figure \ref{Fig:ImP} but for the hydrodynamic approximation to the polarization functions \eqref{PVtRNrH} and \eqref{PVpRN4}.}
\label{Fig:ImP_Hyd}
\end{center}
\end{figure}

To extend the previous discussion to the quantitative level, we show in figure \ref{Fig:RDP_Hyd} the relative difference between figures \ref{Fig:ImP} and \ref{Fig:ImP_Hyd}, together with the region in the $(\omega,k)$ plane which gives a sizable contribution to the radiative coefficients. We consider a range of neutrino energies which is typical of  transport in a neutron star $E_\n = 0-100\,\text{MeV}$ \cite{Oertel:2020pcg}. For this range of neutrino energies, the dominant coefficients are given by the neutrino emissivity $j_{e^-}$ and the anti-neutrino absorption $1/\bar{\l}_{e^-}$. The region which contributes to the calculation of the radiative coefficients is indicated by the purple polygon in figure \ref{Fig:RDP_Hyd}.

Figure \ref{Fig:RDP_Hyd} shows that, in the region relevant for the calculation of $j_{e^-}$ and $1/\bar{\l}_{e^-}$, the accuracy of the hydrodynamic approximation is of about $0$ to $50\%$ for the transverse part, and $0$ to $100\%$ for the longitudinal part. Also, the largest deviation from the hydrodynamic result is consistently reached in the corners of the plots, that is for the largest values of $\omega$  and $k$. Note that the region where the hydrodynamic approximation is the best, is different for the transverse and longitudinal parts : whereas it is located near the line $\omega = k$ for the transverse part, it is close to the location of the diffusive peak $\omega = D\vec{k}^2$ for the longitudinal part. This observation is consistent with the respective leading order of the corrections in the hydrodynamic expansion, as shown in \eqref{H2} and \eqref{H4}. It also indicates that the first corrections in $\omega$ and $\vec{k}^2$ have opposite signs.

\begin{figure}[h]
\begin{center}
\includegraphics[width=0.49\textwidth]{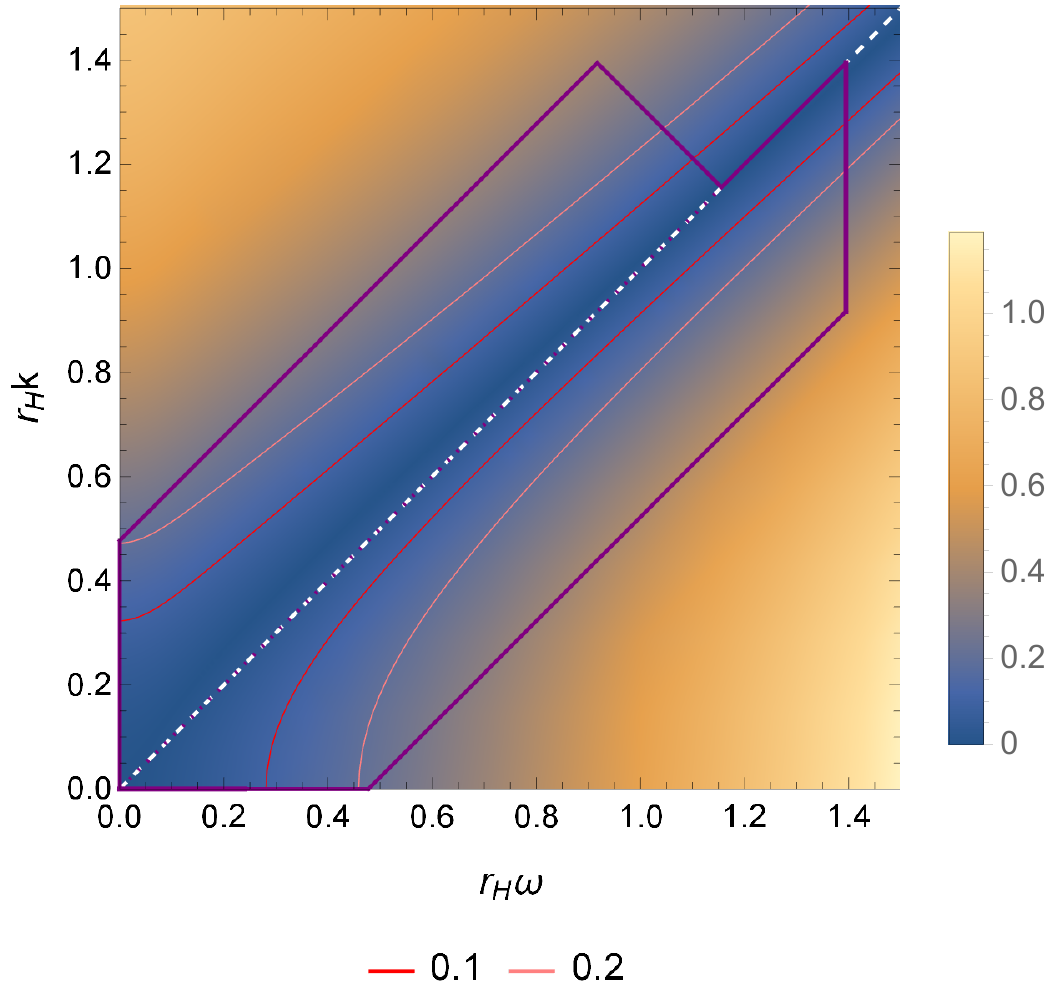}
\includegraphics[width=0.49\textwidth]{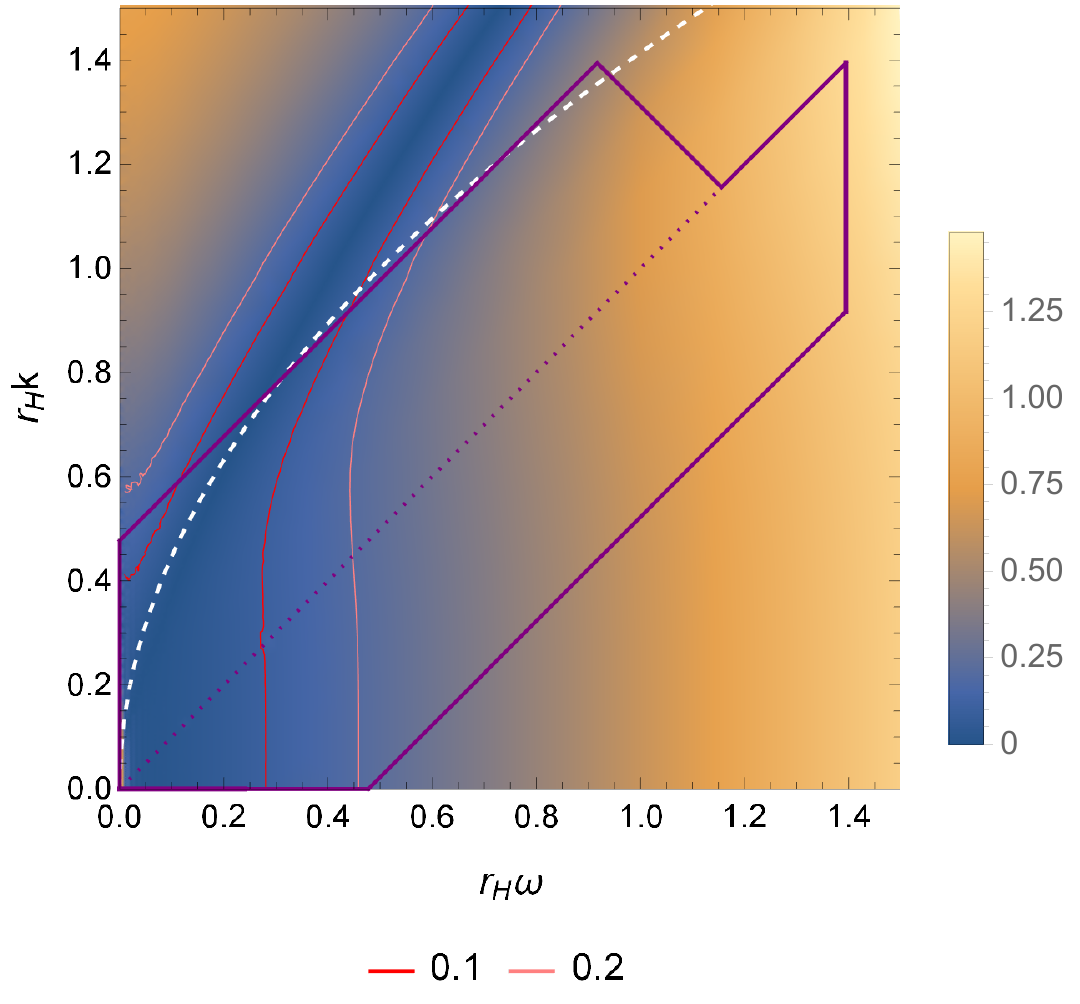}
\caption{Relative difference of the imaginary part of the transverse (\textbf{Left}) and longitudinal (\textbf{Right}) charged current retarded polarization functions with the hydrodynamic approximation. The relative difference is defined as the absolute value of the difference divided by the hydrodynamic approximation. The parameters of the medium are the same as in figures \ref{Fig:ImP} and \ref{Fig:ImP_Hyd}. The red and pink contours correspond respectively to the $10\%$ and $20\%$ lines, and the purple polygon encloses the area that is relevant for the calculation of $j_{e^-}$ and $1/\bar{\l}_{e^-}$. The way this region is determined is detailed in Section \ref{Sec:App}. The white dashed lines indicate the location where the relative difference typically goes to 0. For the transverse part (\textbf{Left}) it corresponds to the line of lightlike momenta $\omega=k$, and for the longitudinal part (\textbf{Right}) to the location of the diffusive peak $\omega  = Dk^2$.}
\label{Fig:RDP_Hyd}
\end{center}
\end{figure}

All in all, the numerical results for the polarization functions presented in this subsection indicate that, at $n_B = 1\,\mathrm{fm^{-3}}$ and for the region in the $(\omega,k)$ plane which is relevant for neutrino transport, the hydrodynamic approximations \eqref{H2} and \eqref{H4} not only reproduce the qualitative features of the exact numerical result, but are also quite good quantitatively. This is especially true for the transverse part of the correlator. This suggests that, in the calculation of the radiative coefficients \eqref{j}-\eqref{lb}, replacing the retarded 2-point function by its leading order hydrodynamic approximation may give a rather good approximation to the coefficients. In the following, we investigate the validity of this statement for a whole range of baryon densities $n_B$ and neutrino energies $E_\n$.

\subsubsection{Radiative coefficients}

We now turn to the analysis of the radiative coefficients themselves, that are the emissivities and absorptions listed in \eqref{j}-\eqref{lb}. More specifically, we shall be analyzing the opacities defined as in \eqref{N1}. We  consider a range of baryon number densities, $n_B$, between $10^{-3}$ and $1\,\mathrm{fm^{-3}}$, and neutrinos energies between $0$ and $100\,\mathrm{MeV}$. These are typical values for neutrinos scattering in a cooling neutron star \cite{Oertel:2020pcg}.

For the parameters of interest, the neutrino chemical potential is positive and large compared with the temperature, so that the emission of anti-neutrinos \eqref{jbe_hD} is suppressed. The two quantities that will be the object of our analysis are therefore
\be
\label{N2} \bar{\kappa}_e(E_\n) \equiv \bar{\ka}_{e^-}(E_\n) + \bar{\ka}_{e^+}(E_\n) = \frac{1}{\bar{\l}_{e^-}(E_\n)} + \frac{1}{\bar{\l}_{e^+}(E_\n)} \, ,
\ee
\be
\nn \kappa_e(E_\n) \equiv \ka_{e^-}(E_\n) = j_{e^-}(E_\n)+ \frac{1}{\l_{e^-}(E_\n)} \, .
\ee

We first present in figure \ref{Fig:op12} plots of the opacities at the two extreme values of baryonic density that we considered,
\be
 n_B^{(1)} = 10^{-3}\,\mathrm{fm^{-3}}\quad \mathrm{and}\quad n_B^{(2)} = 1\,\mathrm{fm^{-3}}\, .
\ee
which correspond to the values of the chemical potentials
\be
 \mu_1 \simeq 47\,\text{MeV} \quad \mathrm{and}\quad \mu_2 \simeq 650\,\text{MeV} \, ,
\ee
respectively. We also show in figure \ref{Fig:opnB} the full density dependence of the opacity at zero neutrino energy. The qualitative behavior is essentially dictated by statistics, so it is the same that was observed in previous works, for example\footnote{Note that the neutrino chemical potential was negative in \cite{Oertel:2020pcg}, whereas it is positive here. This implies that the role of neutrinos and anti-neutrinos are exchanged with respect to \cite{Oertel:2020pcg}.} in \cite{Oertel:2020pcg}. The anti-neutrino opacity increases with the baryon density and the neutrino energy. As for the neutrino opacity, it also increases with density, but has a different behavior as a function of the neutrino energy: it is more or less a constant until a threshold located near $E_\n \sim \mu_\n$, where it decreases in relative value to a number of order $\OO(T/\mu)^2$. An estimate of the typical magnitude and parameter dependence of the opacities at $E_\n \lesssim T$ is given by \eqref{A13}, which was derived within the diffusive approximation.

\begin{figure}[h]
\begin{center}
\includegraphics[width=0.49\textwidth]{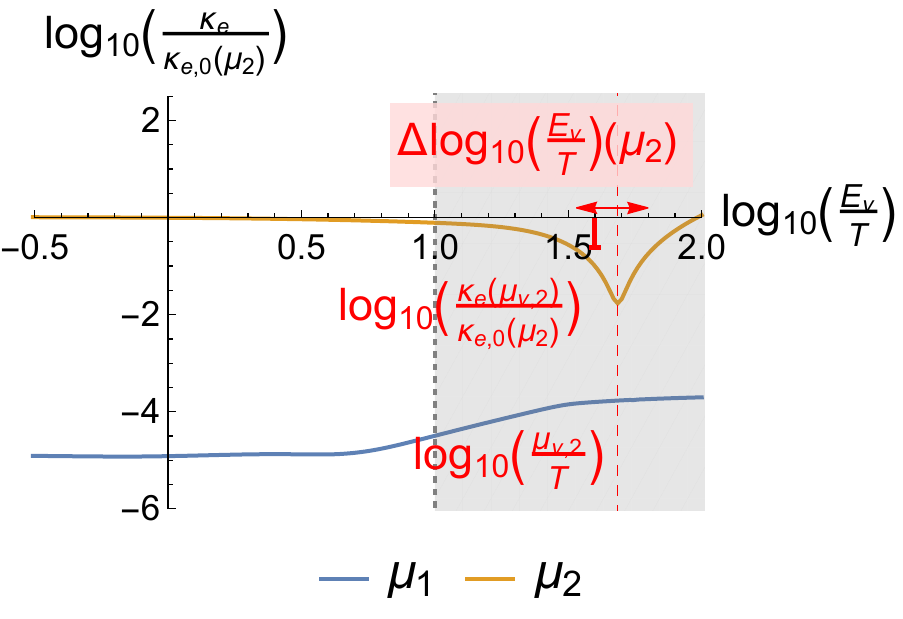}
\includegraphics[width=0.49\textwidth]{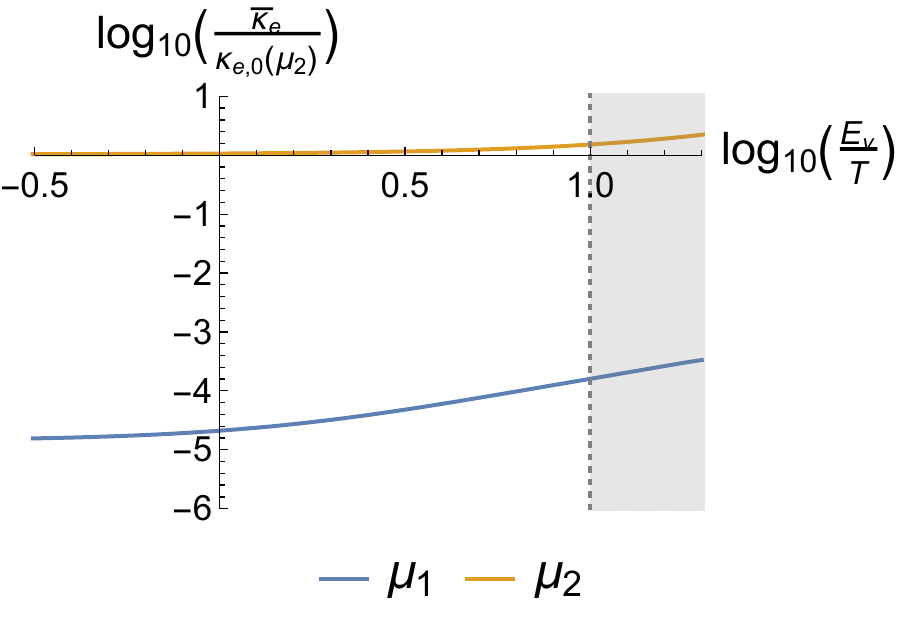}
\caption{Neutrino (\textbf{Left}) and anti-neutrino (\textbf{Right}) opacities as a function of the neutrino energy $E_\n$. We show two values of the baryon density, $n_B^{(1)} = 10^{-3}\,\mathrm{fm^{-3}}$ and $n_B^{(2)} = 1\,\mathrm{fm^{-3}}$, associated to the corresponding chemical potentials $\m_1$ and $\m_2$.  The opacities are normalized by the typical value at $E_\n = 0$ \eqref{A13}, evaluated for the largest density $n_B^{(2)}$. In the plot for the neutrino opacity, the typical scales that control the location, depth \eqref{A9} and width \eqref{A10} of the observed dip are indicated, where we defined $\D\log_{10}(E_\n/T)(\m) \equiv \log_{10}(\m_\n + \D E_\n(\m)/2)-\log_{10}(\m_\n - \D E_\n(\m)/2)$. The region in gray corresponds to neutrino energies larger than $100\,\mathrm{MeV}$, which is not expected to be relevant for transport in a neutron star. It is included in the plot in order to show the complete qualitative behavior of the neutrino opacities as a function of $E_\n$, including the threshold at $E_\n = \m_\n$.}
\label{Fig:op12}
\end{center}
\end{figure}
\begin{figure}[h]
\begin{center}
\includegraphics[width=0.75\textwidth]{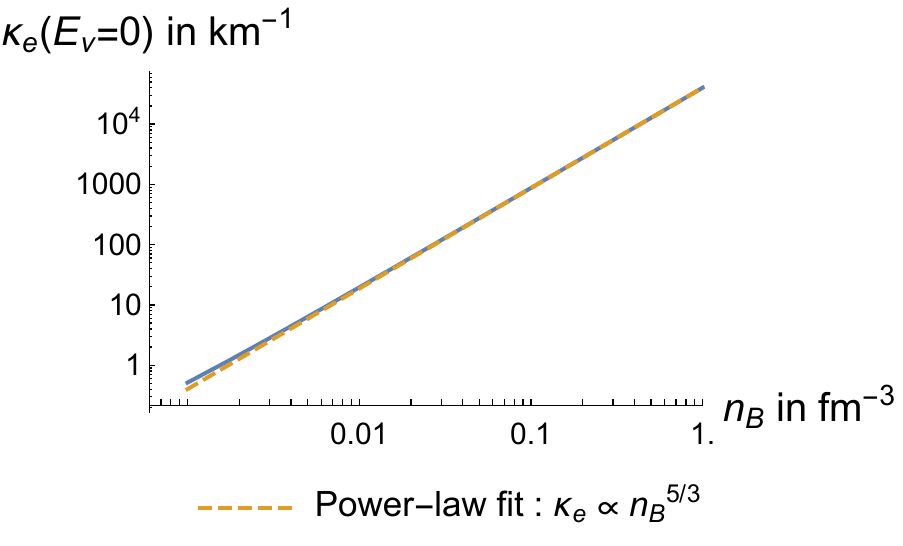}
\caption{Neutrino opacity at zero neutrino energy, as a function of the baryon density $n_B$. At $E_\n=0$, the opacities are the same for neutrinos and anti-neutrinos. The dashed orange line shows the asymptotic power-law behavior at $n_B \gg T^3$, $\kappa_e \sim n_B^{5/3}$ (see equation \eqref{C2} in the next subsection).}
\label{Fig:opnB}
\end{center}
\end{figure}

To estimate quantitatively the accuracy of the approximations presented in Section \ref{Sec:App}, the opacities were numerically computed and compared to the results from the approximations over the whole 2-dimensional parameter space of baryon number density $n_B = 10^{-3} - 1\,\mathrm{fm^{-3}}$ and neutrino energy $E_\n = 0-100\,\mathrm{MeV}$. We discuss in turn the degenerate, hydrodynamic, diffusive and diffusive and degenerate approximations, analyzing each time both the neutrino and anti-neutrino opacities.

\begin{figure}[h]
\begin{center}
\includegraphics[width=0.49\textwidth]{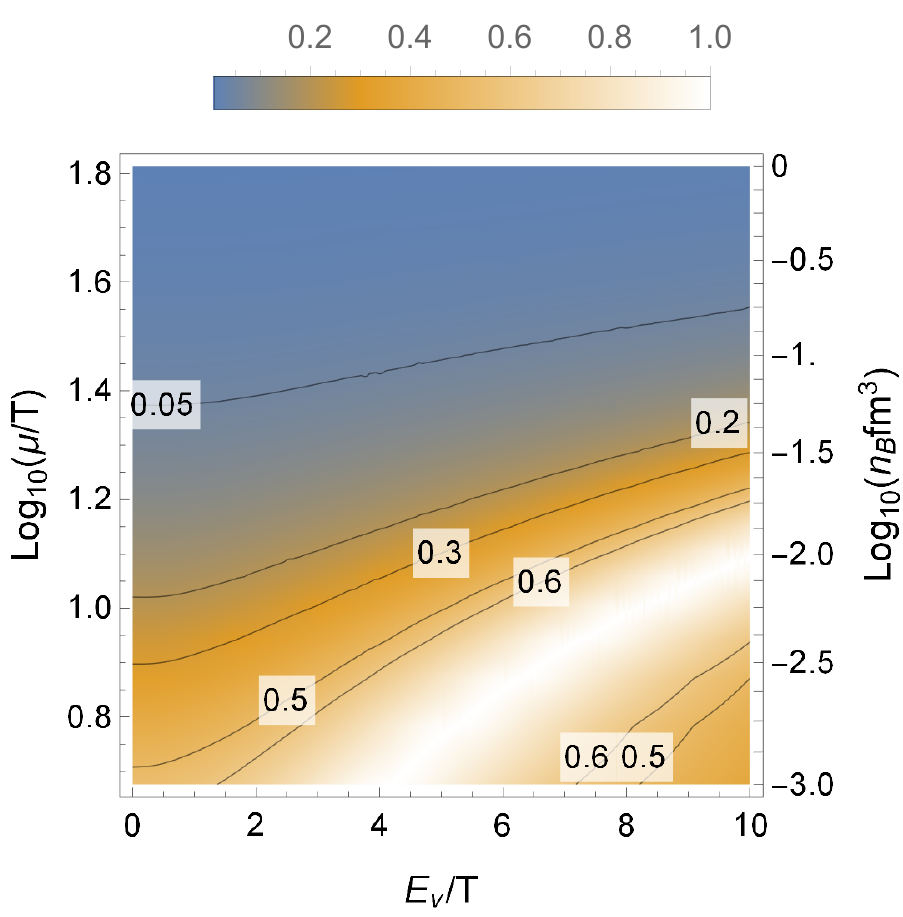}
\includegraphics[width=0.49\textwidth]{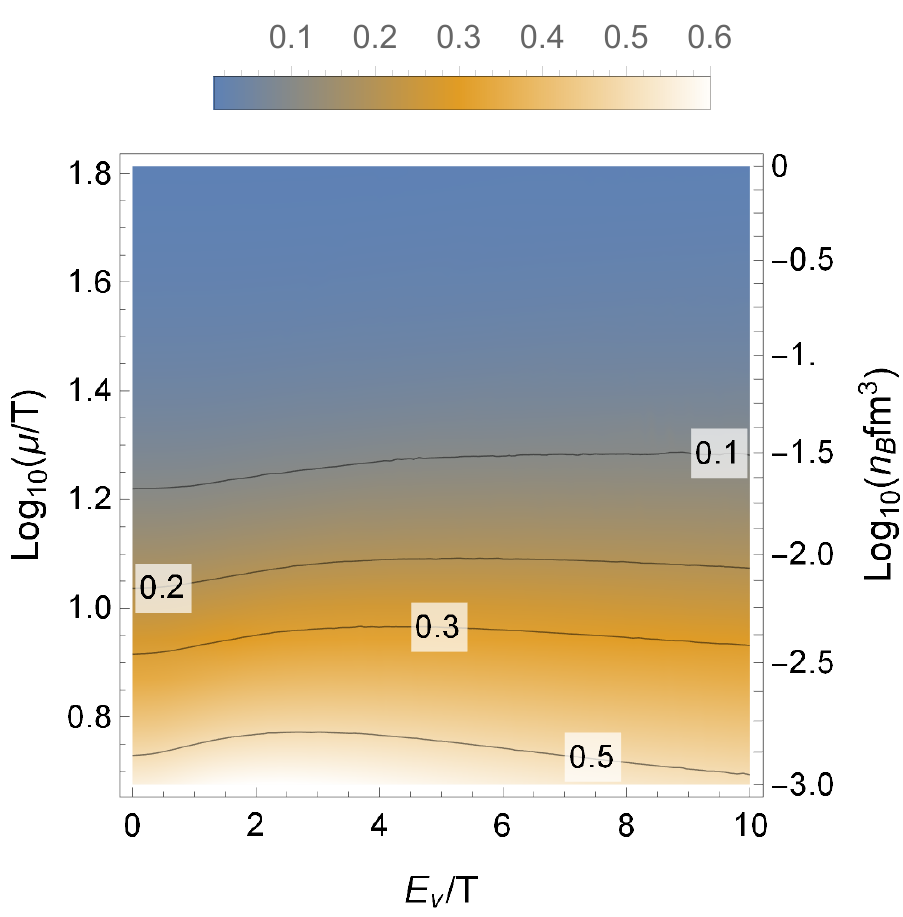}
\caption{Relative difference between the degenerate approximation and the exact opacities, for neutrinos (\textbf{Left}) and anti-neutrinos (\textbf{Right}). }
\label{Fig:RDdegen}
\end{center}
\end{figure}

Figure \ref{Fig:RDdegen} shows the relative difference between the exact opacities calculated numerically and the opacities calculated within the degenerate approximation. In the case of the neutrino opacity, the approximation is worst on the curve $E_\n = \m_\n$, where the degenerate approximation goes to 0 whereas the exact result remains finite. Apart from that curve, the magnitude of the error is of about $5$ to $30\%$ over most of the parameter space, and it reduces when $|E_\m - \m_\n|$ increases, that is both at larger baryon density and larger neutrino energy.

For anti-neutrinos, the degenerate approximation becomes very good at high density, and it reaches less than $5\%$ of error at $n_B = n_B^{(2)}$. On the contrary, the approximation becomes unreliable for the lowest densities, reaching more than $50\%$ at $n_B = n_B^{(1)}$. The error depends only marginally on the neutrino energy over the range investigated.

\begin{figure}[h]
\begin{center}
\includegraphics[width=0.49\textwidth]{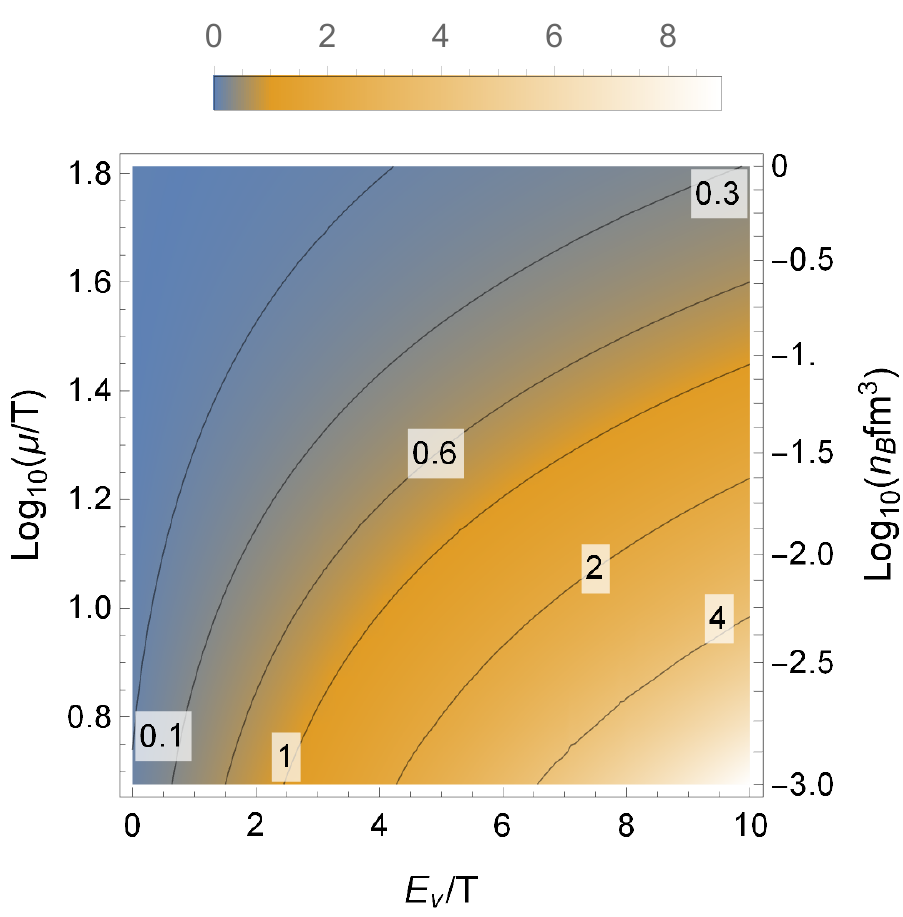}
\includegraphics[width=0.49\textwidth]{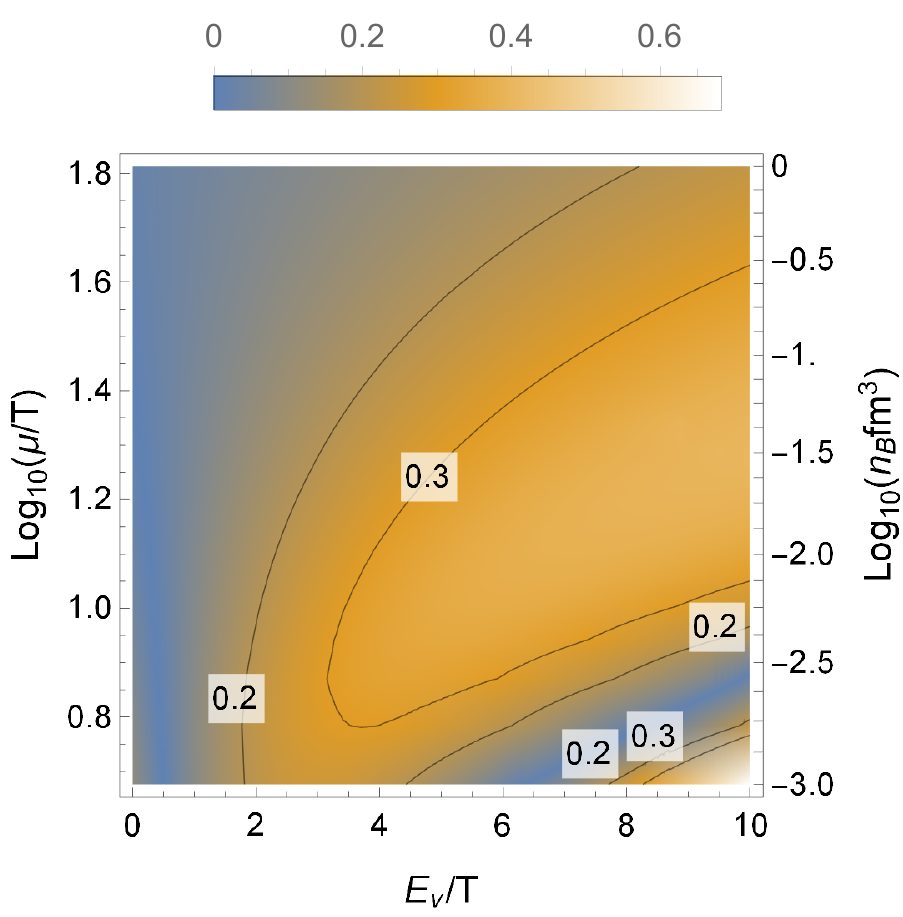}
\caption{Relative difference between the hydrodynamic approximation and the exact transverse opacities, for neutrinos (\textbf{Left}) and anti-neutrinos (\textbf{Right}). }
\label{Fig:RDhydT}
\end{center}
\end{figure}

\begin{figure}[h]
\begin{center}
\includegraphics[width=0.49\textwidth]{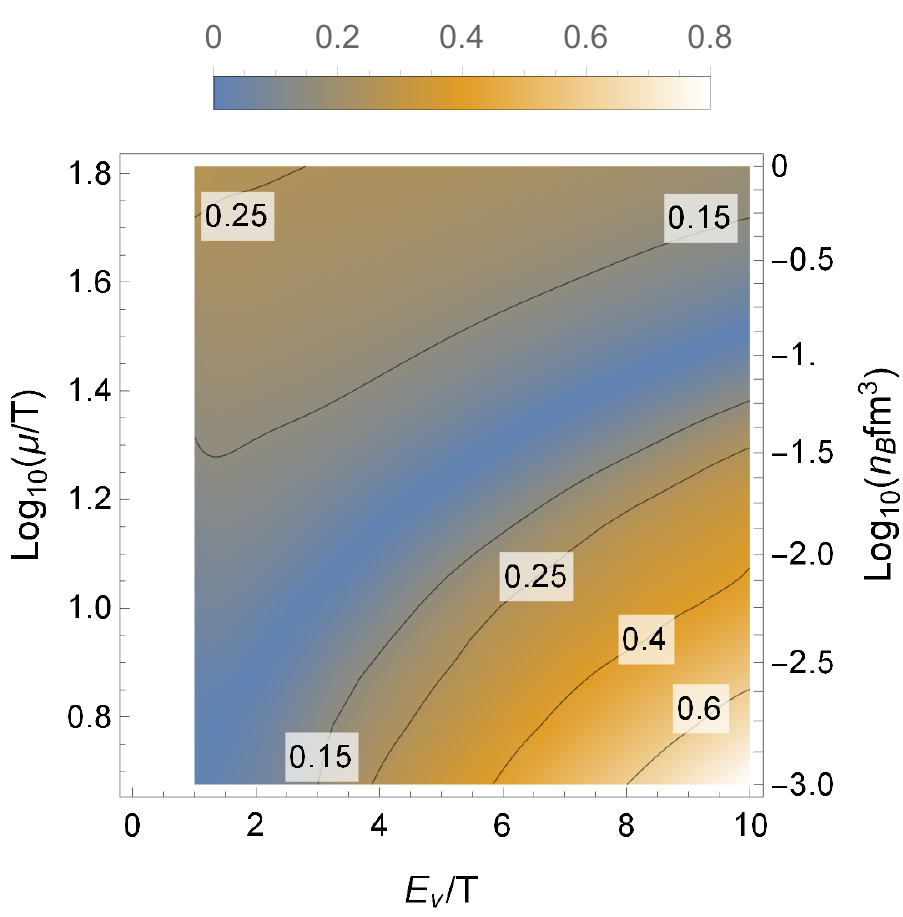}
\includegraphics[width=0.49\textwidth]{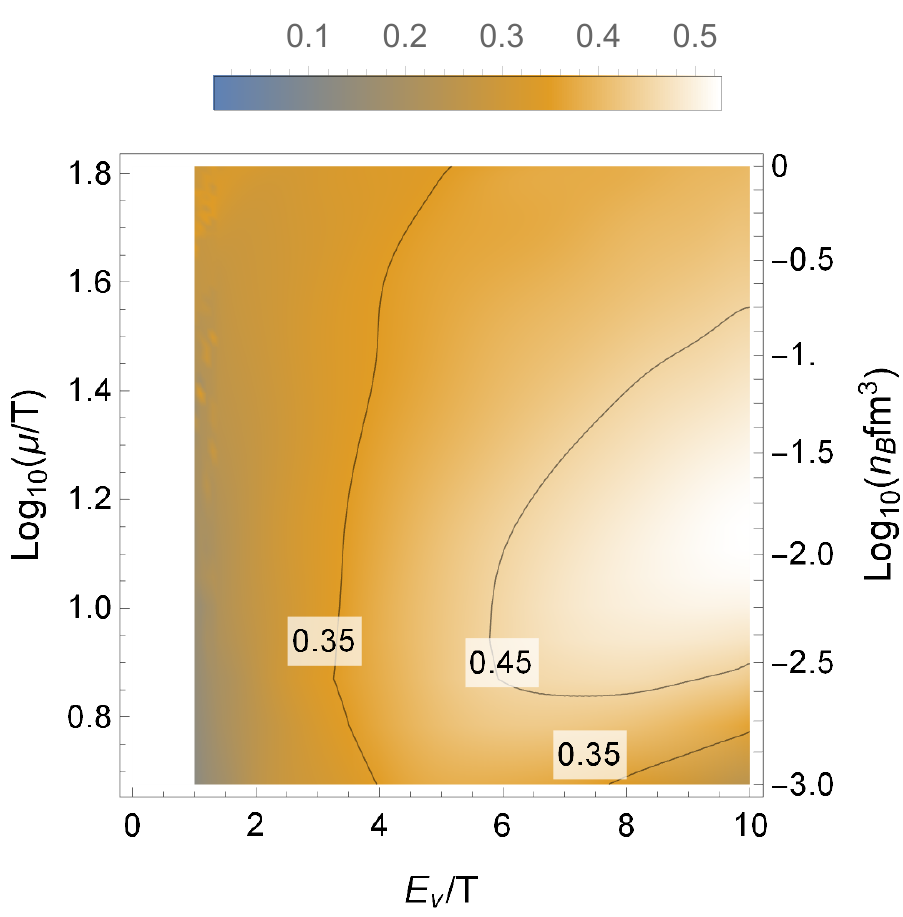}
\caption{Relative difference between the hydrodynamic approximation and the exact longitudinal opacities, for neutrinos (\textbf{Left}) and anti-neutrinos (\textbf{Right}). We do not show the region of small $E_\n$ since there are large numerical errors there. The reason is that the longitudinal opacity becomes very small in this region, of order $\OO(m_e/\m_\n)^4$, with $m_e$ the electron mass.}
\label{Fig:RDhydL}
\end{center}
\end{figure}

\begin{figure}[h]
\begin{center}
\includegraphics[width=0.49\textwidth]{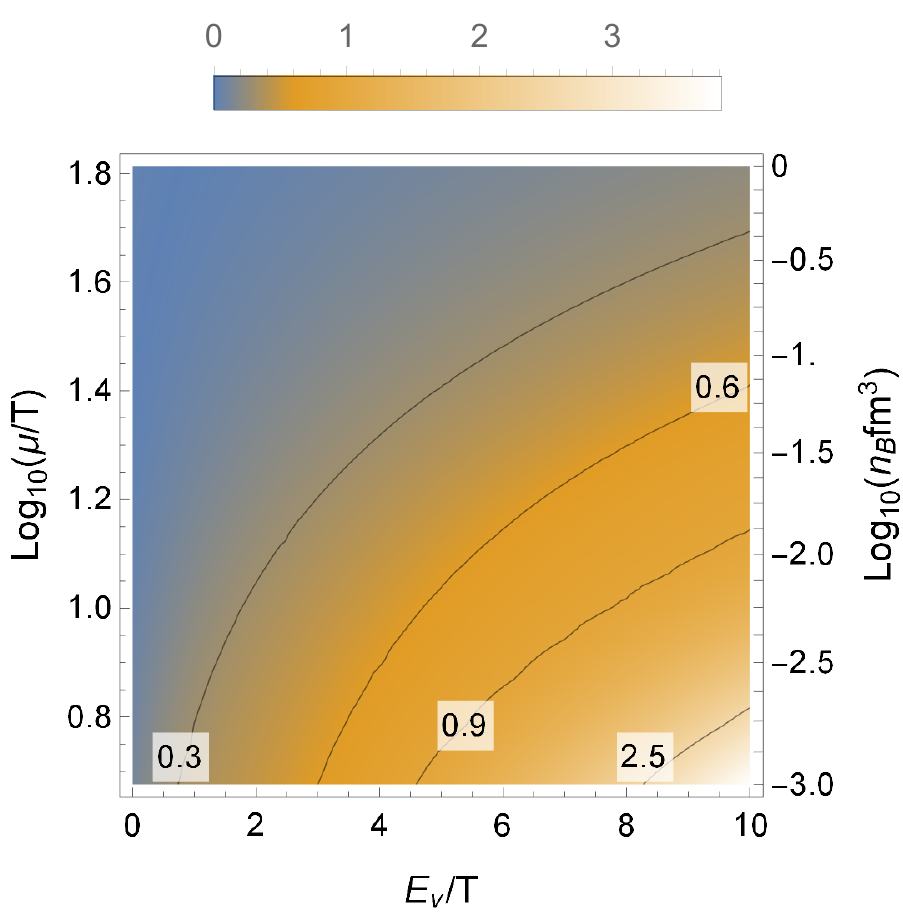}
\includegraphics[width=0.49\textwidth]{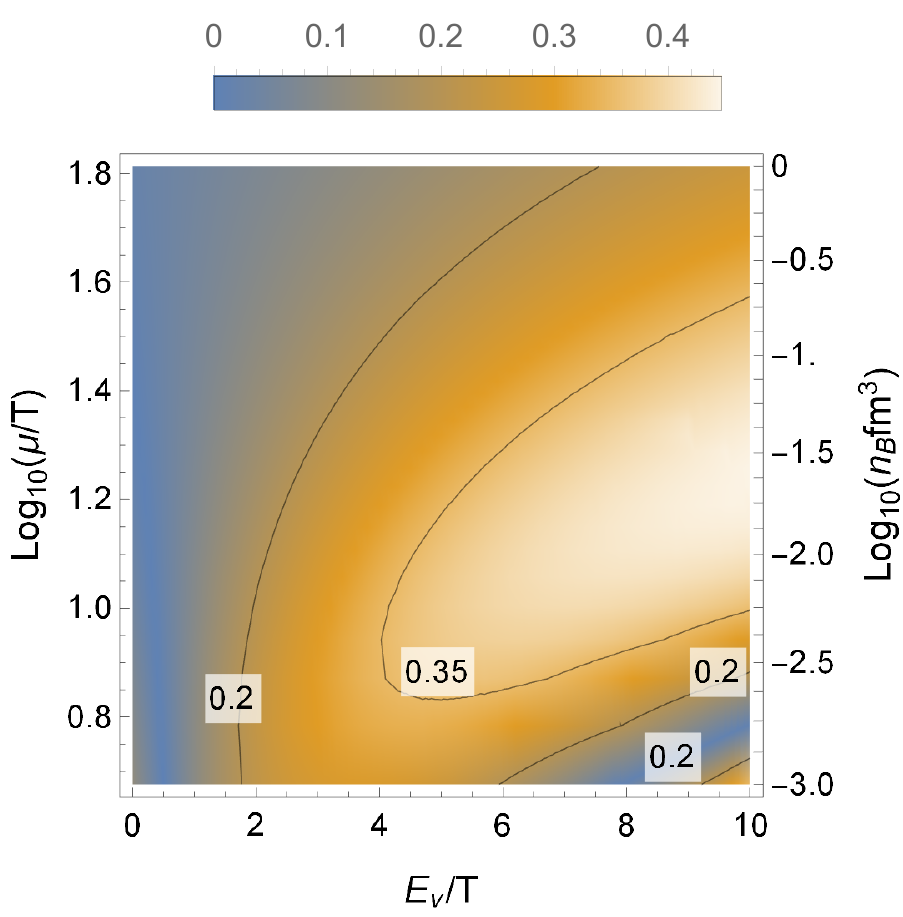}
\caption{Relative difference between the hydrodynamic approximation and the exact total opacities, for neutrinos (\textbf{Left}) and anti-neutrinos (\textbf{Right}). }
\label{Fig:RDhyd}
\end{center}
\end{figure}

We now focus on the hydrodynamic approximation. We start by discussing separately the transverse and longitudinal components, for which the relative difference between the hydrodynamic and the exact opacity is respectively shown in figures \ref{Fig:RDhydT} and \ref{Fig:RDhydL}.

The qualitative behavior for the transverse part of the opacity is similar for neutrinos and anti-neutrinos: the error becomes larger at higher neutrino energy and smaller at higher density. There is an exception to this trend in the case of anti-neutrinos, for which the approximation crosses the exact result on a curve at low density and high energy (which appears as a blue line in the bottom-right corner of right Figure \ref{Fig:RDhydT}). In both cases, the error is smaller than $30\%$ for $n_B = n_B^{(2)}$ or $E_\n < T$. However, the error grows large at low density and high energy, reaching more than $800\%$ at $n_B = n_B^{(1)}$ and $E_\n = 10T$ in the case of neutrinos, and about $70\%$ for anti-neutrinos. That being said, we note that the error is smaller than $40\%$ over most of the parameter space for anti-neutrinos, whereas the error on the neutrino opacity is already larger than $60\%$ for a wide range of energies at $n_B < 0.1\,\mathrm{fm^3}$.

In short, the main information from figure \ref{Fig:RDhydT} is that the hydrodynamic approximation to the transverse part of the opacities is reasonably good at high density and low $E_\n$, but becomes mostly unreliable at low density and high $E_\n$. The situation turns out to be better for the anti-neutrino opacity, which remains quite good over the whole parameter space that was investigated. This last point is most probably accidental, and we would expect no significant difference were we to take into account all possible values of energies and densities.

The error from the hydrodynamic approximation to the longitudinal part of the opacities is shown in figure \ref{Fig:RDhydL}. The qualitative behavior of the error on the neutrino opacity is essentially similar to the transverse case, apart from the presence of a curve where the error vanishes (in blue in the left of figure \ref{Fig:RDhydL}). Quantitatively, the error is smaller than the transverse case: at high density it is of about $20\%$, and although it grows large at low density and high energy, it remains below $80\%$. As for the error on the anti-neutrino opacity, it is very similar to the transverse case. It depends marginally on the parameters over the range that is investigated, and is typically comprised between $30\%$ and $50\%$. All in all, the comparison between figures \ref{Fig:RDhydL} and \ref{Fig:RDhydT} indicate a similar qualitative behavior, but with the approximation to the longitudinal opacity typically more accurate than for the transverse opacity.

Finally, figure \ref{Fig:RDhyd} shows the error from the hydrodynamic approximation for the total opacity, which is the sum of the transverse and longitudinal parts. On the whole, figure \ref{Fig:RDhyd} looks similar to the figure for the transverse part (figure \ref{Fig:RDhydT}). This indicates that the latter contributes more to the opacity than the longitudinal part over a large part of the parameter space. This is true for sure at small neutrino energy, where the longitudinal part becomes very small, of order $\OO(m_e/\m_\n)^4$. The contribution from the longitudinal part implies that the error on the total opacity is lower than for the transverse part.

On the whole, figure \ref{Fig:RDhyd} indicates that the hydrodynamic approximation to the opacities is reasonably accurate at high density and/or low neutrino energy, whereas it becomes unreliable in the opposite limit. More precisely, the following quantitative results are observed
\begin{itemize}
\item The error from the hydrodynamic approximation to the neutrino opacity is between $0$ and $40\%$ for densities $n_B>10^{-1}\,\mathrm{fm^{-3}}$ or neutrino energies $E_\n < 20\,\text{MeV}$. The error exceeds $100\%$ for densities typically smaller than $3\times 10^{-3}\,\mathrm{fm^{-3}}$ and $E_\n > 60\,\text{MeV}$. Some additional information can be extracted from the comparison of figure \ref{Fig:RDhyd} with the plot for the degenerate approximation (figure \ref{Fig:RDdegen}). In figure \ref{Fig:RDdegen}, the white line where the error is equal to $100\%$ corresponds to the place where $E_\n = \m_\n$. In the degenerate limit, this line separates the region where the neutrino opacity is dominated by emissivity $j_{e^-}$ from that where it is dominated by the absorption $1/\l_{e^-}$ (see Table \ref{tab:cEn}). Since the location of the $E_\n = \m_\n$ line is close to the contour at $90\%$ accuracy in the right of figure \ref{Fig:RDhyd}, this means that the hydrodynamic approximation describes reasonably well the neutrino emissivity, whereas it is very rough as far as the absorption is concerned.

\item The accuracy of the hydrodynamic approximation to the anti-neutrino opacity is between $10$ and $40\%$ over the range of parameters that was considered.
\end{itemize}

\begin{figure}[h]
\begin{center}
\includegraphics[width=0.49\textwidth]{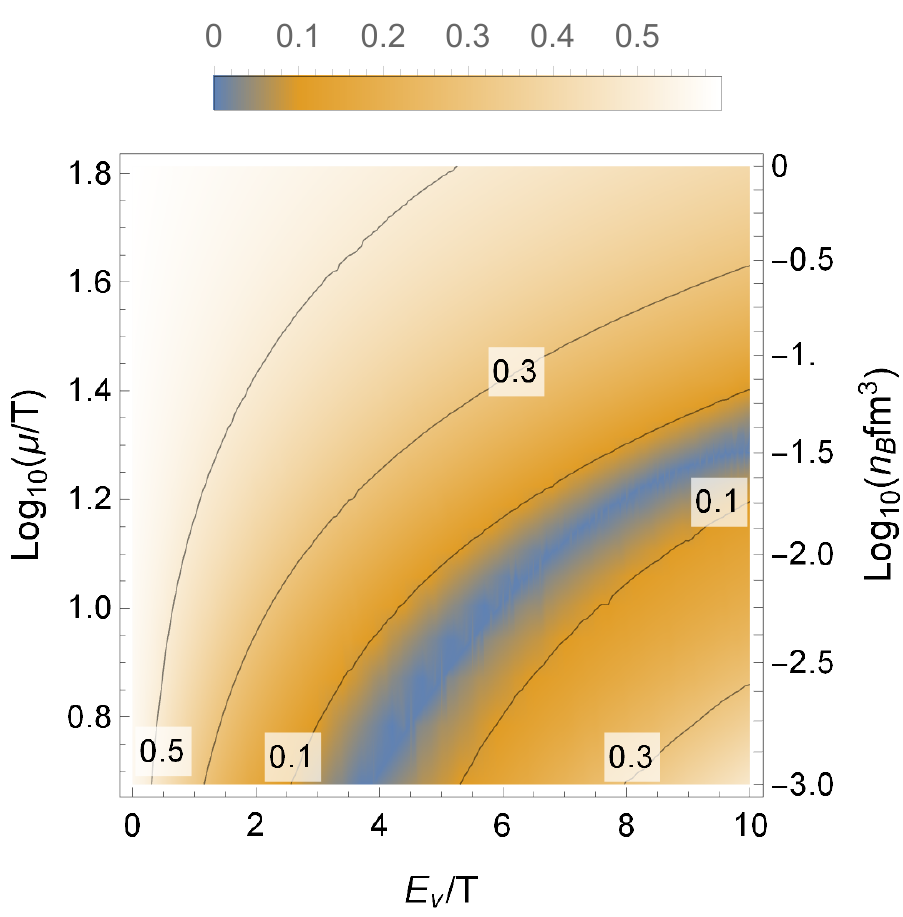}
\includegraphics[width=0.49\textwidth]{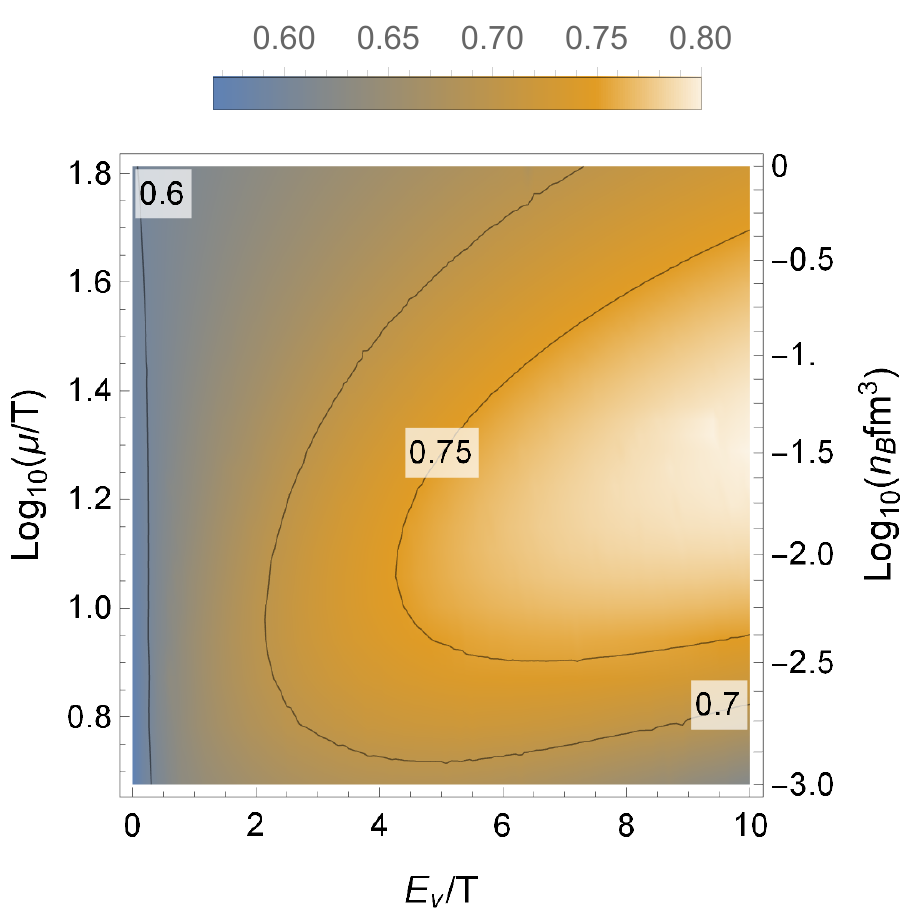}
\caption{Relative difference between the diffusive approximation and the exact opacities, for neutrinos (\textbf{Left}) and anti-neutrinos (\textbf{Right}). }
\label{Fig:RDdiff}
\end{center}
\end{figure}

\begin{figure}[h]
\begin{center}
\includegraphics[width=0.49\textwidth]{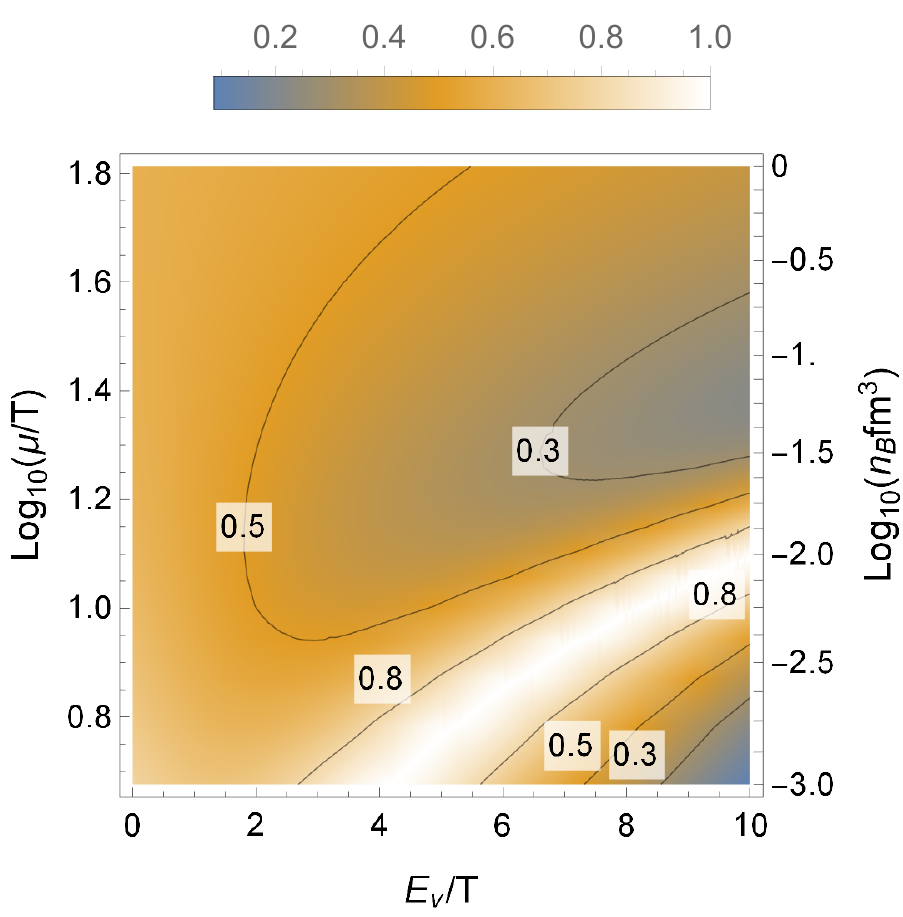}
\includegraphics[width=0.49\textwidth]{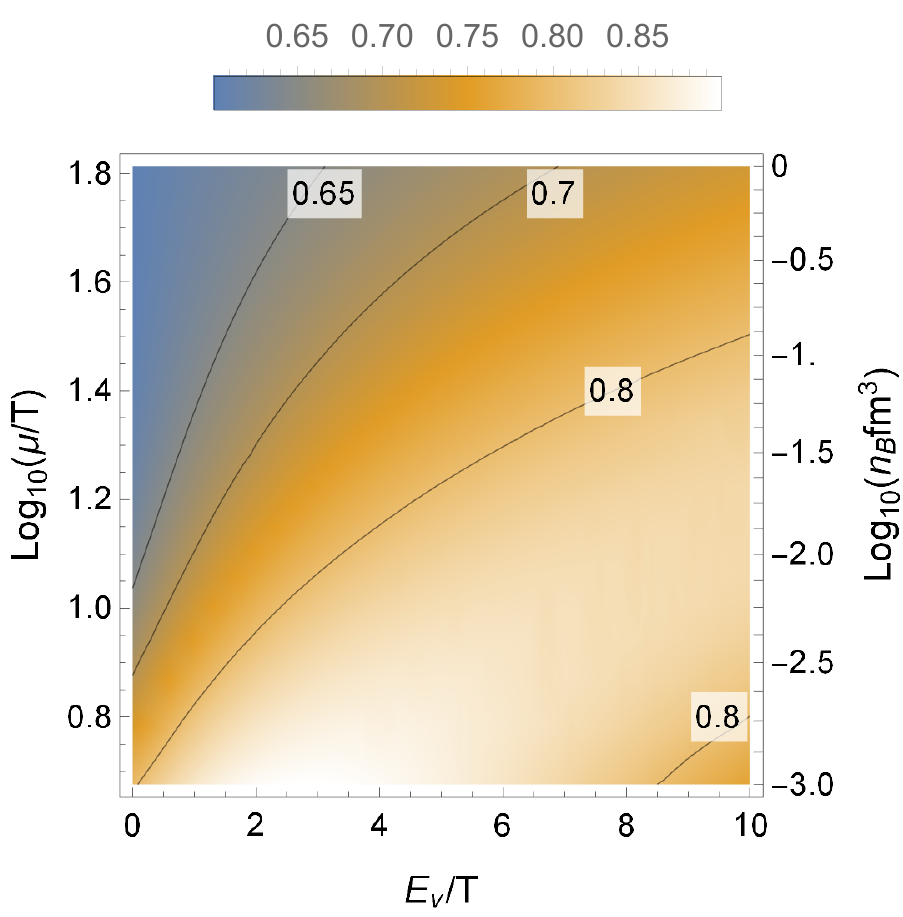}
\caption{Relative difference between the diffusive and degenerate approximation and the exact opacities, for neutrinos (\textbf{Left}) and anti-neutrinos (\textbf{Right}). }
\label{Fig:RDdiffdegen}
\end{center}
\end{figure}

The last approximation that we investigate is the so-called diffusive approximation, in which only the contribution from the time-time component of the current-current correlators is included in the hydrodynamic approximation. The error for this approximation
is shown in figure \ref{Fig:RDdiff}. On the whole, the error for neutrinos is observed to take values typically between $20$ and $60\%$, whereas for anti-neutrinos the range is between $60$ and $80\%$. For neutrinos, the difference is observed to vanish on one curve in the parameter space, which is located close to the curve $E_\n=\m_\n$ (the white curve at $100\%$ error in figure \ref{Fig:RDdegen}). Comparing with the previous analysis of the hydrodynamic approximation, we see that the accuracy of the diffusive approximation is worse for anti-neutrinos. In the case of neutrinos, the comparison depends on the parameters: although the diffusive approximation is found to be less accurate at low energy or high density, the error is actually much smaller at high energy and low density.

To conclude, the general outcome observed in figure \ref{Fig:RDdiff} is that, as expected, the diffusive approximation is essentially less accurate than the hydrodynamic approximation. However, there is an exception in the case of the neutrino opacity, at low density and high energy. The latter is due to accidental cancellations, which are not expected to occur for general setups.

For completeness, we also analyze the accuracy of the crudest approximation to the opacities, which is obtained by combining the degenerate approximation with the diffusive approximation. This specific approximation is the one that is used to derive the approximate expressions \eqref{A8} and \eqref{A11}-\eqref{A12}. The error for this diffusive and degenerate approximation is shown in figure \ref{Fig:RDdiffdegen}. As expected, the error is dominated by the degenerate approximation at low density, whereas the main cause of error comes from the diffusive approximation at high density.

\begin{figure}[h]
\begin{center}
\includegraphics[width=0.49\textwidth]{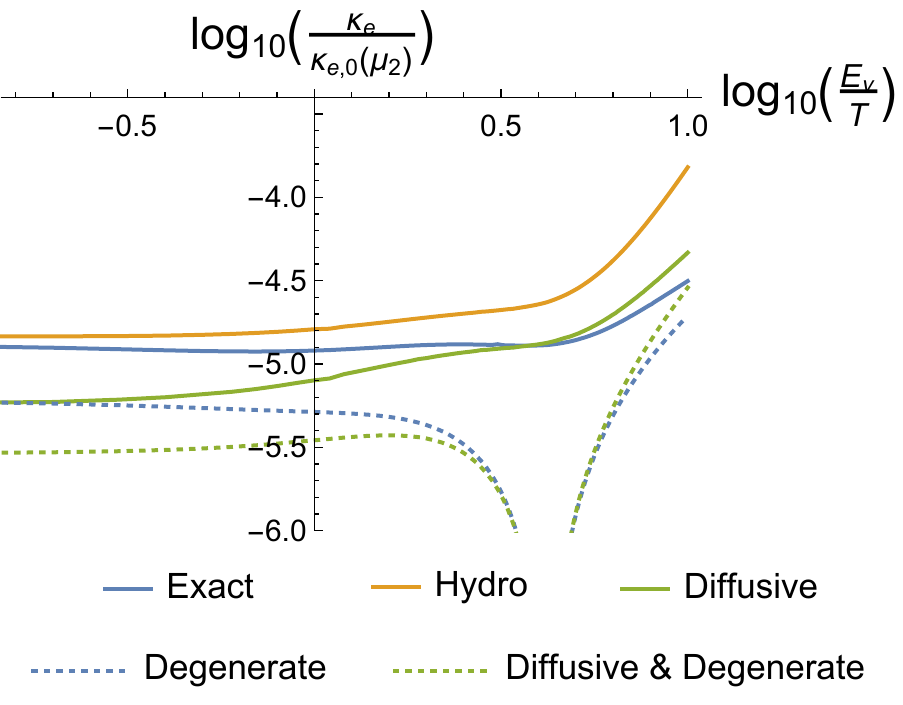}
\includegraphics[width=0.49\textwidth]{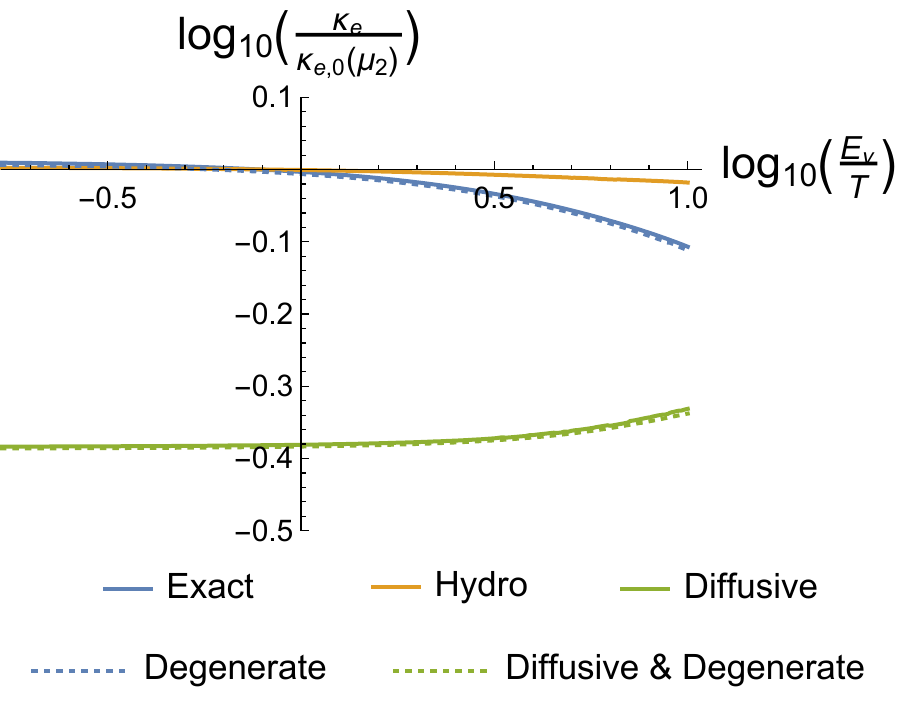}
\caption{Neutrino opacity $\kappa_{e^-}$ and its various approximations as a function of the neutrino energy $E_\n$, for $n_B = n_B^{(1)}$(\textbf{Left}) and $n_B = n_B^{(2)}$ (\textbf{Right}). The opacity is normalized by the approximate value at $E_\n = 0$ \eqref{DA23}, evaluated at $n_B=n_B^{(2)}$.}
\label{Fig:AppsK}
\end{center}
\end{figure}

\begin{figure}[h]
\begin{center}
\includegraphics[width=0.49\textwidth]{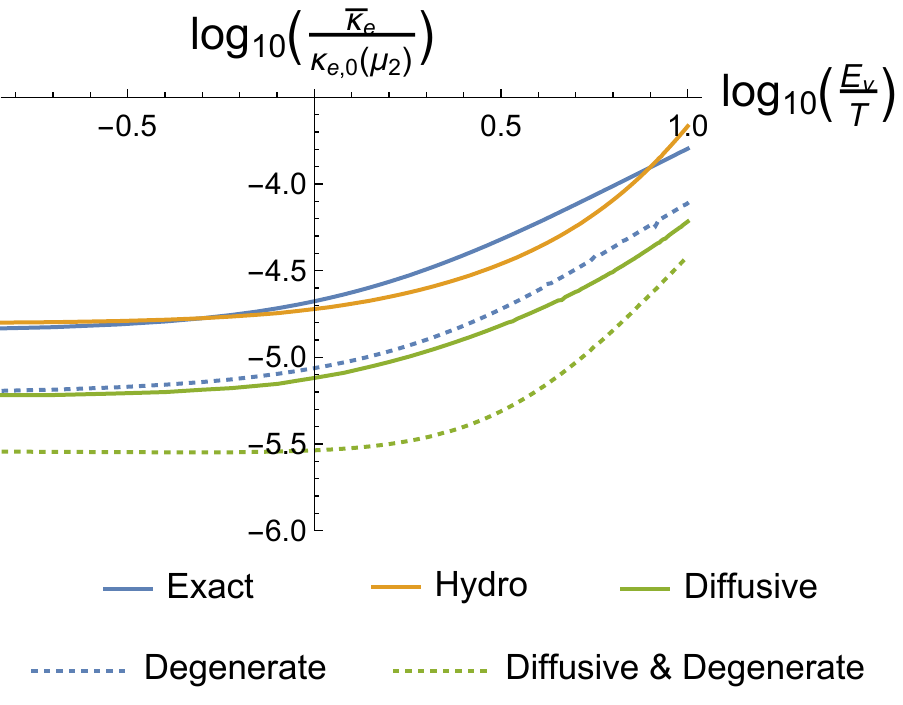}
\includegraphics[width=0.49\textwidth]{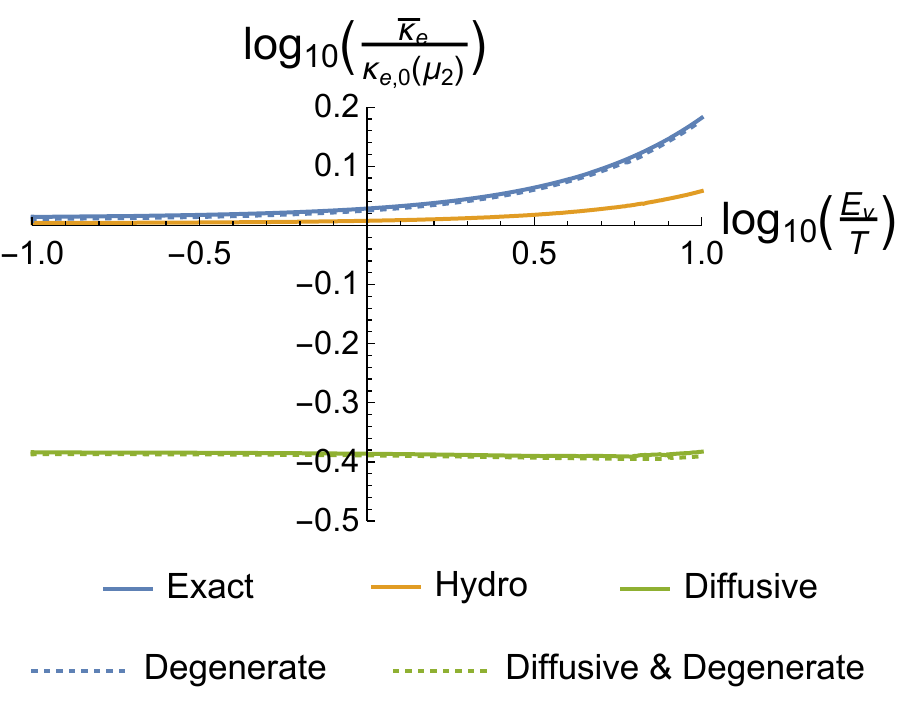}
\caption{Anti-neutrino opacity $\bar{\kappa}_{e^-}$ and its various approximations as a function of the neutrino energy $E_\n$, for $n_B = n_B^{(1)}$(\textbf{Left}) and $n_B = n_B^{(2)}$ (\textbf{Right}). The opacity is normalized by the approximate value at $E_\n = 0$ \eqref{DA23}, evaluated at $n_B=n_B^{(2)}$.}
\label{Fig:AppsKb}
\end{center}
\end{figure}

In order to give another view on the above analysis, we compare in figures \ref{Fig:AppsK} and \ref{Fig:AppsKb} the exact opacity computed numerically with the various approximations at the two extreme values considered for the baryon density. Figure \ref{Fig:AppsK} shows the result for neutrinos and figure \ref{Fig:AppsKb} for anti-neutrinos. These figures illustrate the general results from the analysis of this section. The degenerate approximation is good at high density and becomes unreliable at low density. At high density $n_B = n_B^{(2)}$, the hydrodynamic approximation is within a few tens of percents of error from the exact result, whereas the diffusive one is off by about a factor of 2. At low density $n_B = n_B^{(1)}$, both approximations tend to lose accuracy. However, they appear in some cases to be quite close to the exact result over the range of neutrino energies considered here, which is due to accidental crossings with the exact result. This happens for the diffusive approximation in the case of neutrinos, and for the hydrodynamic approximation in the case of anti-neutrinos.

The main conclusion from the analysis of this subsection is that, for neutrino energies of a few times the temperature, the accuracy of the hydrodynamic approximation depends on the baryonic density. At $n_B \gtrsim 10^{-1}\,\mathrm{fm^{-3}}\big(T/(10\,\text{MeV})\big)^3$, the holographic opacities are quite well approximated by using the leading order hydrodynamic expressions of the correlators \eqref{H2} and \eqref{H4}, whereas higher order corrections become large at lower densities. In practice, this means that, for the highest densities realized in neutron stars, computing only the leading order flavor transport coefficients $\s$ and $D$ from the holographic model is a sufficient input to obtain a good estimate of the neutrino opacities. At lower densities, higher order transport coefficients are required to produce a reasonable approximation. Eventually, as $\m/T$ becomes of order 1, $r_H E_\n$ becomes significantly larger than 1 for $E_\n \gtrsim T$. When this is the case, the hydrodynamic expansion cannot be used anymore, and the full holographic calculation of the chiral current 2-point function is needed to compute the opacities.

\subsection{Comparison with other calculations}

We conclude this section by comparing the results obtained for the neutrino opacities with other calculations from the literature. We start by evaluating the typical order of magnitude of the opacities and compare it with other references. The leading process of neutrino emission is known to be qualitatively different in quark matter compared with nuclear matter. Indeed, whereas the direct Urca process is kinematically suppressed in nuclear matter, it can be realized in quark matter \cite{Iwamoto:1980eb,Burrows:1980ec}. Since our calculations take place in deconfined matter, the second part of this subsection brings the focus on the comparison with previous results in quark matter.

In Appendix \ref{Sec:DiffApp}, approximate expressions for the opacities are derived within the diffusive and degenerate approximation (defined in Section \ref{Sec:App}). In particular, at zero neutrino energy the final result is given by \eqref{DA23}
\be
\label{C1} \kappa_{e,0}(n_B) = \frac{G_F^2|M_{ud}|^2}{2304} \left(3\pi^4\right)^{\frac{1}{6}} N_c^{\frac{1}{2}} (M\ell)^7w_0^{25/3}\big(\mu(n_B)\big)^5 + \OO(\e^4) \, ,
\ee
where $\e$ is the small parameter of the hydrodynamic expansion \eqref{A5}. We compare \eqref{C1} with the exact value of the opacity at $E_\n = 0$ in figure \ref{Fig:ka0}, which shows that \eqref{C1} reproduces the correct order of magnitude for the opacity at $n_B > 10^{-2}\,\mathrm{fm^{-3}}$. Substituting the numerical values of the parameters results in the following number
\be
\label{C2} \kappa_{e^-}(0) \simeq 6.2\times 10^2\, \mathrm{km^{-1}}\, \left(\frac{n_B}{0.1\,\mathrm{fm^{-3}}}\right)^{\frac{5}{3}} \,   \left(\frac{(M\ell)^3}{(M\ell)^3_{\text{free}}}\right)^{-\frac{1}{2}}\left(\frac{w_0^2(M\ell)^3}{(w_0^2(M\ell)^3)_{\text{free}}}\right)^{\frac{5}{6}} \, ,
\ee
where we also used \eqref{mueq2} to express $\m$ as a function of the baryon density $n_B$.
\begin{figure}[h]
\begin{center}
\includegraphics[scale=1.]{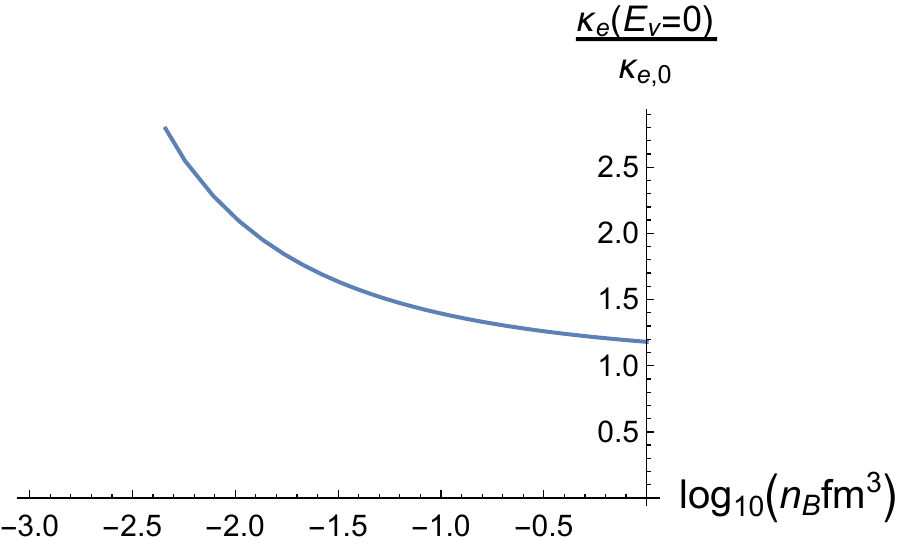}
\caption{The ratio of the neutrino opacity at $E_\n = 0$ over the approximate expression from the hydrodynamic and degenerate approximation \eqref{C1}. The ratio is shown as a function of the baryon density in $\mathrm{fm^{-3}}$.}
\label{Fig:ka0}
\end{center}
\end{figure}

It is interesting to compare \eqref{C2} with the values of the opacities that are currently used in numerical simulations of neutrino transport. The most common rates that are used to describe neutrino transport in nuclear matter are based on mean-field calculations \cite{Reddy:1997yr}, often completed by the random phase approximation to include some degree of nucleon-nucleon correlations \cite{Burrows:1998cg,Reddy:1998hb}.  The latest results using these methods in the non-relativistic regime are summarized in \cite{Oertel:2020pcg}. To compare with \eqref{C2}, we note that the opacities computed in \cite{Oertel:2020pcg} at $E_\n = 0$ and $n_B = 0.11\,\mathrm{fm^{-3}}$ are between 1 and about $30\,\mathrm{km^{-1}}$. The value that we obtained \eqref{C2} is therefore about one order of magnitude larger than the largest opacities from \cite{Oertel:2020pcg}. Also, in \cite{Oertel:2020pcg}  multiplying the baryon density by 10 results in opacities which are about 100 times larger. This dependency is close to the behavior in $n_B^{5/3}$ from our conformal result \eqref{C2}.

We now consider the comparison with references that address the calculation of the neutrino radiative coefficients in quark matter \cite{Iwamoto:1980eb,Burrows:1980ec,Iwamoto:1982zz,Dai:1995uj,Schafer:2004jp}. The approach considered in those references is qualitatively very different from ours, since the calculations are done in perturbative QCD. The comparison with our results will therefore indicate how much our strongly-coupled calculation differs from the weakly-coupled result. More specifically, the results that are most readily compared with those presented in this work are derived in \cite{Iwamoto:1982zz}. The neutrino opacity $\ka_{e^-}(E_\n)$ for degenerate neutrinos is given by equation (6.27) of \cite{Iwamoto:1982zz}
\begin{align}
\nn \ka^{\text{pQCD}}_{e^-}(E_\n) = &\frac{4G_F^2}{\pi^3}|M_{ud}|^2 p_F(d)^2 p_F(\n) \times \\
\label{C3} &\times \left[1 + \frac{1}{2} \frac{p_F(\n)}{p_F(d)} + \frac{1}{10}\left(\frac{p_F(\n)}{p_F(d)}\right)^2\right] \Big[(E_\n-\m_\n)^2 + (\pi T)^2\Big]  \, ,
\end{align}
where the $p_F(f_i)$ refer to the Fermi momentum of the corresponding species. We considered the case $|p_F(d)-p_F(\n)|\geq |p_F(u)-p_F(d)|$, since it is the right ordering in isospin symmetric matter. The weakly coupled quark matter is described by a Fermi liquid, for which the relation between the Fermi momentum and the chemical potential is given by
\be
\label{C4} p_F(u) = \m_u \left(1 + \OO(\a_s) \right) \sp p_F(d) = \m_d \left(1 + \OO(\a_s) \right)  \, ,
\ee
where $\a_s \equiv g^2/(4\pi)$, $g$ being the strong interaction Yang-Mills coupling. At leading order $\OO(\a_s^0)$, \eqref{C3} becomes
\be
\label{C5} \ka^{\text{pQCD}}_{e^-}(E_\n) = \frac{8G_F^2}{\pi^3}|M_{ud}|^2 \m^2\m_\n \Big[(E_\n-\m_\n)^2 + (\pi T)^2\Big] \left[1 + \frac{1}{2} \frac{\m_\n}{\m} + \frac{1}{10}\left(\frac{\m_\n}{\m}\right)^2 \right]  \, .
\ee

The comparison between the perturbative result \eqref{C5} and the exact neutrino opacity computed from our holographic model is shown in figure \ref{Fig:kpQCD}. We consider fixed values of the temperature $T = 10\, \text{MeV}$ and baryon density $n_B =  0.11\,\mathrm{fm^{-3}}$, whereas the neutrino energy $E_\n$ is varied. Figure \ref{Fig:kpQCD} shows that, although the qualitative behavior of the two opacities is the same, the perturbative opacity is about two orders of magnitude larger than the result from our calculation.

Note that the widths of the dip in opacity at $E_\n = \m_\n$ scale differently in the energy scales for the two results. Whereas the estimate for the width in our calculation is given by \eqref{A10}, which is controlled by the neutrino and baryonic chemical potentials, it is clear from \eqref{C5} that the width of the perturbative result is controlled by the temperature
\be
\label{C8} \D E_\n \simeq \sqrt{\frac{3N_c}{w_0^2}}\frac{\m_\n^2}{\m} \sp \D E_\n^{\text{pQCD}} = \pi T \, .
\ee
\begin{figure}[h]
\begin{center}
\includegraphics[scale=1.]{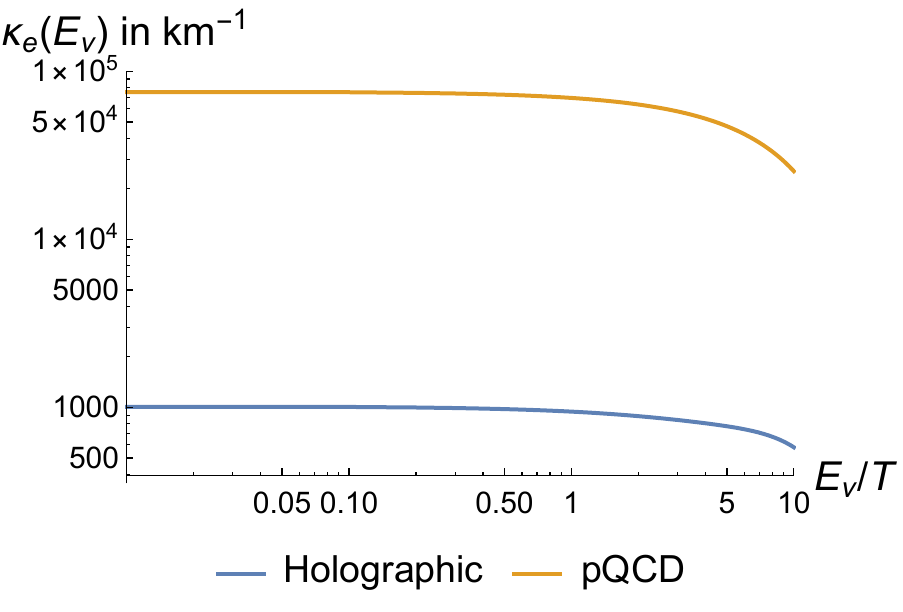}
\caption{Neutrino opacity from our holographic result (blue) compared with the perturbative QCD result \cite{Iwamoto:1982zz} (orange), at $n_B=0.11\,\mathrm{fm^{-3}}$ and $T=10\,\text{MeV}$. The opacity is expressed in $\mathrm{km^{-1}}$.}
\label{Fig:kpQCD}
\end{center}
\end{figure}

We conclude this comparative analysis by a numerical estimate of the two opacities, which should support the results shown in figure \ref{Fig:kpQCD}. The typical order of magnitude of $\ka_{e^-}$ will be estimated from its value at $E_\n = 0$. For our holographic result the estimate is given by \eqref{C2}. For the perturbative result, the corresponding number can be inferred from equation (6.28) of \cite{Iwamoto:1982zz}, together with (7.27)
\be
\label{C9} \ka_{e^-}^{\text{pQCD}}(0) \simeq 7.9\times 10^4\, \mathrm{km^{-1}}\, \left(\frac{n_B}{0.1\,\mathrm{fm^{-3}}}\right)^{\frac{5}{3}} \, ,
\ee
This shows that the perturbative opacity is about two orders of magnitude larger than our holographic calculation \eqref{C2}. The dependence on the baryon number density $n_B$ is given by the same power $5/3$, which is a consequence of the degenerate limit.

All in all, the analysis of this section indicates that the magnitudes of the rates computed from our strongly-coupled holographic model are larger than the results from approximate calculations in nuclear matter, but still much smaller than the perturbative result, by about two orders of magnitude . Since the holographic calculation was done in the deconfined phase, it is rather surprising that we obtain rates that are much closer in magnitude to the nuclear result. The quark matter that we considered is also such that the direct Urca process is not kinematically suppressed, so we might have expected the resulting opacities to be closer to the result from \cite{Iwamoto:1982zz}. The lesson from this comparison is that the neutrino emissivity is highly suppressed when taking full account of the non-perturbative nature of the strong interaction.

\section*{Acknowledgements}\label{ACKNOWL}
\addcontentsline{toc}{section}{Acknowledgements}

We would like to thank  D. Brattan, T. Demircik, C. Ecker, B. Gouteraux, C. Hoyos, N. Jokela,
T. Konstandin, A. Kurkela, D. Mateos, S. Reddy, J. Sonnenschein and  A. Vuorinen. We especially thank D. Arean and R. Davison for sharing their earlier holographic calculations with us, and  M. Oertel for giving us a lot of input on the problem addressed in this paper.

 This work was supported in part  by the Advanced ERC grant SM-GRAV, No 669288 and by a CNRS contract IEA199430.
MJ has been supported by an appointment to the JRG Program at the APCTP through the Science and Technology Promotion Fund and Lottery Fund of the Korean Government. He has also been supported by the Korean Local Governments -- Gyeong\-sang\-buk-do Province and Pohang City -- and by the National Research Foundation of Korea (NRF) funded by the Korean government (MSIT) (grant number 2021R1A2C1010834). FN and EP would like to
thank the Asian Pacific Center for Theoretical Physics in Pohang, Korea, for hospitality  while  this work was in progress.

\newpage

\appendix
\renewcommand{\theequation}{\thesection.\arabic{equation}}
\addcontentsline{toc}{section}{Appendix\label{app}}
\section*{Appendix}

\section{Details of the formalism}

We review in this appendix a few results of QFT at finite temperature, that are useful for the formalism which describes neutrino transport.

\subsection{Bosonic correlators at equilibrium}

\label{Sec:eq}

In this subsection, we review in more detail the derivation of the equilibrium properties obeyed by the chiral current real-time two-point functions. Specifically, we explain how the results \eqref{GRA3} and \eqref{KMSG<p} are obtained. Instead of the chiral-current 2-point function, we consider the case of a scalar operator
\be
\label{GBa} iG_B(x_1,x_2) \equiv \left<T_C J(x_1)J(x_2)\right> \, ,
\ee
where $J$ is a scalar hermitian operator. The case of a vector such as the chiral current is completely analogous.

As mentioned in the text, the first result \eqref{GRA3} is related to the invariance of the system at equilibrium under time-translation. At the level of the bosonic 2-point function \eqref{GBa}, this invariance implies that
\be
\label{GDta} G_B(t_1,t_2) = G_B(\D t,0) \equiv G_B(\D t) \sp \D t \equiv t_1 - t_2 \, .
\ee
The expressions for the retarded and advanced propagators are the equivalent of \eqref{GRGADt}
\be
\label{GRGADta} iG_B^R(\D t) = \theta(\D t) \left< \left[J(\D t), J(0) \right] \right> \sp iG^A(\D t) = -\theta(-\D t) \left< \left[ J(0), J(-\D t) \right] \right> \, .
\ee
Taking the complex conjugate of the retarded correlator gives
\be
\label{GRsDta} -i(G_B^R(\D t))^* = \theta(\D t) \left< \left[J(0), J(\D t) \right] \right> = -iG_B^A(-\D t) \, ,
\ee
so that
\be
\label{GRsGAa} (G_B^R(\D t))^* = G_B^A(-\D t) \, .
\ee
In momentum space, \eqref{GRsGAa} reads
\be
\label{GRsGApa}(G_B^R(p^0))^* = G_B^A(p^0) \, .
\ee
When combined with the property (which does not require equilibrium)
\be
\label{GRA2a} G_B^A(p^0) = G_B^R(-p^0) \, ,
\ee
\eqref{GRsGAa} fixes the behavior of the retarded (and advanced) 2-point function under a change of sign of $p^0$
\be
\label{GRA3a} \text{Im}\, G_B^R(-p^0) = -\text{Im}\, G_B^R(p^0) \sp \text{Re}\, G_B^R(-p^0) = \text{Re}\, G_B^R(p^0) \, .
\ee
Likewise, time-translation invariance implies that the time-ordered and anti-time-ordered propagators are related  by
\be
(G_B^F(p^0))^* = -G_B^{\bar{F}}(p^0) \, .
\ee

We now focus on the result \eqref{KMSG<p}, which is a consequence of the KMS symmetry. Because the Hamiltonian $\hat{H}$ is the generator for time translation, the operator $J$ can be shifted with an imaginary time according to
\be
\label{stpa} J(t) = \ex^{-\b \hat{H}} J(t-i\b) \ex^{\b \hat{H}} \, .
\ee
If we substitute this equality in the $G_B^<$ propagator, we obtain
\begin{align}
\nn iG_B^<(\D t) &= \left< J(0) J(\D t) \right>\\
\nn &= \text{Tr}\left[ \ex^{-\b(\hat{H}-\mu \hat{N})}J(0) \ex^{-\b \hat{H}} J(\D t-i\b) \ex^{\b \hat{H}}  \right] \\
\nn &= \text{Tr}\left[ \ex^{\b\mu \hat{N}}J(0) \ex^{-\b \hat{H}} J(\D t-i\b) \right] \\
\nn &= \text{Tr}\left[ \ex^{-\b\mu \hat{N}}J(\D t-i\b)\ex^{\b\mu \hat{N}}J(0) \ex^{-\b (\hat{H}-\mu\hat{N})} \right] \\
\nn &= \text{Tr}\left[ \ex^{\b\mu}J(\D t-i\b)J(0) \ex^{-\b (\hat{H}-\mu\hat{N})} \right] \\
\nn &= \ex^{\b\mu}\text{Tr}\left[ \ex^{-\b (\hat{H}-\mu\hat{N})} J(\D t-i\b)J(0) \right] \\
\label{KMSG<ta} &= \ex^{\b\mu} iG_B^>(\D t -i\b) \, ,
\end{align}
where $\hat{N}$ is the boson number operator, and $\mu$ the associated chemical potential. We used the cyclicity of the trace to go from the second to the third line, and the fact that $[\hat{N},J] = J$ to go from the third to the fourth line. In momentum space \eqref{KMSG<ta} becomes
\be
\label{KMSG<pa} G_B^<(p) = \ex^{-\b(p^0-\mu)} G_B^>(p) \, .
\ee
Note that the bosonic chemical potentials are equal to zero in the nuclear matter. This is why no chemical potential appears in \eqref{KMSG<p}.

\subsection{Free fermion propagator}

\label{Sec:G0}

We review here the derivation of the equilibrium free fermion propagator $G^0(x_1,x_2)$ at finite temperature and density. The expression for the latter is
\be
\label{defG0} iG^0_{\a\b}(x_1,x_2) \equiv \frac{\text{Tr}\left[\rho_0T_C\{\psi_\a(x_1)\bar{\psi}_\b(x_2)\}\right]}{\text{Tr}\left[\rho_0\right]} \, ,
\ee
where $\rho_0\equiv \exp{\left(-\b (\mathcal{H}_0-\mu N_0)\right)}$ is the equilibrium grand canonical density matrix. The equilibrium Hamiltonian and particle number operators can be expressed in terms of the fermion and anti-fermion creation and annihilation operators as
\be
\label{NH0a} N_0 = \int \frac{\mathrm{d}^3p}{(2\pi)^3 2E_p} \, \mathrm{\left(a^\dagger_p a_p - b^\dagger_p b_p\right)} \sp \mathcal{H}_0 = \int \frac{\mathrm{d}^3p}{(2\pi)^3 2E_p} \,E_p\, \mathrm{\left(a^\dagger_p a_p + b^\dagger_p b_p\right)} \, ,
\ee
where $E_p = \sqrt{\vec{p}^2+m^2}$ is the fermion on-shell energy. We dropped the zero-point energy, which yields the same factor in the numerator and denominator of expectation values such as  \eqref{defG0}. Starting from the canonical anti-commutation relations of the fermionic creation and annihilation operators
\be
\label{acab} \{\mathrm{a_p,a_{p'}^\dagger}\} = (2\pi)^3 2E_p \d\big( \vec{p} - \vec{p}'\big) \sp \{\mathrm{b_p,b_{p'}^\dagger}\} = (2\pi)^3 2E_p \d\big( \vec{p} - \vec{p'} \big) \, ,
\ee
we obtain that
\be
\label{eFa}  \{\ex^{-\b(\mathcal{H}_0-\mu N_0)},\mathrm{a_p}\} = \ex^{-\b(\mathcal{H}_0-\mu N)}\left( 1 + \ex^{-\b(E_p-\mu)} \right) \mathrm{a_p}  \, ,
\ee
\be
\label{eFb} \{\ex^{-\b(\mathcal{H}_0-\mu N_0)},\mathrm{b_p}\} = \ex^{-\b(\mathcal{H}_0-\mu N)}\left( 1 + \ex^{-\b(E_p+\mu)} \right) \mathrm{b_p} \, .
\ee
These commutators can then be used to compute the expectation values
\be
\label{TreFaa} \frac{\mathrm{Tr}\left[\rho_0\,\mathrm{a_p^\dagger a_{p'}}\right]}{\mathrm{Tr}\left[\rho_0\right]} = (2\pi)^3 2E_p n_f(E_p-\mu) \d\big( \vec{p} - \vec{p'} \big) \, ,
\ee
\be
\label{TreFbb} \frac{\mathrm{Tr}\left[\rho_0\,\mathrm{b_p^\dagger b_{p'}}\right]}{\mathrm{Tr}\left[\rho_0\right]} = (2\pi)^3 2E_p n_f(E_p+\mu) \d\big( \vec{p} - \vec{p'} \big) \, ,
\ee
where $n_f$ is the Fermi-Dirac distribution
\be
\label{nF} n_f(E) \equiv \frac{1}{\ex^{\b E} + 1} \, .
\ee
In terms of the creation and annihilation operators, the fermionic spinor field in the interaction picture reads
\be
\label{Pab} \psi_\a(x) = \sum_{s=\pm}\int\frac{\mathrm{d}^3p}{(2\pi)^3\mathrm{2E_p}}\left[\mathrm{a_{s,p}}u_{s,\a}(\vec{p})\ex^{-i\mathrm{p}\cdot x} + \mathrm{b_{s,p}^\dagger}v_{s,\a}(\vec{p})\ex^{i\mathrm{p}\cdot x} \right] \, .
\ee
Substituting \eqref{Pab} into \eqref{defG0} and using \eqref{TreFaa}-\eqref{TreFbb} gives the final result\footnote{Recall the spinor sum rules which in this normalization read $\sum_{s=\pm} u_{s,\a}(\mathbf{p})\bar{u}_{s,\b}(\mathbf{p}) = (\slashed{p}+m)_{\a\b}$ and $\sum_{s=\pm} v_{s,\a}(\mathbf{p})\bar{v}_{s,\b}(\mathbf{p}) = (\slashed{p}-m)_{\a\b}$ . } for the free fermionic propagator at finite temperature and density
\begin{align}
\nn iG_{\a\b}^0(x_1,x_2) &= \int\frac{\mathrm{d}^3p}{(2\pi)^3\mathrm{2E_p}}\left[ (\slashed{p}_+ +m)\left(\theta_C(x_1^0-x_2^0)-n_f(E_p-\mu)\right)\ex^{-ip\cdot(x_1-x_2)}\right. \\
\label{G0fa} &\left. \qquad\qquad\qquad+ (\slashed{p}_- +m)\left(\theta_C(x_2^0-x_1^0)-n_f(E_p+\mu)\right) \ex^{ip\cdot(x_1-x_2)} \right] \, ,
\end{align}
where $\theta_C$ is the Heaviside function on the CTP and we defined
\be
\label{defppm} \slashed{p}_\pm = \pm E_p\g_0 -\vec{p}\cdot\vec{\g} \, .
\ee
In particular, the Wightman functions for the free quasi-particles are given in momentum space by
\begin{align}
\nn iG^{0,<}(k) = -(\slashed{k}+m+\mu\g_0)\frac{\pi}{E_p}&\Big[ n_f(E_p-\mu)\d(E_p-k^0-\mu)- \\
\label{G0<a} &- (1-n_f(E_p+\mu) )\d(E_p+k^0+\mu) \Big] \, ,
\end{align}
\begin{align}
\nn iG^{0,>}(k) = (\slashed{k}+m+\mu\g_0)\frac{\pi}{E_p}&\Big[ (1-n_f(E_p-\mu))\d(E_p-k^0-\mu) -\\
\label{G0>a} &- n_f(E_p+\mu) \d(E_p+k^0+\mu) \Big] \, .
\end{align}
To go from \eqref{G0fa} to \eqref{G0<a}-\eqref{G0>a}, we used the fact that the 0-component of the quasi-particle momentum is shifted with respect to that of the particle as
\be
\label{qpdisp} k^0 = p^0 - \m \, .
\ee

\subsection{A discussion of  the ``quasi-particle" approximation}

\label{Sec:qp}

In this sub-appendix, we discuss in more detail the underlying assumptions of what is referred to as the quasi-particle approximation (introduced in section \ref{QPA}). The starting point of this approximation is more of a \emph{near-equilibrium} approximation. At equilibrium, the fermionic Wightman functions obey relations similar to \eqref{J<ReJR}-\eqref{J>ReJR}
\be
\label{NLO1} G^<_{\a\b}(p_\n) = -n_f(p_\n^0-\m_\n) \rho_{\a\b}(p_\n) \, ,
\ee
\be
\label{NLO2} G^>_{\a\b}(p_\n) = (1-n_f(p_\n^0-\m_\n)) \rho_{\a\b}(p_\n) \, ,
\ee
where $\rho_{\a\b}$ is the equilibrium spectral density. The approximation then assumes that the neutrinos are sufficiently close to equilibrium for the Wightman functions to be parametrized as
\be
\label{NLO6} G_{\a\b}^<(X,p_\n) = -F(X,p_\n^0,\vec{p}_\n^{\,\,2}) \rho_{\a\g}(X,p_\n^0,\vec{p}_\n)\frac{\d_{\g\b}-\g^5_{\g\b}}{2} \, ,
\ee
\be
\label{NLO7} G_{\a\b}^>(X,p_\n) = (1-F(X,p_\n^0,\vec{p}_\n^{\,\,2})) \frac{\d_{\a\g}-\g^5_{\a\g}}{2}\rho_{\g\b}(X,p_\n^0,\vec{p}_\n) \, ,
\ee
where $F$ is the neutrino distribution function, and the spectral density $\rho$ is further assumed to be close to its equilibrium value, up to corrections in the coupling. Note the presence of the projectors $(1-\g^5)/2$, that implement the fact that the Standard Model neutrinos are left-handed.

Upon assuming the ansatz \eqref{NLO6}-\eqref{NLO7}, the quasi-particle approximation is then a consequence of the weakly-coupled nature of the neutrino interactions. That is, the equilibrium spectral density is equal to the free spectral density, up to corrections from the interactions
\be
\label{NLO8} \rho(p_\n^0,\vec{p}_\n) = \frac{\pi}{E_\n}\slashed{p}_\n\big( \d(p_\n^0- E_\n) - \d(p_\n^0+E_\n) \big) + \OO(G_F^2) \sp E_\n \equiv |\vec{p}_\n| \, .
\ee
Another consequence of the weak coupling is that, at leading order, the energy of the neutrino quasi-particles is shifted with respect to that of the neutrinos as in \eqref{qpdisp}
\be
\label{NLO3} k_\n^0 = p_\n^0 - \m_\n  + \OO(G_F^2) \, ,
\ee
so that the Wightman functions for the quasi-neutrinos are given by
\be
\label{NLO4} G^<_{\n,\a\b}(X,k_\n) = -F_\n(X,k_\n^0,\vec{k}_\n^{\,\,2}) \rho_{\a\g}(X,k_\n^0+\m_\n,\vec{k}_\n)\frac{\d_{\g\b}-\g^5_{\g\b}}{2} \, ,
\ee
\be
\label{NLO5} G^>_{\n,\a\b}(X,k_\n) = (1-F_\n(X,k_\n^0,\vec{k}_\n^{\,\,2})) \frac{\d_{\a\g}-\g^5_{\a\g}}{2}\rho_{\g\b}(X,k_\n^0+\m_\n,\vec{k}_\n) \, .
\ee
The quasi-neutrino distribution function $F_\n$ is split into neutrino and anti-neutrino distributions as
\be
\label{NLO9} F_\n(X,k_\n^0,\vec{k}_\n^{\,\,2}) = f_\n(X,k_\n^0,\vec{k}_\n^{\,\,2})\theta(k_\n^0+\m_\n) + \big(1 - f_{\bar{\n}}(X,k_\n^0,\vec{k}_\n^{\,\,2})\big) \theta(-k_\n^0-\m_\n) \, .
\ee

The ansatz \eqref{NLO4} is then substituted in the Kadanoff-Baym (KB) equation \eqref{KB<b}, that we reproduce here for convenience,
\be
\label{NLO10} i\partial^{X}_\mu\text{Tr}\left\{\g^\mu G^<(X,k)\right\} =  -\text{Tr}\left\{G^>(X,k)\Sigma^<(X,k) - \Sigma^>(X,k)G^<(X,k)\right\} \, .
\ee
There is a similar equation for the other Wightman function
\be
\label{NLO11} i\partial^{X}_\mu\text{Tr}\left\{\g^\mu G^>(X,k)\right\} =  -\text{Tr}\left\{G^>(X,k)\Sigma^<(X,k) - \Sigma^>(X,k)G^<(X,k)\right\} \, .
\ee
The difference of the two KB equations implies that a specific trace of the spectral function is preserved by the kinetic evolution\footnote{In deriving \eqref{NLO12} and \eqref{NLO13}, we use the fact that the trace of $\g^5\g^{\m_1}\dots \g^{\m_n}$ is zero for all $n\leq 3$.  }
\be
\label{NLO12} i\partial^{X}_\mu\text{Tr}\left\{\g^\mu \rho(X,k)\right\} = 0 \, .
\ee

The equation for the distribution functions is then given by
\begin{align}
\nn i\partial^{X}_\mu F_\n(X,k)\text{Tr}\left\{\g^\mu \rho(X,k)\right\} =  -\text{Tr}&\big\{F_\n(X,k)\rho(X,k)\Sigma^<(X,k) +\\
\label{NLO13} &+ (1-F_\n)\Sigma^>(X,k)\rho(X,k)\big\} \, .
\end{align}
At leading order in the weak coupling, this reduces to the Boltzmann equations for the neutrino and anti-neutrino distribution functions, \eqref{dtfn} and \eqref{dtfnb}.

In the general case, the ansatz for the Wightman functions contains several fields, that are organized according to the expansion in the generators of the Clifford algebra, with the condition of SO(3) symmetry

\begin{align}
\nn G^>(X,k_\n^0,\vec{k}_\n^2) =&\,\,  \mathcal{S}(X,k_\n^0,\vec{k}_\n^2) + i\g^5\mathcal{P}(X,k_\n^0,\vec{k}_\n^2) + \\
\nn &+\g^0\mathcal{V}_{(0)}(X,k_\n^0,\vec{k}_\n^2) + k_i\g^i\mathcal{V}_{(1)}(X,k_\n^0,\vec{k}_\n^2)+\\
\nn & +  \g^5\g^0\mathcal{A}_{(0)}(X,k_\n^0,\vec{k}_\n^2) + k_i\g^5\g^i\mathcal{A}_{(1)}(X,k_\n^0,\vec{k}_\n^2) +\\
\label{NLO14} &+ k_i[\g^0,\g^i] \mathcal{D}(X,k_\n^0,\vec{k}_\n^2) \, .
\end{align}
Imposing that the Standard Model neutrinos are left-handed further reduces the form of the ansatz
\be
\label{NLO15} G^>(X,k_\n^0,\vec{k}_\n^2) = \frac{1-\g^5}{2}\Big(\g^0\mathcal{V}_{(0)}(X,k_\n^0,\vec{k}_\n^2) + k_i\g^i\mathcal{V}_{(1)}(X,k_\n^0,\vec{k}_\n^2)\Big) \, .
\ee
This can be rewritten in terms of an effective neutrino chemical potential out of equilibrium $\m_{\text{eff}}$
\be
\label{NLO16} G^>(X,k_\n^0,\vec{k}_\n^2) = \frac{1-\g^5}{2}\mathcal{F}(X,k_\n^0,\vec{k}_\n^2)\Big( \slashed{k} + \m_{\text{eff}}(X,k_\n^0,\vec{k}_\n^2) \g^0 \Big) \, ,
\ee
where the field $\mathcal{F}$ contains information about the spectrum and the distribution of neutrinos. The quasi-particle approximation assumes that $\m_{\text{eff}}$ is close to the chemical potential of $\b$-equilibrium, and that $\mathcal{F}$ can be factorized as a distribution function, times the sum of Dirac delta functions that appear in the leading order equilibrium spectral density \eqref{NLO8}. Instead of a single Boltzmann equation for the neutrino distribution function, the general transport problem with the ansatz \eqref{NLO16} will involve two coupled equations for the two scalar fields $\m_{\text{eff}}$ and $\mathcal{F}$. The formalism and computational setup we have in this paper, allows then the evaluation of the two independent distributions far from equilibrium.

\section{Weak vertices for neutrino interactions}

\label{Sec:nSE_corr}

The charged current neutrino self-energy at leading order in the electroweak couplings is computed from the diagram in figure \ref{Fig:SEW}. The expression for the weak vertices appearing in this diagram is given by the Glashow-Weinberg-Salam (GSW) theory. We are interested in energy scales much lower than the W boson mass, in which regime the weak vertices are well described by a low energy effective theory, where the W and Z boson exchanges are replaced by weak current contact interactions. For easy reference, the purpose of this appendix is to give a review of the form that the effective vertices take.

After electroweak symmetry breaking, the GSW Lagrangian contains terms that couple the electroweak vector bosons to the fermion electroweak currents
\be
\label{EFT1} S_{eff}=\int \left[-\frac{g_W}{\sqrt{2}}W^{\pm}_{\m}(J_{\pm}^{\m}+\bar J_{\pm}^{\m})+ \frac{g_W}{\cos{\theta_W}} Z_{\m}(J_0^{\m}+\bar J_{0}^{\m}) + e A_\m (J^\m_{\text{em}}+\bar{J}^\m_{\text{em}})\right]
\ee
with $g_W$ the electroweak coupling constant, $\theta_W$ the Weinberg angle and $e = g_W \sin{\theta_W}$ the elementary charge. From left to right, the currents that appear are respectively the charged, neutral and electromagnetic currents. The currents without bar are the lepton currents and those with the bar are the QCD currents made out of quarks.

Integrating out (classically) the weak bosons we get the quadratic effective action
\begin{align}
\nn S_{eff'}=\int \bigg[\frac{g_W^2}{2}(J_{+}^{\m}+\bar J_{+}^{\m})D^W_{\m\n}(J_{-}^{\n}+\bar J_{-}^{\n})+h.c.&+\frac{g_W^2}{\cos^2{\theta_W}}(J_0^{\m}+\bar J_{0}^{\m})D^Z_{\m\n}(J_0^{\n}+\bar J_{0}^{\n})+\\
&+ e A_\m (J^\m_{\text{em}}+\bar{J}^\m_{\text{em}}) +{\cal O}(e G_F^2)\bigg]
\end{align}
where $D^{W,Z}$ are the tree-level gauge boson propagators in a given gauge. The higher order corrections come from the cubic and quartic interactions between the electroweak gauge bosons, as well as Higgs interactions. They start at order $\OO(e G_F^2)$, with the leading contribution coming from the $WW\g$ vertex. In the limit where all momenta are much smaller than $M_W$, the $W,Z$ propagators are replaced by Dirac deltas and we obtain
\begin{align}
\nn S_{eff'}=\int \bigg[2\sqrt{2}G_F\Big( (J_{+}^{\m}+\bar J_{+}^{\m})\eta_{\m\n}(J_{-}^{\n}+\bar J_{-}^{\n})+h.c. &+2(J_0^{\m}+\bar J_{0}^{\m})\eta_{\m\n}(J_0^{\n}+\bar J_{0}^{\n})\Big)+\\
&+ e A_{\m} (J^\m_{\text{em}}+\bar{J}^\m_{\text{em}}) + \OO(e G_F^2) \bigg]
\end{align}
This can be decomposed as
\be
S_{eff'}=S_W+S_{S}+S_{J^2}+S_{\g}
\ee
with
\be
S_W= 2\sqrt{2}G_F\int \left[J_{+}^{\m}J_{-\m}+h.c. + 2 J_0^{\m}J_{0\m} \right]+{\cal O}(e G_F^2)
\ee
\be
S_S= 4\sqrt{2}G_F\int\left[ J_{+}^{\m}\bar J_{-\m}+h.c. +2 J_0^{\m}\bar J_{0\m} \right]+{\cal O}(e G_F^2)
\ee
\be
S_{J^2}= 2\sqrt{2}G_F\int\left[ \bar J_{+}^{\m}\bar J_{-\m}+h.c. + 2\bar J_0^{\m}\bar J_{0\m} \right]+{\cal O}(e G_F^2)
\ee
and
\be
S_\g = \int\left[ e A_{\m} (J^\m_{\text{em}}+\bar{J}^\m_{\text{em}}) \right]+{\cal O}(e G_F^2)
\ee
$S_W$ is the standard leptonic Fermi interaction and its neutral counterpart, whereas $S_\g$ gives the electromagnetic interaction of the fermions. $S_S$ is the interaction of leptons with the QCD weak current, and $S_{J^2}$ is the weak interaction of the strong currents. It is small and is therefore not expected to play an important role in the strongly-coupled quark-gluon plasma. It can be included if necessary in the holographic calculation by changing the boundary conditions of the gauge fields, according to the standard double-trace dictionary, \cite{Witten01}. The higher order corrections that start at order ${\cal O}(e G_F^2)$ include tree level terms, as well as electroweak loop corrections. These terms include higher-point couplings between the weak currents, as well as couplings of the charged current to the photon.

We now want to compute the effective interaction for neutrinos in the strong plasma and for this we must compute
\be
e^{-W_{eff}}=\langle e^{-S_{S}}\rangle
\ee
where the expectation value is obtained in the state (or ensemble)  of strongly coupled matter.
This can be expanded as
\begin{align}
\nn W_{eff}=4\sqrt{2}G_F\int \bigg[ & 2 J_{+}^{\m}\langle \bar J_{-\m}\rangle +h.c. +4 J_0^{\m}\langle \bar J_{0\m}\rangle  + \\
 &+ 4\sqrt{2}G_F  \Big(J_{+}^{\m}J_{-\n} \langle \bar J_{-\m}\bar J^{\n}_+\rangle +h.c. + 2J_{0}^{\m}J_{0\n} \langle \bar J_{0\m}\bar J_0^{\n}\rangle \Big) + \OO(e G_F^2)\bigg]
\end{align}
The contact interactions with the one point functions contribute to the neutrino chemical potential $\m_\n$, so they can be absorbed in the definition of $\m_\n$. Then, the effective action for the neutrino interactions contains the term quadratic in the neutrino current plus higher order corrections
\begin{align}
\nn W_{eff}=32G_F^2\int \bigg[J_{+}^{\m}J_{-\n} \langle \bar J_{-\m}\bar J^{\n}_+\rangle +h.c. + 2J_{0}^{\m}J_{0\n} \langle \bar J_{0\m}\bar J_0^{\n}\rangle + \OO(e G_F^2)\bigg]
\end{align}

\section{Background solution}

\label{Sec:bgd}

We review in this appendix the derivation of the Reissner-Nordström background solution \eqref{ds2RN}-\eqref{AhA3s2}.

The equations of motion from the action \eqref{Sb} are the Einstein-Yang-Mills equations
\begin{align}
\nn R_{MN} - \frac{1}{2}\left(R + \frac{12}{\ell^2}\right)g_{MN} = -\frac{w_0^2\ell^2}{4N_c} \text{Tr}\bigg\{\mathbf{F}^{(L)}_{\hphantom{(L)}MP} \mathbf{F}^{(L)P}_{\hphantom{(L)P}N} + \frac{1}{4}\mathbf{F}^{(L)}_{PQ}&\mathbf{F}^{(L)PQ}g_{MN}  +\\
\label{Ba1} & + (L \leftrightarrow R) \bigg\} \, ,
\end{align}
\be
\label{Ba2} D^{(L/R)}_M\Big(\sqrt{-g}F^{(L/R)MN}\Big) = 0 \, ,
\ee
with $D^{(L/R)}_M$ the Yang-Mills covariant derivatives
\be
\label{Ba3} D^{(L)}_M \equiv \partial_M - i [L_M,\, . \,\,] \sp D^{(R)}_M \equiv \partial_M - i [R_M,\, . \,\,] \, .
\ee
The background solution is found by starting from the ansatz
\be
\label{Ba4} \mathrm{ds}^2 = \ex^{2A(r)}\left( -f(r)\mathrm{dt}^2 + f(r)^{-1}\mathrm{dr}^2 + \vec{\mathrm{d x}}^2 \right) \, ,
\ee
\be
\label{Ba5} \mathbf{R}_M = \mathbf{L}_M = \frac{1}{4}\d^{0}_M \hat{V}_0(r)\mathbb{I}_2  \, .
\ee
This ansatz fixes the gauge for the gauge field, up to a shift by a constant. As we shall see below, the regular boundary conditions in the IR \eqref{Ba9} remove this degeneracy.

Substituting the ansatz \eqref{Ba4}-\eqref{Ba5} into the equations of motion \eqref{Ba1}-\eqref{Ba2} results in the following system of equations for the ansatz fields
\be
\label{Ba6} \partial^2_r A - (\partial_r A)^2 = 0 \, ,
\ee
\be
\label{Ba6b} \partial_r A\big(\partial_r f + 4\partial_r A f(r)\big) - \frac{4}{\ell^2}\ex^{2A(r)} + \frac{w_0^2\ell^2}{48N_c}\ex^{-2A(r)} (\partial_r \hat{V}_0)^2 = 0 \, ,
\ee
\be
\label{Ba7} \partial_r\Big(\ex^{A(r)} \partial_r \hat{V}_0\Big) = 0  \, .
\ee
The two integration constants of \eqref{Ba6} correspond to translations and rescalings of $r$. We fix the definition of the coordinate $r$ by writing the solution as
\be
\label{Ba8} A(r) = \log{\left(\frac{\ell}{r}\right)} \, ,
\ee
which implies in particular that the boundary is located at $r=0$. We look for a solution with a horizon at $r=r_H$, where the blackening function $f(r)$ vanishes. For the gauge field to be regular at the horizon, the time component should vanish
\be
\label{Ba9} \hat{V}_0(r_H) =  0 \, .
\ee
This implies that the solution of \eqref{Ba7} is given by
\be
\label{Ba10} \hat{V}_0 = 2\m\left(1 - \left(\frac{r}{r_H}\right)^2\right)  \, ,
\ee
with the boundary source $\m$  corresponding  to the quark number chemical potential. Finally, the solutions for the gauge fields and the scale factor $A(r)$ can be substituted in \eqref{Ba6b}, which yields an equation for $f(r)$
\be
\label{Ba11} \partial_r f -\frac{4}{r}(f(r)-1) - \frac{w_0^2\m^2}{3N_c} \left(\frac{r}{r_H}\right)^4 r = 0 \, .
\ee
The solution takes the form
\be
\label{Ba12} f(r) = 1 - \left(\frac{r}{r_H}\right)^4\left(1 + \frac{w_0^2}{6N_c}\m^2 r_H^2\right) + \frac{w_0^2}{6N_c}\m^2 r_H^2\left(\frac{r}{r_H}\right)^6 \, .
\ee
To avoid a conical singularity of the Euclidean solution at finite temperature, the derivative of $f(r)$ at the horizon should be related to the field theory temperature
\be
\label{Ba13} f'(r_H) = -4\pi T \, .
\ee
This condition results in an equation for the horizon radius $r_H$
\be
\label{Ba14} \frac{w_0^2}{6N_c} \m^2r_H^2 = 2(1 - \pi T r_H) \, ,
\ee
whose solution determines the location of the black-hole horizon as a function of the chemical potential $\m$ and the temperature
\be
\label{Ba15} r_H(T,\m) = \frac{2}{\pi T} \left( 1 + \sqrt{1 + \frac{w_0^2}{3N_c}\frac{\m^2}{\pi^2T^2}} \right)^{-1} \, .
\ee
Note that \eqref{Ba14} allows to rewrite $f(r)$ in the form presented in the text
\be
\label{Ba16} f(r) = 1 - \left(\frac{r}{r_H}\right)^4\left(1 + 2\left(1 - \pi T r_H\right)\right) + 2\left(1 - \pi T r_H\right)\left(\frac{r}{r_H}\right)^6  \, .
\ee

\section{Parameters of the bulk action}

\label{Sec:param}

The bulk action \eqref{Sb} possesses two parameters: the five-dimensional Planck mass $M\ell$, which controls the overall normalization of the action, and the flavor coupling $w_0$. We detail in this appendix how the values of these parameters are fixed by matching to QCD data.

First, the 5-dimensional Planck mass $M\ell$ is fixed by imposing that the zero-chemical potential limit of the pressure be that of a free quark-gluon plasma
\be
\label{Pa1} p = \frac{\pi^2 N_c^2}{45}T^4 \left(1 + \frac{7N_f}{4N_c}\right) \, .
\ee
Lattice results \cite{Borsanyi13} indicate that, for temperatures equal to a few times the deconfining temperature, the pressure in the quark-gluon plasma is already close (within about $20\%$) to the ideal result \eqref{Pa1}. Setting $(M\ell)$ to match \eqref{Pa1} will therefore ensure that the thermodynamics of the holographic model is close to that of QCD in the deconfined phase. The pressure of the holographic model is computed from the grand-canonical potential \eqref{SosRN}. At $\mu \ll T$, it is given by
\be
\label{Pa2} p = (M\ell)^3\left[N_c^2(\pi T)^4  + \frac{1}{4}N_fN_c w_0^2(\pi T)^2\m^2  + \OO(\mu^4)\right]  \, .
\ee
\eqref{Pa2} matches \eqref{Pa1} at $\mu = 0$ if $(M\ell)^3$ is equal to
\be
\label{Pa3} (M\ell)_{\text{free}}^3 = \frac{13}{6}\frac{1}{45\pi^2} \, ,
\ee
where the number of flavors was set to $N_f=2$, and that of colors to $N_c=3$.

As far as the parameter $w_0$ is concerned, it can be fixed such that the baryon number susceptibility at zero density agrees with the ideal Fermi gas result. As for the pressure, it was observed to give a good approximation to the exact result for the quark-gluon plasma on the lattice \cite{Borsanyi11}. The baryon number susceptibility is defined as the first non-trivial cumulant of the pressure at $\mu = 0$
\be
\label{Pa4} \chi_B = \frac{\partial^2 p}{\partial^2 \mu_B}\bigg|_{\mu_B = 0} \, .
\ee
From \eqref{Pa2} it is equal to
\be
\label{Pa4b} \chi_B = \frac{N_f}{2N_c}w_0^2(M\ell)^3(\pi T)^2 \, ,
\ee
whereas the ideal Fermi gas result is
\be
\label{Pa5} \chi_{B,free} = \frac{N_f}{3N_c} T^2  \, .
\ee
Matching the two results fixes the value of $w_0$ to be
\be
\label{w0lat} \big(w_{0}^2(M\ell)^3\big)_{\text{free}} = \frac{2}{3\pi^2} \, .
\ee

In the numerical calculations done in this paper, we used the values of the parameters given by \eqref{Pa3} and \eqref{w0lat}. For comparison, we discuss below another choice for the value of the parameter $w_0$.

\paragraph*{UV limit of the two-point function} Fixing $w_0$ as in \eqref{w0lat} also implies that the holographic two-point function agrees with the perturbative QCD result in the UV limit~\cite{V3,V4}. Here, we will prove explicitly this result. So we now consider the Euclidean version of the correlator \eqref{PVt} and \eqref{PVp} in the UV limit where $\omega^2 + \vec{k}^2$ goes to infinity. In this case the temperature and chemical potential become irrelevant and the computation can be equivalently performed in (Euclidean) AdS space-time. It follows that the Lorentz invariance of the theory is effectively restored and the Euclidean correlator can be written as
\be
\label{pE} \left<J^{(L)}_\l J^{(L)}_\s\right>^E(k) = P_{\l\s}(k) \Pi^E_{(L)}(k)  \, ,
\ee
where the 4-dimensional projector transverse to $k$ was defined in \eqref{P4}. The function $\Pi(k)$ is computed following the standard holographic method, starting from the perturbation
\be
\label{ansVhUV} \d L_\mu(r;x) = \int\frac{\intd k^4}{(2\pi)^4}\ex^{ik.x} \mathcal{L}^{(0)}_{\mu,k}\psi(r) \, ,
\ee
which obeys the equation of motion
\be
\label{EoMVUV} \partial_r^2\psi- \frac{1}{r} \partial_r\psi - k^2 \psi = 0 \, ,
\ee
together with a boundary condition fixing the normalization of the perturbation
\be
\label{bcpV} \psi(0) = 1 \, .
\ee
The solution of this differential problem can be expressed in terms of a modified Bessel function of the second type
\be
\label{solpV} \psi(r) = kr K_1(kr) \, ,
\ee
and the Euclidean on-shell action is
\be
\label{SosbUV} S_{on-shell} = -\frac{1}{8\ell}(M\ell)^3w_0^2 N_c \int_{r=\e} \mathrm{d}^4k \frac{\ell}{r} \mathcal{L}_\mu^{(0)}(-k)P^{\m\n}(k)\mathcal{L}_\nu^{(0)}(k) \partial_r\psi \psi \, .
\ee
Near the boundary, $K_1$ has the following behavior
\be
\label{K1b} K_1(kr) \underset{r\to 0}{\sim} \frac{1}{kr} +\frac{1}{2} kr\log{(kr)} + C_0 kr  \, ,
\ee
where $C_0$ is a constant that does not depend on $k$. This implies that the on-shell action has a logarithmic divergence which is removed by the appropriate counter-term. This leaves the renormalized on-shell action
\be
\label{SosbUVr} S_{ren} = -\frac{1}{16}(M\ell)^3w_0^2 N_c \int_{r=\e} \mathrm{d}^4k  \mathcal{L}_\mu^{(0)}(-k)P^{\m\n}(k)\mathcal{L}_\nu^{(0)}(k) k^2\log{(k^2)} + \mathcal{O}(k^2)  \, ,
\ee
and
\be
\label{PiEs} \Pi^E_{(L)}(k) = -\frac{N_c}{8} (M\ell)^3w_0^2 k^2\log(k^2) \, .
\ee
Identifying with the perturbative QCD result
\be
\label{PiQCD} \Pi^E_{(L),\text{QCD}}(k) = -\frac{N_c}{8}\frac{2}{3\pi^2}k^2\log(k^2) \, ,
\ee
fixes the value of $w_0$ in terms of $M\ell$
\be
\label{w0UV} w_0^2 = \frac{2}{3\pi^2 (M\ell)^3} \, ,
\ee
which agrees with the value derived from the susceptibility \eqref{w0lat}.

\section{Eddington-Finkelstein coordinates}

The general solution to the equations of motion \eqref{EoMVht} and \eqref{EoMEp} behaves near the horizon as the superposition of an infalling and outgoing waves
\be
\label{EF1} \mathcal{L}^\perp = c_1 (r_H-r)^{-i\g}  + c_2 (r_H-r)^{i\g} \sp E^\parallel = d_1 (r_H-r)^{-i\g}  + d_2 (r_H-r)^{i\g} \sp \g \equiv \frac{\omega}{4\pi T} \, .
\ee
Imposing infalling boundary conditions amounts to setting $c_2 = d_2 = 0$, and the infalling solution can be rewritten as
\be
\label{EF2} \mathcal{L}^\perp = c \exp{\left(-i\g \log{\left( 1 - \frac{r}{r_H}\right)}\right)}  \sp E^\parallel = d \exp{\left(-i\g \log{\left( 1 - \frac{r}{r_H}\right)}\right)}\, ,
\ee
which makes it apparent that the solution oscillates very fast near the horizon as soon as $\omega$ in non-zero. When solving the equations of motion numerically, such fast oscillating solutions require high numerical precision to obtain  good accuracy for the behavior of the solution near the boundary. To avoid working with such solutions, it is more convenient to do the numerical calculations in the natural coordinates for infalling solutions, that are the infalling Eddington-Finkelstein coordinates. The change of coordinate is given by
\be
\label{EF3} \vec{x} \to \vec{x} \sp  t \to u = t - r^*(r) \sp r \to r \, ,
\ee
where the tortoise coordinate $r^*(r)$ is such that
\be
\label{EF4} \frac{\intd r^*}{\intd r} = \frac{1}{f(r)} \, .
\ee
Then, the Fourier transform of the gauge field perturbation transforms as
\be
\label{EF5} \mathcal{L}_{\mu,k}(r) \to \ex^{i\omega r^*(r)} \mathcal{L}_{\mu',k}(r) \, ,
\ee
which can be decomposed into
\be
\label{EF6} \mathcal{L}^\perp(r) \to \ex^{i\omega r^*} \mathcal{F}^\perp(r) \sp E^\parallel(r) \to \ex^{i\omega r^*} \mathcal{F}^\parallel(r) \, .
\ee
Notice that, since near the horizon the tortoise coordinate behaves as
\be
\label{EF7} r^*(r) \sim -\frac{1}{4\pi T} \log{\left( 1 - \frac{r}{r_H} \right)} \, ,
\ee
the fields $\mathcal{F}^\perp$ and $\mathcal{F}^\parallel$ do not oscillate near the horizon, and are instead analytic at $r = r_H$\footnote{The infalling Eddington-Finkelstein coordinates are well-defined beyond the horizon, and there exists a solution to the equations of motion which is perfectly regular at $r=r_H$ in these coordinates.}.

Applying the transformation \eqref{EF5} to the equations of motion \eqref{EoMVht} and \eqref{EoMEp} gives the differential equations obeyed by the gauge-fields in Eddington-Finkelstein coordinates, $\mathcal{F}^\perp$ and $\mathcal{F}^\parallel$
\be
\label{EF8} \partial_r^2\mathcal{F}^\perp + \left(\frac{f'(r)+i\omega}{f(r)} - \frac{1}{r}\right)\partial_r\mathcal{F}^\perp - \frac{k^2}{f(r)}\mathcal{F}^\perp = 0 \, ,
\ee
\be
\label{EF9} \partial_r^2\mathcal{F}^\parallel + \left(\frac{f'(r)+i\omega}{f(r)} - \frac{1}{r} + \frac{k^2f'(r)}{\omega^2 - f(r)k^2} \right)\partial_r\mathcal{F}^\parallel - \frac{k^2}{f(r)}\mathcal{F}^\parallel = 0 \, .
\ee
These equations can be solved numerically by shooting from the horizon, where two boundary conditions are imposed. The first condition fixes the normalization of the solution
\be
\label{EF10} \mathcal{F}^\perp(r_H) = \mathcal{F}^\parallel(r_H) = 1 \, ,
\ee
and the second one selects the infalling solution at the horizon. In Eddington-Finkelstein coordinates, this corresponds to requiring that the fields $\mathcal{F}^\perp$ and $\mathcal{F}^\parallel$ are regular at $r = r_H$. By analyzing the equations of motion \eqref{EF8}-\eqref{EF9} near the horizon, 
for the regular solution we find the following relation between the fields and their first derivative at the horizon
\be
\label{EF11} \partial_r\mathcal{F}^\perp(r_H) = \frac{k^2}{i\omega - 4\pi T} \mathcal{F}^\perp(r_H) \sp \partial_r\mathcal{F}^\parallel(r_H) = \frac{k^2}{i\omega - 4\pi T} \mathcal{F}^\parallel(r_H) \, .
\ee

\section{Analysis of the diffusive approximation}

\label{Sec:DiffApp}

This appendix presents an analysis of the diffusive approximation, where the time-time component of the two-point function is assumed to give the largest contribution to the opacities. We first investigate the validity of this approximation, and then use it to derive approximate expressions for the radiative coefficients as a function of the various parameters.

\subsection{Radiative coefficients in the hydrodynamic limit}

This subsection presents the analysis of the radiative coefficients in the hydrodynamic regime $r_H\mu_e, r_H\mu_\n, r_H E_\n \ll 1$, which results in the scalings \eqref{A7}. The results justify the validity of the diffusive approximation in the degenerate and hydrodynamic limit.

A small parameter $\e$ is introduced, and we consider the following scaling of the parameters
\be
\label{DA1} r_H E_\n \sim r_H\m_\n = r_H \m_e = \e \sp r_H T  = \OO(\e^a) \, ,
\ee
where we take $a \gg 1$.
This ensures that the temperature is much smaller than all other energy scales in the problem, such that the degenerate expression of the statistical distributions can be used \eqref{nB_hD} and \eqref{nehD}. We then consider the integrals over the loop electron momentum which define the radiative coefficients \eqref{j}-\eqref{lb}. We focus on the neutrino emissivity $j_{e^-}$ for concreteness, but the others are analogous. The integral over $k_e$ can be rewritten as an integral over the energy $q_{e\n}^0$
\be
\label{DA2} j_{e^-}(E_\nu) = -\frac{G_F^2}{8\pi^2}\int_0^\pi\intd\theta\sin{\theta}\int_0^{\mu_\n-E_\n}\intd\omega\frac{\omega + E_\n}{E_\nu}L_e^{\l\s}\, \text{Im}\,G^R_{c,\s\l}\big(\omega,k(\omega,\theta)\big)\, ,
\ee
\be
\nn \omega \equiv q_{e\n}^0 = E_e - E_\n \, ,
\ee
\be
\nn k(\omega,\theta) \equiv \sqrt{(\omega + E_\n)^2 + E_\n^2 - 2(\omega+E_\n)E_\n \cos{\theta}} \, ,
\ee
where we neglected the mass of the electrons\footnote{The mass of the electron is equal to about $0.5\,\mathrm{MeV}$, which is much smaller than the temperature $T = 10\,\mathrm{MeV}$.}, and the boundaries of the energy integral are fixed by the Fermi-Dirac and Bose-Einstein distributions. Substituting the hydrodynamic expression of the correlators gives
\begin{align}
\label{DA4} j_{e^-}(E_\nu) = \frac{G_F^2}{8\pi^2}&\int_0^\pi\intd\theta\sin{\theta}\int_0^{\mu_\n-E_\n}\intd\omega\frac{\omega + E_\n}{E_\nu}L_e^{\l\s} \times \\
\nn &\times \s\omega\left(P_{\l\s}^{\perp}(\omega,k(\omega,\theta)) + P_{\l\s}^{\parallel}(\omega,k(\omega,\theta)) \frac{\omega^2-k^2(\omega,\theta)}{\omega^2 + D^2k^4(\omega,\theta)}\right)\, ,
\end{align}
which is the sum of several components
\begin{align}
\label{DA5} j_{e^-}^\perp(E_\nu) = \frac{2G_F^2\s}{\pi^2} \int_0^\pi\intd\theta\sin{\theta}(1-\cos{\theta})&\int_0^{\mu_\n-E_\n}\intd\omega\, \omega(\omega + E_\n)^2 \times \\
\nn &\times \left(1 + \frac{(\omega+E_\n)E_\n}{k^2(\omega,\theta)}(1+\cos{\theta}) \right) \, ,
\end{align}
\begin{align}
\label{DA6} j_{e^-}^{(00)}(E_\nu) = \frac{G_F^2\s}{\pi^2} \int_0^\pi\intd\theta\sin{\theta}(1+\cos{\theta})&\int_0^{\mu_\n-E_\n}\intd\omega\, \omega(\omega + E_\n)^2 \times \\
\nn &\times \frac{k^2(\omega,\theta)}{\omega^2 + D^2k^4(\omega,\theta)} \, ,
\end{align}
\begin{align}
\label{DA7} j_{e^-}^{(0i)}(E_\nu) = -\frac{2G_F^2\s}{\pi^2} \int_0^\pi\intd\theta\sin{\theta}(1+\cos{\theta})&\int_0^{\mu_\n-E_\n}\intd\omega\, \omega^2(\omega + E_\n)^2 \times \\
\nn &\times \frac{\omega}{\omega^2 + D^2k^4(\omega,\theta)} \, ,
\end{align}
\begin{align}
\label{DA8} j_{e^-}^{\parallel,(ij)}(E_\nu) = \frac{G_F^2\s}{\pi^2} \int_0^\pi\intd\theta\sin{\theta}(1+\cos{\theta})&\int_0^{\mu_\n-E_\n}\intd\omega\, \omega^3(\omega + E_\n)^2 \times \\
\nn &\times \frac{\omega^2}{k^2(\omega,\theta)(\omega^2 + D^2k^4(\omega,\theta))} \, .
\end{align}
The diffusive approximation assumes that the component $j_{e^-}^{(00)}$ dominates over the other contributions. We now proceed to investigate the validity of this statement in the hydrodynamic limit. This will be done by determining the hydrodynamic scaling (the scaling in $\e$) for each component of the emissivity \eqref{DA5}-\eqref{DA8}.

For a given angle $\theta$, the integration over $\omega$ meets the diffusive peak $\omega = Dk^2$ when $\omega$ is equal to
\begin{align}
\nn \omega^{*}(E_\n,\theta) &\equiv \frac{1+2DE_\n\cos{\theta} - \sqrt{(1+2DE_\n\cos{\theta})^2 - 4D(E_\n+DE_\n^2)}}{2D} - E_\n \\
\label{DA10}  &= 2DE_\n^2(1 - \cos{\theta})(1+\OO(\e)) \, .
\end{align}
From \eqref{DA1}, $\omega^{*}$ is of order $\OO(\e^2)$ whereas the upper bound of the integral $\mu_\n - E_\n$ is of order $\OO(\e)$. The integrals are therefore such that, over most of the integration region, $\omega$ is much larger than $Dk^2$. Specifically, the integrals can be split into two parts as
\be
\label{DA11} \int_0^{\m_\n-E_\n} = \int_0^{A\omega^*}  + \int_{A\omega^*}^{\mu_\n - E_\n} \, ,
\ee
where $A$ is a number much larger than one which is independent of $\e$. The first part contains the contribution from the diffusion peak, whereas $\omega \gg Dk^2$  in the second part.

We now investigate the scaling of the first part of the integral in \eqref{DA11} that we label with the subscript ``diff". Since the transverse integrand \eqref{DA5} does not depend on $r_H$, its hydrodynamic scaling is easily derived
\be
\label{DA9} j_{e^-,\text{diff}}^\perp = \OO(\e^6) \, .
\ee
The longitudinal integrands require a more careful study since $r_H$ appears via the diffusion constant $D = \frac{1}{2}r_H$. The hydrodynamic scaling of the longitudinal emissivity coming from the first part of the integral in \eqref{DA11} can be found by determining appropriate upper and lower bounds. The lower bound is determined according to the following
\begin{align}
&\label{DA11b} \int_0^{A \omega^*}\intd\omega F(\omega,\theta) \frac{k^2(\omega,\theta)\omega}{\omega^2 + D^2 k^4(\omega,\theta)} > \int_0^{2 \omega^*}\intd \omega F(\omega,\theta) \frac{k^2(\omega,\theta)\omega}{\omega^2 + D^2 k^4(\omega,\theta)} \\
\nn &\approx \int_0^{2 \omega^*}\intd \omega F(\omega,\theta) \frac{\omega^*\omega}{D\left(\omega^2 + (\omega^*)^2\right)} > \int_0^{2 \omega^*}\intd\omega F(\omega,\theta) \frac{1}{2D}\left(1 - \left(\frac{\omega}{\omega^*}-1\right)^2\right) \, ,
\end{align}
where $F(\omega,\theta) > 0$ and the sign $\approx$ means that the two expressions are equal up to a factor $1+\OO(\e)$. To write the first expression on the second line, we used the fact that $D \omega$ is of order $\OO(\e^2)$ over the integration interval, so that $k^2(\omega,\theta)=\omega^*/D (1+ \OO(\e))$.  The upper bound is obtained by replacing the fraction in \eqref{DA11b} by its maximum value reached at $\omega = \omega^*$
\be
\label{DA12} \int_0^{A \omega^*}\intd\omega F(\omega,\theta) \frac{k^2(\omega,\theta)\omega}{\omega^2 + D^2 k^4(\omega,\theta)} < \int_0^{A \omega^*}\intd \omega \frac{F(\omega,\theta)}{2D} \, .
\ee
Then, for each component of the longitudinal emissivity \eqref{DA6}-\eqref{DA8}, the bounds on the contribution from the region around the diffusion peak are obtained by replacing $F(\omega,\theta)$ in \eqref{DA11} and \eqref{DA12} by the appropriate expression, and performing the integral. This results in the following bounds
\be
\label{DA13} j_{e^-,\text{diff}}^{(00)} = \a_{(00)}\frac{4G_F^2\s}{3\pi^2} E_\n^4(1 + \OO(\e)) = \OO(\e^4)\, ,
\ee
\be
\label{DA14} j_{e^-,\text{diff}}^{(0i)} = -\a_{(0i)}\frac{32G_F^2\s}{3\pi^2} D^2 E_\n^6(1 + \OO(\e)) = \OO(\e^6) ,
\ee
\be
\label{DA15} j_{e^-,\text{diff}}^{\parallel,(ij)} = \a_{(ij)}\frac{64G_F^2\s}{5\pi^2} D^4 E_\n^8(1 + \OO(\e)) = \OO(\e^8) ,
\ee
\be
\nn \a_{(00)}\in \left(\frac{4}{3},A\right) \sp \a_{(0i)}\in \left(\frac{4}{5},\frac{A^3}{6}\right) \sp \a_{(ij)}\in \left(\frac{32}{21},\frac{A^5}{10}\right) \, .
\ee

We now discuss the contribution from the second part of the integral in \eqref{DA11}, where $\omega \gg Dk^2$. Since this contribution includes essentially the region where $\omega$ and $k$ are of the same order $\omega \sim k = \OO(\e)$, we  label it with the subscript ``lin". The scaling of the transverse part is again easily derived
\be
\label{DA9b} j_{e^-,\text{lin}}^\perp = \OO(\e^4) \, .
\ee
For the longitudinal part, the integrals are computed by neglecting $D^2k^4$ in the denominator of the correlator, which results in integrands that are independent of the diffusion constant $D$. As for the transverse part, the hydrodynamic scaling of the $(0i)$ \eqref{DA7} and $(ij)$ \eqref{DA8} components are then easily derived to be
\be
\label{DA16} j_{e^-,\text{lin}}^{(0i)} = \OO(\e^4)\, ,
\ee
\be
\label{DA17} j_{e^-,\text{lin}}^{\parallel,(ij)} = \OO(\e^4)\, .
\ee
The (00) component is somewhat more subtle since the integrand contains a term that goes as $\omega^{-1}$. This implies that the time-time component contains a term of order $\OO(\e^4\log{(\e^2)})$
\be
\label{DA19} j_{e^-,\text{lin}}^{(00)} = \frac{8G_F^2\s}{3\pi^2}E_\n^4 \log{\left(\frac{D(\m_\n - E_\n)}{D^2 E_\n^2}\right)} + \OO(\e^4) \, .
\ee
As long as $\m_\n - E_\n$ is much larger than $\OO(\e^2)$, this term dominates all the other contributions to the neutrino emissivity. When $E_\n$ is so close to the neutrino chemical potential that $\m_\n - E_\n$ is smaller than $\OO(\e^2)$, the integral includes only the diffusive part in \eqref{DA9} and \eqref{DA13}-\eqref{DA15}. Since in both cases the time-time component dominates, the conclusion of this analysis is that the diffusive approximation is valid in the hydrodynamic limit.

Note that there is one exception to this argument which happens in the limit where $E_\n$ goes to zero. In this limit, the only remaining scale is the neutrino chemical potential, such that all contributions \eqref{DA5}-\eqref{DA8} behave like $\m_\n^4$. This means that the diffusive approximation does not apply when $E_\n \ll \m_\n$.

\subsection{Approximate expressions for the radiative coefficients}

In this subsection, we take the hydrodynamic limit \eqref{DA1} to derive approximate expressions for the neutrino radiative coefficients. We  consider the degenerate limit of the distribution functions \eqref{nB_hD} and \eqref{nehD}, and assume in a first time that $|\m_\n-E_\n| \gg \OO(\e^2)$.

According to the analysis of the previous subsection, the radiative coefficients are dominated by the log term coming from the time-time component  of the 2-point function. For the opacities we obtain
\be
\label{DA20b} \kappa_{e^-}(E_\n) = \frac{8G_F^2\s}{3\pi^2}E_\n^4 \log{\left(\frac{D|\m_\n - E_\n|}{D^2 E_\n^2}\right)} + \OO(\e^4) \, ,
\ee
\be
\label{DA21b} \bar{\kappa}_{e^-}(E_\n) = \frac{8G_F^2\s}{3\pi^2}E_\n^4 \log{\left(\frac{D(\m_\n + E_\n)}{D^2 E_\n^2}\right)} + \OO(\e^4) \, .
\ee
\eqref{DA20b} is valid as long as $|\mu_\n - E_\n|$ is much larger than $\OO(\e^2)$. In particular, instead of diverging, the
opacity goes to zero at $E_\n = \m_\n$ in the degenerate limit.

From Table \ref{tab:cEn}, the anti-neutrino opacity also receives a contribution from the positronic processes. The latter is obtained from \eqref{DA21b} by replacing $\m_\n + E_\n$ by $\m_\n+E_\n - \m_e = E_\n$
\be
\label{DA21d} \bar{\ka}_{e^+}(E_\n) = \frac{8G_F^2\s}{3\pi^2}E_\n^4 \log{\left(D E_\n\right)} + \OO(\e^4) \, .
\ee

Note that the leading order log term in \eqref{DA20b}-\eqref{DA21d} vanishes when $E_\n$ goes to zero, such that the opacities become of order $\OO(\e^4)$ in this limit. The expressions of the opacities at $E_\n = 0$ are of particular interest, since they set the typical opacity scale at a given value of the baryonic density $n_B$. The latter can easily be computed from \eqref{DA5}-\eqref{DA8}
\be
\label{DA22} \ka_{e,0}\equiv \kappa_{e^-}(0) = \bar{\kappa}_{e^-}(0) = \frac{G_F^2\s}{\pi^2}\m_\n^4 + \OO(\e^5) \, .
\ee
Substituting the expressions for the conductivity $\s$ \eqref{sf}, and the neutrino chemical potential \eqref{A1}, we obtain the dependence of the zero-energy opacities on $\m$ and the parameters of the bulk action $(M\ell)$ and $w_0$
\be
\label{DA23} \kappa_{e,0}(n_B) = \frac{G_F^2|M_{ud}|^2}{2304} \left(3\pi^4\right)^{\frac{1}{6}} N_c^{\frac{1}{2}} (M\ell)^7w_0^{25/3}\big(\mu(n_B)\big)^5 + \OO(\e^4) \, ,
\ee
which is valid in the degenerate and hydrodynamic limit.

We are now interested in the regime where $|\m_\n - E_\n|$ is smaller than $\OO(\e^2)$. When $|\m_\n - E_\n|$ goes to zero, both the neutrino emissivity and absorption calculated with the degenerate limit of the distribution functions go to zero. This implies the existence of a dip in the log of the neutrino opacity at $E_\n = \m_\n$, which is clearly visible on figure \ref{Fig:op12}. Here we would like to understand what are the typical width and depth of this dip. As soon as $|\m_\n - E_\n|$ becomes of order $\OO(\e^2)$, the integral over energies that appears in the neutrino emissivity \eqref{DA4} contains only the first part in \eqref{DA11}, where $\omega$ and $Dk^2$ are of the same order. Then, for $T \ll \m_\n - E_\n = \OO(\e^2)$, the neutrino opacity is bounded from above by
\be
\label{DA25} \kappa_{e^-}(E_\n) < \frac{G_F^2\s}{D\pi^2} \int_0^{|\m_\n - E_\n|}\intd \omega (\omega + E_\n)^2 = \frac{G_F^2\s}{D \pi^2}\m_\n^2 |\m_\n - E_\n|\big(1+\OO(\e)\big) \, .
\ee
This becomes much smaller than the leading contribution from the diffusion peak \eqref{DA13} when
\be
\label{DA26} \frac{1}{D}\m_\n^2|\m_\n-E_\n| \ll E_\n^4 \iff |\m_\n-E_\n| \ll D \m_\n^2 .
\ee
Replacing the diffusion constant $D$ by its expression $D=r_H/2$, we find that the typical width of the dip in opacity at $E_\n = \m_\n$ is given by
\be
\label{DA27} \D E_\n = \sqrt{\frac{3N_c}{w_0^2}}\frac{\m_\n^2}{\m}  \, .
\ee
The depth of the dip is controlled by the value of the opacity at $E_\n = \m_\n$. At zero temperature, the opacity will be exactly zero at $E_\m = \m_\n$, and the depth of the dip infinite. At finite temperature, the finite value of the opacity comes from the order $\OO(T/\m)^2$ corrections in \eqref{nB_hD} and \eqref{nehD}. The latter are evaluated as derivatives in energy at the point $E_\n = \m_\n$
\be
\label{DA28} \kappa_{e^{-}}(\m_\n) = \frac{2G_F^2\s}{3}T^2\!\int_0^\pi\! \intd\theta \sin{\theta}(1+\cos{\theta})\partial_\omega\!\left[\omega(\omega+\m_\n)^2\frac{k^2(\omega,\theta)}{\omega^2 + D^2k^4(\omega,\theta)}\right]\!\!\bigg|_{\omega = 0}\!\!\!\!\!\!\!\!(1+\OO(\e)) .
\ee

The angular integral in \eqref{DA28} is singular, which translates the appearance of a non-analytic behavior in the temperature $\sim T^2\log T$. The latter can be traced back to the divergence of the retarded 2-point function at $\theta = 0$, which is due to the forward scattering of soft electrons at $\m_e = \m_\n$. As expected, the divergence is regularized when taking into account that the electron mass $m_e$ is finite
\be
\label{DA29} \kappa_{e^{-}}(\m_\n) = \frac{8G_F^2\s}{3}T^2 r_H^{-2} \left[4\log{\left(\frac{2\m_\n}{m_e}\right)} - 1\right](1+\OO(\e)+\OO(m_e/\m_\n)^2\big) \, .
\ee

\newpage

\end{document}